\documentclass[fleqn,usenatbib]{mnras}

\usepackage[T1]{fontenc}
\usepackage{ae,aecompl}

\usepackage{longtable}

\usepackage{graphicx}	
\usepackage{amsmath}	
\usepackage{amssymb}	
\usepackage{xcolor}

\newcommand{\nubar}{\tilde\nu}

\def\ec{$\eta$~Car}

\def\hst{{\it HST}}
\def\stis{{\it STIS}}

\def\vlt{{\it VLT}}
\def\herschel{{\it Herschel Space Observatory}}
\def\hso{{\it Herschel}}
\def\iso{{\it ISO}}
\def\kms{km~s$^{-1}$}

\def\spire{{SPIRE}}
\def\pacs{{PACS}}
\def\hifi{{HIFI}}
\def\alma{{\it ALMA}}
\def\APEX{{\it APEX}}

\newcommand{\Lsun}{\hbox{$L_\odot$}}
\newcommand{\Msun}{\hbox{$M_\odot$}}

\newcommand{\Jbar}{{\bar J}}

\title[Far-IR Spectrum of $\eta$\ Car \& Homunculus]{Eta Carinae \& the Homunculus: \\ Far Infrared/Sub-millimeter Spectral Lines \\detected with the \herschel}

\author[T. R GULL et al.]{T. R. Gull,$^{1}$\thanks{E-mail: TedGull@gmail.com }
P. W. Morris,$^{2}$
J. H. Black,$^{3}$
K. E. Nielsen,$^{4}$\newauthor
M. J. Barlow,$^{5}$
P. Royer, $^{6}$
and  B. M. Swinyard $^{5,7,8}$
\\
$^{1}$Code 660, Astrophysics Science Division, Goddard Space Flight Center, Greenbelt, MD 20771 USA\\
$^{2}$California Institute of Technology, IPAC M/C 100-22, Pasadena, CA 91125 USA\\
$^{3}$Department of Space, Earth \& Environment, Chalmers University of Technology, Onsala Space Observatory,\\ SE-43992 Onsala, SWEDEN\\
$^{4}$Department of Physics, IACS, Catholic University of America, Washington, DC 20064 USA\\
$^{5}$Dept of Physics \& Astronomy, University College London, Gower St, London WC1E 6BT, UK\\
$^{6}$K. U. Leuven, Inst of Astronomy, Celestijnenlaan 200 D, BE 3001 Leuven, Belgium\\
$^{7}$ Space Science \& Technology Department, Rutherford Appleton Laboratory, Chilton, Didcot, Oxon, UK\\
$^{8}$ Deceased\\
}

\date{Accepted 2020 October 2. Received 2020 September 30; in original form 2020 March 19.}

\pubyear{2018}

\begin{document}
\label{firstpage}
\pagerange{\pageref{firstpage}--\pageref{lastpage}}
\maketitle

\begin{abstract}

The evolved massive binary star $\eta$ Carinae underwent eruptive mass loss events that formed the complex bi-polar ``Homunculus'' nebula harboring tens of solar masses of unusually  nitrogen-rich gas and dust.  Despite expectations for the presence of a significant molecular component to the gas,  detections have been observationally challenged by limited access to the far-infrared and the intense thermal continuum.  A spectral survey of the atomic and rotational molecular transitions was carried out with the {\it Herschel Space Observatory}, revealing a rich spectrum of broad emission lines originating in the ejecta. Velocity profiles of selected \pacs\ lines correlate well with known substructures:  \ion{H}{i} in the central core; NH and weak [\ion{C}{ii}] within  the Homunculus; and [\ion{N}{ii}] emissions in fast-moving structures external to the Homunculus.  We have identified transitions from [\ion{O}{I}],  \ion{H}{I}, and 18 separate light C- and O-bearing molecules including CO, CH, CH$^+$, and OH, and a wide set of N-bearing molecules, NH, NH$^+$, N$_2$H$^+$, NH$_2$, NH$_3$, HCN, HNC, CN, and N$_2$H$^+$.  Half of these are new detections unprecedented for any early-type massive star environment.  A very low ratio [$^{12}$C/$^{13}$C] $\leq$ 4 is estimated from five molecules and their isotopologues.  We demonstrate that non-LTE effects due to the strong continuum are significant.  Abundance patterns are consistent with line formation in regions of carbon and oxygen depletions with nitrogen enhancements, reflecting an evolved state of the erupting star with efficient transport of CNO-processed material to the outer layers.   The results offer many opportunities for further observational and theoretical investigations of the molecular chemistry under extreme physical and chemical conditions around massive stars in their final stages of evolution.

\end{abstract}




\section{Introduction}

\label {Intro}Massive stars, despite their large reservoirs of hydrogen, exist for rather brief periods of time $-$ millions of years $-$ due to prodigious consumption of nuclear fuel necessary to counteract the huge gravitational forces\footnote{Observations discussed herein were recorded with the \herschel, an ESA space observatory with science instruments provided by European-led Principal Investigator consortia and with important participation from NASA. Herein, called \hso}. The transition from main sequence to the end stage, whether it be a supernova,  a black hole or other,  lasts for even shorter intervals $-$ tens to hundreds of thousands of years. These massive stars  undergo a series of evolutionary  changes, exhibiting specific spectroscopic characteristics \citep{Groh14}. Instabilities occur leading to eruptive ejections that are not easily predicted. The very nearby  \ec\ (distance 2350 pcs; \cite{Smith06}), which underwent a massive ejection in the 1840s and a lesser ejection in the 1890s, may have undergone such a transitional phase to its current luminous blue variable (LBV) in a binary system.

The 1840s event of \ec, when its apparent magnitude brightened to rival that of Sirius, drew the attention of many astronomers \citep{Frew04}. Indeed the energetics approached that of supernovae, but the erupting system survived \citep{Davidson97a}. Modern supernova surveys detect similar events, the so-called supernova imposters, many of which precede actual supernova events by months to  a few years (see for example \cite{Mauerhan13}. However nearly two centuries have passed since the Great Eruption  and \ec\ has not yet produced a supernova. \cite{Smith18c}, based upon analysis of light echoes from the 1840s event, suggest that the Great Eruption was the result of a stellar merger in a triple system leading to the current binary.

\cite{Damineli96} deduced a 5.5-year periodicity defined by an extended high-ionization state and a months-long, low-ionization state of nebular and wind lines that led to the determination that \ec\  is a binary.  Detailed studies of this massive binary, its ejecta and modeling thereof have characterized the current state of the system, but have brought little understanding as to why a supernova has not yet occurred. Indeed, recent studies of the last three periastron passages suggest that the interacting winds of the massive binary were quite stable from 2000 to 2015 \citep{Teodoro16}. Such is confirmed by the recent light curve study by \cite{Damineli19}, who demonstrated that the apparent brightening (~0.1 mag/yr) is due to dissipation of material in line of sight, but close to \ec. 

Spectroscopic studies of the  ejecta at visible and ultraviolet wavelengths revealed that nitrogen is very overabundant with corresponding, greatly-decreased abundances of carbon and oxygen \citep{Davidson82a, Davidson86a}. Follow-up studies of abundances in the Weigelt blobs \citep{Weigelt86}, ionized clumps of gas within 1000 AU of \ec\ that appear to have been ejected in the late nineteenth century \citep{Smith98},
 confirm the overabundance of nitrogen and depletion of carbon and oxygen extending the relative depletion to as much as 50 to 100-fold \citep{Verner05a}. Stellar evolutionary models of massive stars indicate that the ejecting star was in transition from hydrogen-burning, moving towards helium-burning, with mixing between the nuclear core and the extended outer shells, characteristic of stars more massive than 60 \Msun \citep{Ekstrom12}.

The luminosity of \ec, based upon early infrared measurements of the Homunculus \citep{Westphal69} and analysis by \cite{Cox95}, was estimated to be 5$\times$10$^6$\Lsun. Analysis, including continuum measures by {\it Infrared Space Observatory} (\iso), \hso\  and the {\it Atacama Large Millimeter Array} (\alma), suggested a currently lower luminosity of 3.0$\times$10$^6$\Lsun\ \citep{Morris17}. A much more complete study by \cite{Mehner19} recently demonstrated that 1) the bolometric luminosity has remained stable over the past five decades, but 2) short-term variations, likely related to the 5.54 year binary period, could be present.

%
Current  estimates  of the binary  members are that the primary, \ec~A, is a Luminous Blue Variable (LBV), exceeding 90 \Msun\ \citep{Hillier01} with a massive wind (8.5x10$^{-4}$ \Msun yr$^{-1}$ expanding at 420 \kms \citep{Groh12, Madura13}) and the hot secondary, \ec~B, is an O or WN star, ranging between 30 to 50 \Msun\  \citep{Verner05a, Mehner10} with a less massive, but faster wind (1x10$^{-5}\Msun$yr$^{-1}$ expanding at 3000 kms$^{-1}$; \citep{Pittard02}). The least well known  of these parameters is the  total mass of the binary which is inferred from the total luminosity of the Homunculus. A recent analysis of the dynamics of the binary and interacting winds suggests a total mass exceeding 100 \Msun, but is still limited by assumptions about the orbital inclination and relative luminosities of the two stars \citep{Grant20}.

\begin{figure*}
   \centering
     \includegraphics[width=17.5cm,angle=0]{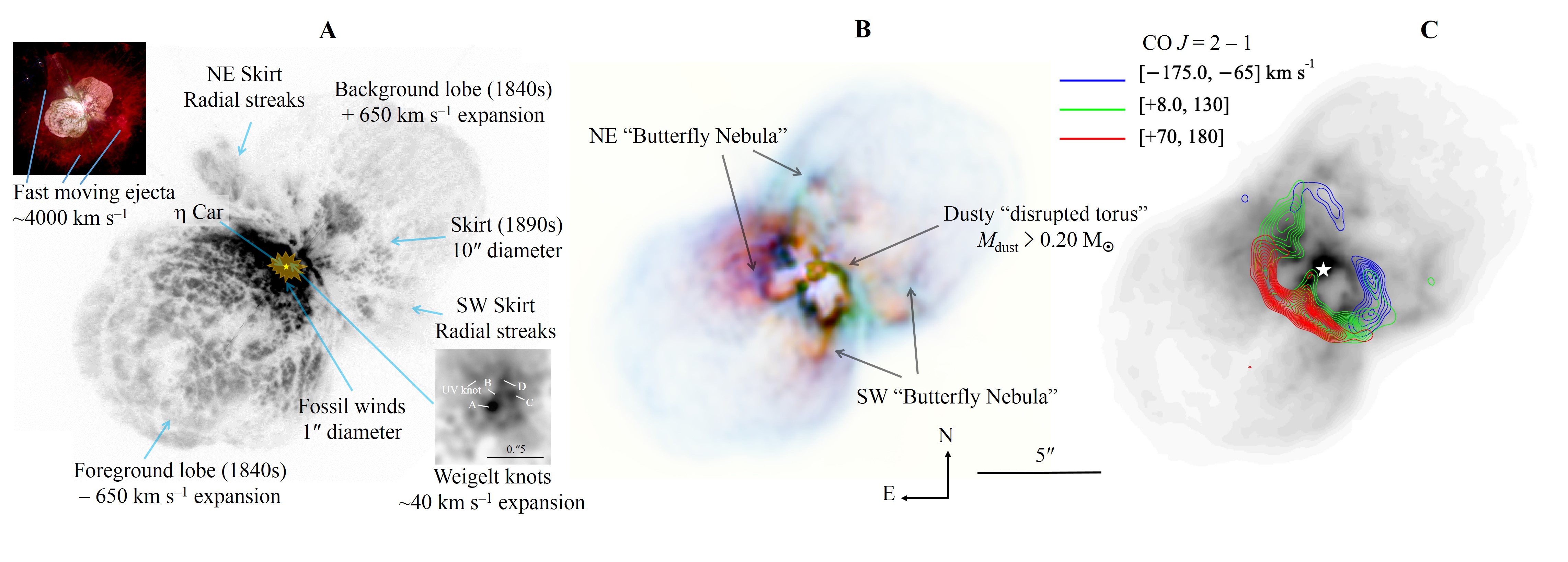}
    \caption{Ejecta structures surrounding \ec\ at visible, infrared, and sub-mm wavelengths on reverse relative intensity scales with major relevant structures identified.  The bipolar Homunculus nebula is tilted out of the sky plane by 45$^\circ$, and subtends 10\arcsec $\times$ 20\arcsec\ in the 2011 epoch. The SE and NW lobes are expanding at $-$600 and $+$600 \kms, respectively.  Much fainter, outer, irregular structures, expand at space velocities approaching 6000 \kms \citep{smith08b,Kiminki16}. (A) Optical \hst/ACS-HRC F660N image \protect\citep{Dorland07}, showing primarily H$\alpha$, [\ion{N} {ii}], and continuum radiation.  The stellar fossil winds and the so-called 'Weigelt knots' shown in the lower right inset of the central $1\farcs3\times 1\farcs3$\ region are described by \protect\cite{Gull16} and references therein. The outer, fast moving ejecta are accentuated in the upper left \hst\ image (credit: NASA, ESA, and the SM4 ERO Team).  (B) Composite \vlt/VISIR imaging of the 13 $\mu$m (red) and 11.2 $\mu$m (green) dust band emission, and 8.98 $\mu$m [\ion{Ar}{iii}] emission (blue), from P. Morris et al. (in prep).  The "Butterfly Nebula" refers to the morphology of the IR emission described by \protect\cite{Chesneau05a}.   (C) Contours of CO $J = $2-1 emission  observed with \alma\ overlaid onto a \vlt/VISIR 7.9 $\mu$m image, from \protect\cite{Morris20}. The emission indicated in red contours accounts for 65\%\ of the total radiated power. (red contours indicate red-shifted 0 to 140 \kms; green contours indicate -40 to 0 \kms;  blue contours indicate -40 to -140 \kms.)}\label{nebular}
\end{figure*}

Our goals with \hso\ observations were to characterize the massive ejecta in the far-IR spectral region by molecular and atomic emission lines, to compare their spatial distributions with respect to the continuum and  to inter-compare spatial and velocity properties. The angular resolution (10 to 30\arcsec) and the very different velocity resolutions of the instruments provide a first, coarse look at a complex system of ejecta, but further studies with higher spatial and spectral resolutions are necessary to quantify contributions of various regions.  

 As a guide to the reader, we display in Figure \ref{nebular} imagery in the visible and infrared spectral regions and list the following structures previously discovered by diverse studies in the visible, ultraviolet and radio spectral regions, all of which might contribute far-IR emissions:
\begin {itemize}
	
	\item Surrounding the massive binary, \ec, which is cloaked by  x-ray generating, interacting wind structures, are hemispherical shells of fossil winds  that extend spatially $\approx$1\arcsec\ diameter with velocities ranging from $-$470 to $+$470 \kms) that persist for at least 3 binary cycles \citep{Gull09, Teodoro13, Gull16}. 
	\item Within the extended fossil structures, at angular distances, 0\farcs2 to 0\farcs3  from \ec, the very slowly moving ($-$40 \kms), dense, partially ionized Weigelt clumps have  proper motions consistent with an origin in the 1890s event \citep{Weigelt12}. Multiple spectral studies demonstrated an excess of nitrogen with very little carbon and oxygen, implying they are highly-processed material  \citep{Verner05a, Mehner10}. Further studies by \cite{Teodoro20} indicate these clumps are bright surfaces of more extensive dusty clumps that  are infrared-bright.
	\item The radio H~II region, several arcseconds in size, is offset a few arcseconds  to the northwest of \ec. It grows and shrinks in synchronization with the 5.5-year binary period \citep{Duncan03}).  
	\item The Butterfly Nebula, a cavity bounded in the near infrared by continuum and line emission $\approx$3 to 4\arcsec\ in diameter, surrounds \ec, the interacting winds and the H~II region \citep{Chesneau05a}.  \cite{Artigau11} found the ionized material, associable with the boundaries of the Butterfly Nebula that expands with velocities ranging from  $-$50 to $+$50 \kms. The Strontium Filament, discovered by \cite{Zethson01a} is one of the brighter rims of this structure moving at -100 \kms. 
	\item A molecular shell  of comparable size was found to be expanding at velocities from $-$100 to $+$100 \kms. \citet{Loinard12} identified a 'sub-arcsecond' molecular emitting region, in observations with the 12 m \APEX, using arguments based upon optical depths and  aperture  constraints.  On the other hand, \cite{Morris17} found the size to be greater than 3\arcsec\ based on self-consistent modeling of the combined \APEX\ (lower excitation)  and \hso/HIFI (higher excitation) CO spectra.  More recently, \cite{Smith18a} found a partial torus in CO (2-1) centered on \ec, looping about 2\arcsec\ to the east. 
	\item The Little Homunculus, an ionized, internal, bipolar structure 4\arcsec\ in size, expands at 200 to 300 \kms\  \citep{Ishibashi03}. 	
	\item A disrupted torus-like equatorial structure, $\approx$ 5\arcsec $\times$ 8\arcsec, observed in continuum mid-infrared  \citep{Morris99, Polomski99, Smith03a}. This structure, internally bounded by the Butterfly cavity, lies between the two Little Homunculus lobes.
	\item The Homunculus (the outer bipolar structures, 10\arcsec$\times$20\arcsec\ in size, as seen by scattered starlight from dust) is mapped in NIR emission lines of H$_2$ and Fe$^+$  expanding at velocities from -600 to +600 \kms \citep{Smith02, Steffen14}, plus the extended skirt structures noted by \cite{Walborn78} to the NE.
	
	\item More distant, rapidly-moving bullets, seen in H$\alpha$\ and [N~II], an arcminute in angular extent, move at  space velocities up to 5000 to 6000 \kms\ \citep{Weis97, smith08b, Kiminki16,Mehner16}. 
	\item  Not ejecta but noticeable as background structures, the extended (2\degr\ diameter), ionized Carina Nebula contributes narrow emission lines of [\ion{C}{ii}], [\ion{O}{ii}] especially notable in the 70 to 200 $\mu$m spectral region of \pacs.
	\end{itemize}

Section \ref{obscal} summarizes the observations accomplished in this program along with discussion on calibrations as this source proved to be one of the brightest objects observed with \hso. Subsection \ref{PACS_obs} describes the observations and calibration of the \pacs\ instrument and includes a list of emission lines identified with the central spaxel of the \pacs. Subsection \ref{SPIRE_obs} describes the observations and calibration of the \spire\ data. Line identifications and spectral plots are included in Appendix \ref{plotsid}.

Section  \ref{results}
presents data analysis of the \pacs\ observations (subsections \ref{back} through \ref{NH}).  Preliminary modeling  of the \spire\ spectra  is in subsections    \ref{molecules}, \ref{Smodel} and \ref{ions} which address identified lines leading to abundance estimates.  We found it useful to include spectral observations of \ec\ done with the \iso/Short Wave Spectrograph {\it SWS} in  subsection \ref{swsH}  that provided information for improved modeling  of the hydrogen recombination spectrum.  
 A summary and conclusion section follows \ref{concl}. Appendix \ref{plotsid} displays the \spire\ spectrum and lists the line identifications.
	
\section{Observations and Calibrations}\label{obscal}
\begin{table*}
	\caption{Log of \hso\ Observations}
	\label{Tbl-1}\begin{tabular}{llllll}\hline
Instrument mode & Date &Obs ID& Phase$^a$ & Spectral Interval/Line  & Comments\\
\hline
 \pacs\ Spectral Scan$^{b}$ & 2011-08-06 & 1342225814 - 822&0.463 & B3A (R1), B2B (R1), B2A (R1)$^{c}$& ON, OFF$^{d}$, OFFCLOSE$^{e}$\\
 \pacs\ Spectral Scan$^{f}$ & 2012-01-26 & 1342238354 - 359&0.549 & B3A (R1), B2B (R1), B2A (R1)& ON; 1,1; 1,2; 2,1; 2,2$^{g}$\\
 \spire\ Spectrometer$^{h}$ & 2011-09-16 &1342228699&0.484 & & 15 Bright$^{h}$\\
  \spire\ Spectrometer$^{h}$ & 2011-09-16 &1342228700&0.484 & &25 Nominal$^{h}$\\
\hline
\end{tabular}
\\
\begin{itemize}
\item[$^a$]Phase refers to binary phase based upon both disappearance of He I emission and X-ray drop with periastron passage on MJD 2456874.4 $\pm$ 1.3 days and orbital period of 2022.7 $\pm$ 0.3 days  \citep{Teodoro16}.
\item[$^{b}$]The 2011-08-06 visit had the \pacs\ array oriented at PA=147.9$\degr$. Nominal angular resolution ranged from 9\arcsec\ at 57$\mu$m to 13\arcsec\ at 180$\mu$m \citep{Bocchio16}.
\item[$^{c}$]For each B scan, a parallel scan was done with  the R grating.\\
\item[$^{d}$]OFF corresponds to offset 15.8\arcmin east  and 9.2\arcmin north.\\
\item[$^{e}$]OFFCLOSE corresponds to offset 53\arcsec east and 46\arcsec north.\\
\item[$^{f}$]The 2012-01-26 visit had the \pacs\ array oriented at PA= 335.4$\degr$ (see Figure \ref{PACS_Fig1}).\\
\item[$^{g}$]The four observations denoted 1,1; 1,2; 2,1; 2,2 are from a set of raster scans recorded with right ascension and declination offsets 4\farcs5 from \ec.\\ 
\item[$^{h}$]Two sets of scans were performed:  25  FTS scans in "nominal" mode were accomplished with a total integration time of 3330 seconds; 15 additional scans in "bright-source" mode of reduced sensitivity were added for 1998 seconds. As no saturation effects were noted, the nominal scans with higher signal to noise ratio were reduced and presented in this discussion. The nominal \spire\ PSF is 18\arcsec\ at 190$\mu$m and expands to 37\arcsec\ at 670$\mu$m \citep{Swinyard10}.
\end{itemize}
\end{table*}

The observations presented in this paper were acquired with the two imaging spectrometers on the \herschel\ \citep{Pilbratt10} in the Open Time program OT1\_tgull\_3 between August 6, 2011, and January 26, 2012, corresponding to orbit phases $\Delta\Phi$  = 0.48 - 0.55 on {\ec}'s 5.54-year period; i.e.,  during apastron passage with a mean separation of $\approx$16 A.U. of the stellar companions.  The grating spectrometer in the Photodetector Array Camera and Spectrometer (PACS; \citealt{Poglitsch10}) provided wavelength coverage of 55 - 190 $\mu$m (1578 - 5454 GHz), employing two Ge:Ga photoconductor arrays.  The Fourier Transform Spectrometer (FTS) of the Spectral and Photometric Imaging REceiver (SPIRE; \citealt{Griffin10}) covered 194 - 671 $\mu$m (447 - 1550 GHz) with two bolometer arrays.  

For our investigation into the behavior of the far-IR/sub-millimeter \ion{H}{I} Rydberg transitions observed with \pacs\ and \spire, we include a mid-IR spectrum obtained with the \iso\ \citep{Kessler96}) Short Wavelength Spectrometer (SWS; \citealt{deGraauw96}).  These data, obtained through large apertures in January 1996 in the S01 grating mode at  spectral resolutions $R \approx$ 1200-1500 over 2.4$-$45.2 $\mu$m (07100250), have been presented by \cite{Morris99, Morris17}, who discuss their calibrations in detail.   Only the 2.4$-$8.0 $\mu$m range covering \ion{H}{I} Brackett, Pfund, Humphreys, and upper state $n$ = 7 series is relevant to the present study.  We also include a previously unpublished high resolution ($R \approx$ 3500) grating  scan of the \ion{H}{I} Br $\alpha$ 4.05 $\mu$m line, obtained in the program {\tt{WINDS1}} (29700104, J. Martin-Pintado, P.I.).  Data processing methods are the same as those described by \cite{Morris17}.

As part of the our scientific rationale, certain frequency ranges were targeted for follow-up, or were augmented by existing calibration observations of \ec, with the high spectral resolution Heterodyne Instrument for the Far Infrared (HIFI; \citealt{Roelfsema12}).  The 490 - 1900 GHz range of HIFI provided good frequency overlap with SPIRE, extending to higher excitation molecular transitions and [\ion{C}{ii}] at the long wavelength end (red band) of PACS.  HIFI observations of $^{12}$CO and $^{13}$CO, and the 450 - 1900 GHz continuum as observed with HIFI and SPIRE have been published in a study of the dust composition and mass in {\ec} by \cite{Morris17}.   
\\

\subsection{ \pacs\ Observations and Calibration}\label{PACS_obs}

\begin{figure*}
    \centering
     \includegraphics[width=18 cm,angle=0]{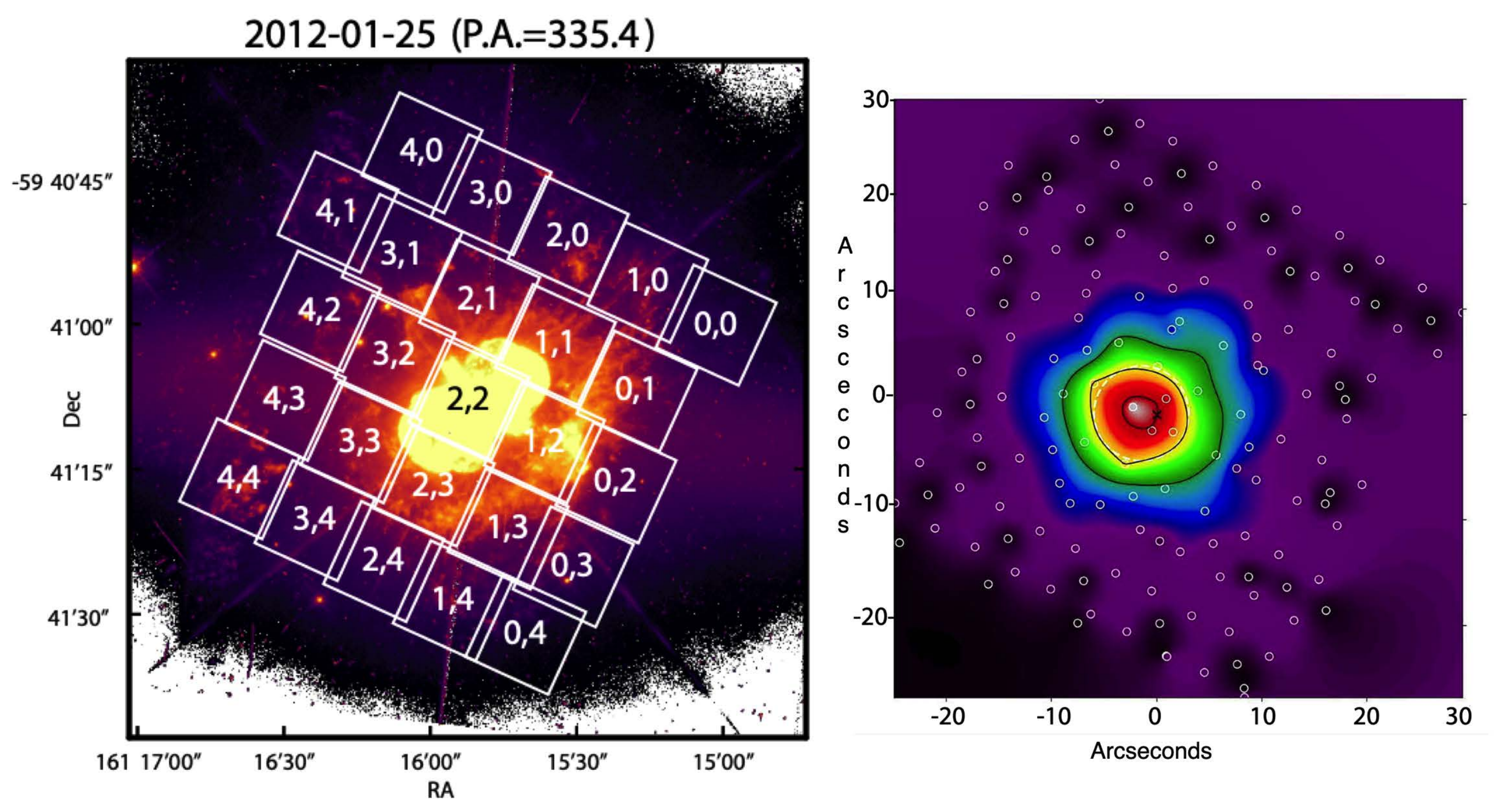}
    \caption{Placement of \pacs\ spaxels for the visit executed 2012 January 25 (left), and a contour map of the continuum at 85 $\mu$m (right).  The \pacs\ footprint is overlaid on an \hst\ image of \ec\ stretched in intensity.  \label{PACS_Fig1} Contour levels are 0.85, 0.5 and 0.25 of the peak value. The  \pacs\ FWHM beam at 85 $\mu$m is 9$\arcsec$ in diameter.  The circles on this map depict the sampling positions of  mappings accomplished during visit 2 with  \pacs. \label{Cont}}

\end{figure*}

The \pacs\ instrument utilized a 5$\times $5 array of spatial pixels (spaxels). Each of the spaxels subtends a 9$\farcs$4 $\times\ $9$\farcs$4 square in an irregularly-spaced array on the sky as displayed in Figure \ref{PACS_Fig1} projected on \ec\ and the Homunculus.  The  full width at half maximum (FWHM) of the \pacs\ beam at 85 $\mu$m is 9$\arcsec$.  Spectral resolutions $\lambda/\Delta\lambda$ vary between 1500 (200 \kms) in the grating red band R1 (103 - 190 $\mu$m) to 2250 (133 \kms) in the two blue bands B2A (55 - 72 $\mu$m) and B2B (70 - 95 $\mu$m), and 4000 (75 \kms) in a third lower sensitivity blue band B3A (55 - 70 $\mu$m).   For comparison, the expansion of the bipolar lobes of the Homunculus is $\pm$600 \kms, while CO associated with the dusty disrupted torus-like structure (see Fig.~\ref{nebular}) has velocities to roughly $\pm$250 \kms (\citealt{Loinard12, Morris17, Smith18a}).

The \pacs\ observations were conducted in two visits designed to address potential issues with saturation and hysteresis of the detectors, and to determine the spatial extent of any detected emissions.  The first visit, executed August 6, 2011, initially observed an OFF position 20\arcmin\  northeast of \ec, then at an OFFCLOSE position 1\arcmin\ northeast of \ec, and finally an ON position centered on \ec\ (see Table \ref{Tbl-1} for the specific offset positions from \ec). The second visit, executed on January 26, 2012, recorded grating scans centered on \ec, then repeated withfour positions in a raster format using 4\farcs5 (half-spaxel dimension) offsets from \ec. 

\begin{table*}

\caption{Line Identifications from the \pacs\  spectra using the averaged central spaxel \label{PACS-T} }

\begin{tabular}{lrrrrrrl}
Identification & Measured     & Wave    & Rest             & Rad Vel$^{c}$ & FWHM$ $ & Flux$^{d}$   & Comments \\
                   & Wave$^{a}$ & number & Wave$^{b}$ &                       &                      & (10$^{-15}$ &    \\
& ($\mu$m) & (cm$^{-1}$)&($\mu$m)  &(km/s) & (km/s)&Wm$^{-2}$) & \\

\hline
$[$\ion{N}{iii}] $^2$P$_{3/2}$--$^2$P$_{1/2}$              & 57.34 & 174.41 & 57.32 & $+$90 & 360 & 22.30 &  Continuum leakage + fringes\\
NH $N$ = 5--4, multiple $J$                                            & 61.52 & 162.54 & 61.51 & $+$85 & 230 & 7.60 & Blend $J$ = 6--5, 5--4, 4--3.\\
$[$\ion{O}{i}] $^3$P$_2$--$^3$P$_1$                           & 63.12 & 158.44 & 63.18 & $-$330 & 620 &  7.70 & Blended  \\
$[$\ion{O}{i}] $^3$P$_2$--$^3$P$_1$                           & 63.17 & 158.30 &63.18 & $-$60 & 125 &  6.57 & Blended \\
$[$\ion{O}{i}] $^3$P$_2$--$^3$P$_1$                           & 63.21 & 158.21 & 63.18 & $+$117 & 85 &  2.14 & Blended \\
$[$\ion{O}{i}] $^3$P$_2$--$^3$P$_1$                           & 63.27 & 158.05 & 63.18 & $+$426 & 410 &  5.90 & Blended \\
CH$_3$OH 7$_4^+$--8$_5^+$ E1  $vt=1-0$                   & 69.00 & 144.93 & 68.98 & $+$65 & 240 & 4.40 & Tentative \\
\ion{H}{i} 12--11                                                           & 69.06 & 144.80 & 69.07 & $-$50 & 210& 7.00 & \dotfill \\
NH $N$ = 4--3, multiple $J$                                            & 76.24 & 131.16 & 76.19 & $+$190 & 370 & 2.41 & Blend $J$ = 4--2, 4--4 \\
NH $N$ = 4--3,  multiple $J$                                           & 76.75 & 130.29 & 76.74 & $+$40 & 600 & 7.73 & Blend $J$ = 5--4, 4--3, 3--2, 3--4 \\
OH $N$ = 3$^-$--2$^+$, $J$ = 7/2--5/2                         & 84.42 & 118.47 & 84.42$^{e}$ & 0 & 125 & 1.34 & \dotfill \\
OH $N$ = 3$^+$--2$^-$, $J$ = 7/2--5/2                         & 84.60 & 118.20 & 84.60$^{e}$ & 0 & 270 & 1.70 & \dotfill \\
$[$\ion{Fe}{ii}] a$^6$D$_{3/2}$--a$^6$D$_{1/2}$         & 87.38 & 114.44 & 87.38  & $-$10 & 200 & 2.60 & \dotfill \\
$[$\ion{O}{iii}] $^3$P$_0$--$^3$P$_1$                          & 88.36 & 113.18 & 88.36 & 0 & NR & 6.18 & Background Carina Nebula\\ 
unidentified                                                                   & 88.65 & 112.79 & \dotfill & \dotfill & 185 & 4.40 & \dotfill \\
\ion{H}{i} 13--12                                                           & 88.75 & 112.68 & 88.75 & $-$10 & 240 & 5.37 & \dotfill \\
$[$\ion{Fe}{i}] a$^5$D$_1$--a$^5$D$_0$                       & 111.24 & 89.90 & 111.18 & $+$150 & 400 & 1.61 & Tentative; weak \\
\ion{H}{i} 14--13                                                           & 111.80 & 89.45 & 111.86 & $-$140 & 520 & 6.60 & \dotfill\\
OH $N$ = 2$^+$--1$^-$, $J$ = 5/2 - 3/2                        & 119.25 & 83.86 & 119.23 & $+$42 & 265 & 1.08 & Partial blend.  \\
OH $N$ = 2$^-$--1$^+$, $J$ = 5/2 - 3/2                        & 119.43 & 83.73 & 119.44 &$-$10 & 290 & 1.37 & Partial blend. \\
$[$\ion{N}{ii} ] $^3$P$_1$--$^3$P$_2$                          & 121.92 & 82.01 & 121.90 & $+$65 & 640 & 17.77 & Extended shell; see Fig.~\ref{NII}.\\
NH$_3$ 4$_1$ -- 3$_1$                                                  & 127.17 & 78.63 & 127.15 & $+$40 & 350 & 0.69 & Hyperfine blend \\
$^{15}$NH$_3$ 4$_3$ -- 3$_3$                                     &  127.48 & 78.44 & 127.54 & $-$140 & NR & 0.19 & Tentative; weak. \\
\ion{H}{i}  19--17                                                          & 132.12 & 75.69 & 132.11 & $+$3.0 & 300 & 0.56 & weak  \\

\ion{H}{i} 15--14                                                           & 138.57 & 72.16 & 138.65 & $ -$170 & 560 & 3.76 & \dotfill \\
$[$\ion{O}{i}] $^3$P$_1$--$^3$P$_0$                           & 145.59 & 68.69 &145.54 & $+$90 & 820 & 0.60 & weak \\
NH $N$ = 2--1, $J$ = 3--2                                             & 153.36 & 65.21& 153.34$^{f}$ & $+$25 & 190 & 1.27 & NH Complex 1\\
NH $N$ = 2--1, $J$ = 3--2                                              & 153.56 & 65.21& 153.34 & $+$450 & 210 & 0.59 & NH Complex 2\\
NH $N$ = 2--1, $J$ = 2--1                                              & 153.11 & 65.31 & 153.09 & $+$25 & 190 & 1.00 & NH Complex 1\\
NH $N$ = 2--1, $J$ = 2--1                                             & 153.33 & 65.31 & 153.09$^{f}$ & $+$450 & 210 & 0.37 & NH Complex 2\\                                

NH$_2$    3$_{22}$ -- 3$_{13}$                                    & 152.85 & 65.42 & 152.92 & $-$140 & NR & 0.18 & \dotfill \\                                           
$[$\ion{C}{ii}] $^2$P$_{1/2}$ -- $^2$P$_{3/2}$            & 157.74 & 63.40 & 157.74 & 0  & 180 & 0.98 & $\sim$40\% background. \\
$[$\ion{C}{ii}] $^2$P$_{1/2}$ -- $^2$P$_{3/2}$            & 157.52 & 63.37 & 157.74 & -420 & 450 & 0.28 & Broad component (blue)\\
$[$\ion{C}{ii}] $^2$P$_{1/2}$ -- $^2$P$_{3/2}$            & 157.52 & 63.37 & 157.74 & +400 & 390 & 0.33 & Broad component (red)\\
NH$_2$  3$_{13}$ -- 2$_{02}$                                       & 159.51 & 63.40 & 159.36 & $+$280 & \dotfill & 0.58 & Blend 159.36, 159.53 $\mu$m\\
NH$_2$  3$_{30}$ -- 2$_{21}$                                       & 169.22 & 59.09 & 169.37 & $-$270 & 120 & 0.70 & Blended with \ion{H}{i} \\
\ion{H}{i} 16-15                                                            & 169.39 & 59.03 & 169.41 & $-$36 & 260 & 0.95 &  Blended with NH$_2$ \\

\hline
\end{tabular}
\begin{itemize}
\item[$^{a}$]Measured wavelengths are accurate to $\pm$0.01$\mu$m (40\kms) from 60 to 100$\mu$m (order 2) and to $\pm$0.02$\mu$m (80\kms) from 100 to 
200$\mu$m (order 1). Spectral resolutions vary with wavelength in each band; see Sec.~\ref{PACS_obs} and the \pacs\ Observer's Manual  http://herschel.esac.esa.int/Docs/PACS/html/pacs\_om.html. 
\item[$^{b}$]Vacuum wavelengths for atomic lines are from the NIST Atomic Spectra Database 
(https://www.nist.gov/pml/atomic-spectra-database); molecular line wavelengths are from the Cologne Database for Molecular 
Spectroscopy \citep{Mueller05} and the molecular database at the Jet Propulsion Laboratory \citep{Pickett98}.
\item[$^{c}$]Radial velocities are measured in the Local Standard of Rest (LSR). NR indicates that the line is not resolved. 
\item[$^{d}$]Formal measurement uncertainties on line fluxes are 10\%; systematic underestimate up to 35\% for lines at $\lambda$ < 100 $\mu$m may be caused by saturation and hysteresis effects (see subsection \ref{PACS_obs}).
\item[$^{e}$]From Brown et al. (1982).
\item[$^{f}$]Measurements of the line centered at 153.34 $\mu$m are based on profile decomposition using the 2-component model shown in Fig.~\ref{NHFIT}. Complex 1 and 2 refer to the model components shown in the figure.
\end{itemize}
\end{table*}

\begin{figure}
\includegraphics[angle=0,width=8.4cm]{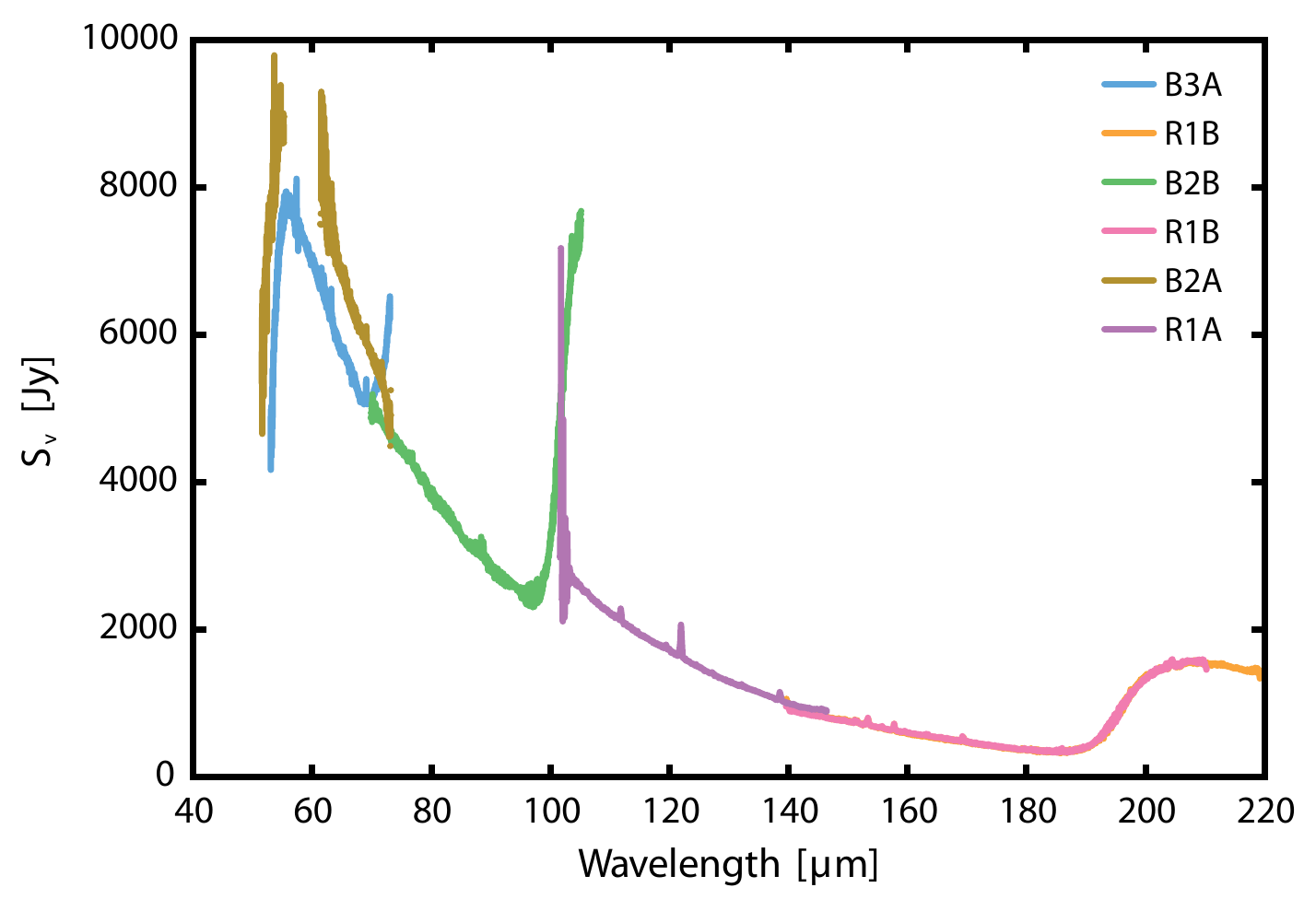}
\caption{The {SED} of \ec\ as measured by the central spaxel (labeled 2,2 in figure \ref{PACS_Fig1}).  Wavelength ranges for each grating order are given in Sec.~\ref{PACS_obs}.  \ec\ and its ejecta are the brightest extrasolar source at 25 $\mu$m. Hence short wavelength continuum dominates in this spectral region, inflicting saturation in B2A and filter leakage longward of 190 $\mu$m in the R2 spectral scan. Likewise a filter leak from third order into second order  is present in B2B from 95 to 105 $\mu$m and leakage from fourth order is present in B3A longward of 70 $\mu$m. The very small peaks against the strong continuum, as seen in the R1A and R1B spectral segments, are the stronger emission lines studied in this paper with \pacs. \label{spec_cen}}
\end{figure}

We summarize the saturation and spectral order leakage effects on the photometric calibrations as follows:

 \begin{itemize}
 \item Only B2A observations were saturated, however B3A data taken at higher spectral resolution over the same wavelength range were not. 
\item Each position was observed by 4 independent scans up and down in wavelength for each selected spectral region.  End portions of each scan are strongly contaminated by inter-order continuum leaks \footnote{See \pacs\ Users Manual https://www.cosmos.esa.int/documents/\\12133/996891} (see Figure \ref{spec_cen}).  We excluded B2A data from our analysis due to the leakage and saturation.   The line survey does not include 90 to 105 $\mu$m spectral interval due to continuum leakage.
\item Agreement of scans in each of B2B and B3A is generally within 5\%, and around 10\% in R1. In the interpolation between the longest B2B and shortest R1 wavelength, the maximum disagreement between these two bands is $\approx$ 30\%.  
\item  The level of agreement between scans in each band is likely dominated by detector dark current drifts arising from signal persistence. We estimate that measured line fluxes are uncertain by 20\%.  Variations in fluxes between scans, as would be expected for a semi-extended point source, were noticeable in the spaxels close to the centre spaxel, labeled 2,2 in Figure \ref{PACS_Fig1}, left, likely due to pointing drifts.  While the continuum flux map  (Figure \ref{PACS_Fig1}) suggests  the 85 $\mu$m peak may be several arcseconds east of the nominal position for \ec, this is within nominal positional accuracy for the \pacs\ beam size and telescope pointing at this phase of the mission.

 \end{itemize}

For reference, we present in Figure \ref{spec_cen} the  observed spectral energy distribution (SED) for the central spaxel (2,2), nominally centered on \ec. The continuum from \ec\ and its ejecta are brightest at about 20 $\mu$m, corresponding to a dusty source of a few hundred K  with a luminosity of $\approx$3-5$\times$10$^6$\Lsun\ \citep{Westphal69,Morris17}. At wavelengths long-ward of 50 $\mu$m, this 1/$\lambda$\ tail of the thermal emission dominates the spectrum. Hence,  the \pacs\ spectral region emission lines are  weak relative to the very bright continuum. 

\subsection{\spire\ Observations and Calibration}\label{SPIRE_obs}

SPIRE-FTS spectral scans of \ec\ were obtained soon after the first visit with PACS as two observations with different sensitivity settings  anticipating high fluxes, particularly at the blue end of the SSW module (303 - 674 $\mu$m).  The ``nominal mode'' was intended for sources with fluxes up to around 500 Jy over any portion of the available wavelength range, and consists of 25 FTS scans with a total integration time of 3330 sec.  Due to the risk of detector saturation at input signals above 500 Jy, a ``bright mode'' observation with a total integration time of 1998 sec over 40 scans was also taken at higher biasing (or 3 to 4 times lower sensitivity).  Comparison of the two sets of spectra indicated that no saturation effects were present in the nominal mode spectrum and, since the nominal mode spectrum has higher S/N, well over 100 at all wavelengths, it is used for our analysis herein. 

The calibrated SPIRE continuum levels were included with \iso/SWS, HIFI, and \alma\ observations in a study of the unique dust spectrum of {\ec} by \cite{Morris17}, who provided a detailed description of the FTS data processing.  We point out here that because of the large beam sizes of SPIRE, 16.5$''$  to 43$''$ FWHM, compared to the size of the Homunculus, only one detector per section (SSWD4 and SLWC3) recorded on-source fluxes; see Figure~2 in \cite{Morris17} for the footprint of the SPIRE detectors during the {\ec} observations.   The remaining off-axis detectors were used to estimate and correct for  background emission.  

The large discontinuity of beam sizes of the SSW and SLW modules where they overlap in wavelength produces an additional small offset in levels in the output from the central detectors, which is to be expected for point source photometric calibrations applied to a semi-extended source as {\ec}.  This remaining offset was removed with SPIRE's semi-extended correction tool, using a Sersic brightness profile for a source 5$''$ in diameter, which is roughly consistent with \alma\ observations of the molecular emitting region \citep{Morris17}.   Details on the calibrations and procedures which have been followed are contained in \cite{Swinyard10, Wu13, Swinyard14, Hopwood15, Valtchanov18}, and the  \spire\ Handbook.\footnote  {https://www.cosmos.esa.int/web/herschel/legacy-documentation-spire}

Following the background and semi-extended source size corrections, we estimate continuum and line fluxes to be uncertain by $\approx$10\% in the SSW section, and somewhat higher at $\approx$20\% in the SLW section due to some departures from uniformity in the background emission.  

The spectra of \ec\ and ejecta are plotted in Figure \ref{Sfig1} through \ref{Sfig13}. Identified lines and measured fluxes are listed in Table~\ref{Tbl-SPIRE}.\\

\section{Results} \label{results}
 
The  \hso\ instruments could provide only limited spatial information through direct imaging of \ec\ because of the angular sizes of structures within the 10$''$ $\times$ 20$''$ Homunculus. Fortunately the \pacs\ and \spire\ spectral surveys are laden with molecular and atomic emission lines, giving us a new,  spectacular view into the chemistry and physical conditions of the bulk structures via emission lines from lower energy transitions.  Many of these far-IR transitions are observed in $\eta$ Car for the first time.

First we confirm that, as anticipated in the observation planning, the central source continuum is very strong in both sets of observations and is emitted  from the massive central torus-like structure dominating mid-IR wavelengths \citep{Morris17, Mehner19}.  Aside from the ensuing calibration issues of inter-order spectral leakage, partial saturations, and stray light during the PACS observations summarized above, the interpretations must take into account that the line and continuum emission arise from sources of varied angular extent.   The coupling of the continuum radiation to the \ion{H}{I} spectrum in the mid-IR observed with \iso, and the far-IR Rydberg states detected in our \hso\ scans, is examined in Sec.~\ref{ions} and \ref{swsH}.  Supported by complementary observations at higher angular and spectral resolutions, we assume for the remainder of this study that the entirety of the continuum is restricted to the central pointing of both instruments for modeling the spectral surveys, which we present below.  

\vspace{1em}
\subsection {\pacs\ Data Analysis} \label{PACS}
\subsubsection{Background, extended and compact source emissions \label{back}}

While the \pacs\ spectral mapping carried out on \ec\ lacks sufficient angular resolution to tie down the locations of the detected emission lines to better than $\sim$5$''$, we can deduce from each spaxel's spectral content whether the emission originates in the background Carina Nebula, or from semi-extended and compact sources within the Homunculus and central region.  The expanding lobes, dusty Butterfly Nebula, and the equatorial skirt each have characteristic velocities as described above, so that the origin of most spectral features is revealed from their observed velocity structure and variations.  The reader must recognize that these structures are rapidly evolving, and that the fluxes are qualitative in the {\hso\ epoch of these observations.

Emission line measurements derived from the spaxel centered on \ec\ (2,2) are listed in Table~\ref{PACS-T}.  As described above, many lines originate from extended structures, fast-moving ejecta and/or background ionized gas.   The lines are relatively weak compared to continuum, and were measured in each of the six scans centered or with sub-spaxel offsets from  \ec.  We applied the standard line detection condition for the measured intensities, namely 3$\sigma$ = $3 \times \sigma_{\rm{RMS}} \sqrt{2 \Delta v_{\rm{bin}} \Delta v_{\rm{FWHM}}} $, where $\sigma_{\rm{RMS}}$ is the RMS noise of the line-free baseline,  $\Delta v_{\rm{bin}}$ is the channel binning width in km~s$^{-1}$, and $\Delta v_{\rm{FWHM}}$ is the full width at half maximum intensity of the spectral line fitted as a Gaussian profile.

The line fluxes listed in Table \ref{PACS-T} are measured using the 9\farcs4$\arcsec$$\times$9\farcs4 spaxel sampling the \pacs\ point spread function (PSF) that changes in each band.  The formal measurement uncertainties are around 10\%.  Because of near saturation due to the strong continuum, however, we are unable to provide a total flux for these emission lines since the flux for the central spaxel may be depressed by up to 35\%, especially in the blue bands, compared to the fluxes in adjacent spaxels.  

Examination of the spectral line distribution in the PACS maps readily reveals the origin on the scales of the major features of \ec :   
\begin{itemize}
\item The atomic emission lines of [\ion{O}{i}], [\ion{O}{iii}], [\ion{C}{ii}] and [\ion{N}{ii}] contribute background emission, as expected from many previous studies of the Carina Nebula;
\item Emission from the outer, fast moving 'bullets' is detected in [\ion{N}{ii}];
\item Semi-extended emission of [\ion{O}{i}], [\ion{O}{iii}], [\ion{C}{ii}], [\ion{Fe}{ii}], and [\ion{N}{ii}] in the Homunculus;
\item H recombination emission from the central region;
\item Molecular emission from OH, NH, NH$_2$,  NH$_3$ and tentatively CH$_3$OH from the central region and Homunculus.
\end{itemize}

We discuss the observed spectral characteristics of these features next.


\subsubsection{[\ion{O}{i}] 63.18 $\mu$m and  [\ion{O}{iii}] 88.36 $\mu$m}

The [\ion{O}{iii}] line provides a useful reference for consistency between spaxels (Figure \ref{H12}), as it originates entirely from the background Carina Nebula extending over several degrees. Measures of the line flux, peak wavelength, and FWHM are nearly constant over the entire 45$\arcsec\times$ 45$\arcsec$ map, except on the central spaxel where the line flux drops nearly two-fold compared to the other 24 spaxels. This suppression may be caused by near-saturation of the central spaxel, particularly at the blue end of the spectrum.  As a consequence, we cannot judge if there is any contribution from \ec.  However, nebular [\ion{O}{iii}] is not expected in the Homunculus since:  (1) no [\ion{O}{iii}] nor [\ion{O}{ii}] lines have been detected at visible wavelengths from  the Homunculus \citep{Davidson86a, Verner05a, Zethson12}; and (2) \cite{Smith04d} found no visible [\ion{O}{iii}] near the Homunculus, but increasing amounts detected at tens of arcseconds distance.} This decrease in the infrared [\ion{O}{iii}] centered on \ec\ is likely the result of the  massive winds and ejecta, both greatly depleted of oxygen, clearing oxygen from the  immediate surroundings.

\begin{figure}
\includegraphics[angle=0, width=7.8cm]{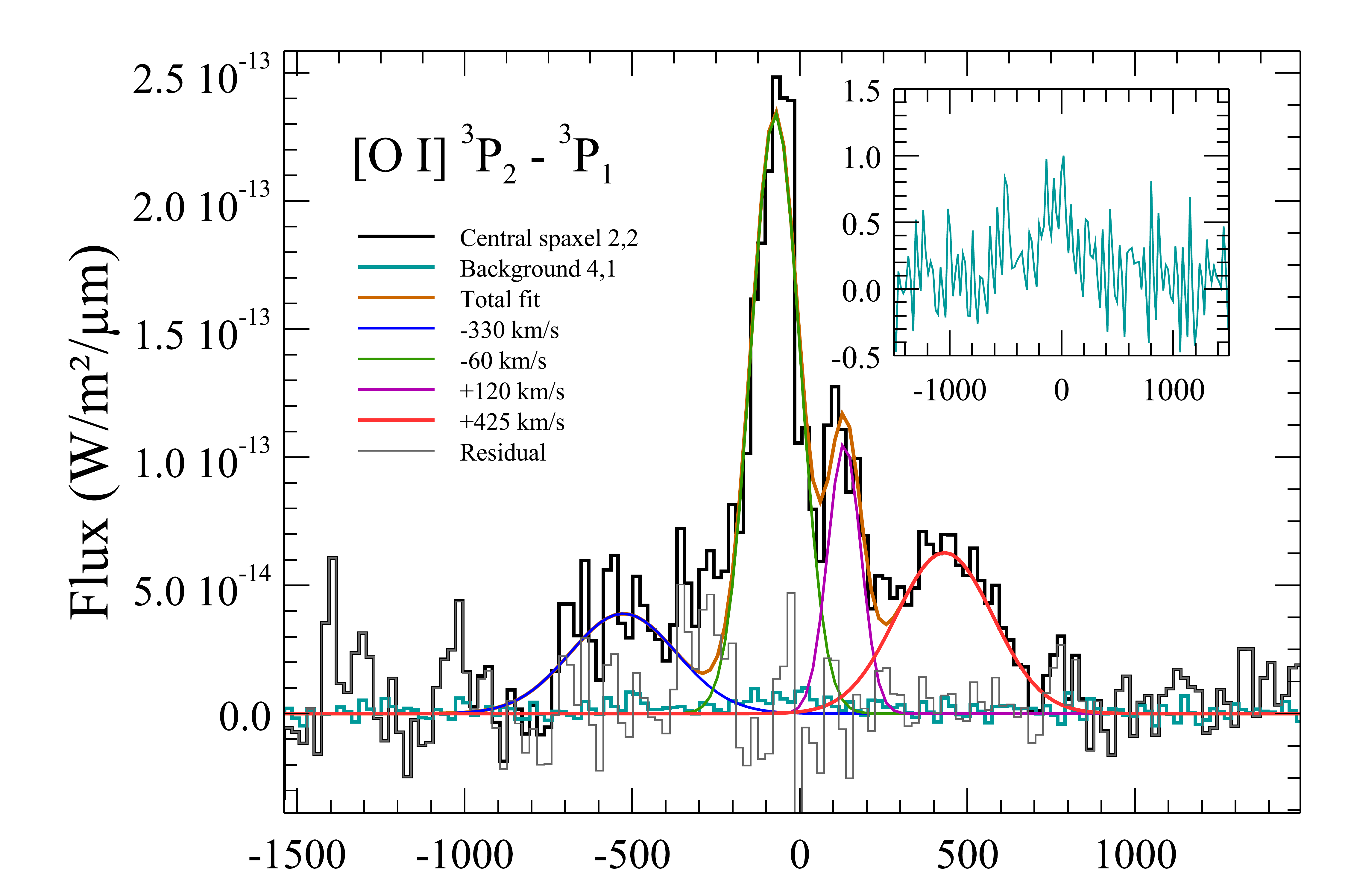} \\
\includegraphics[angle=0,width=7.8cm]{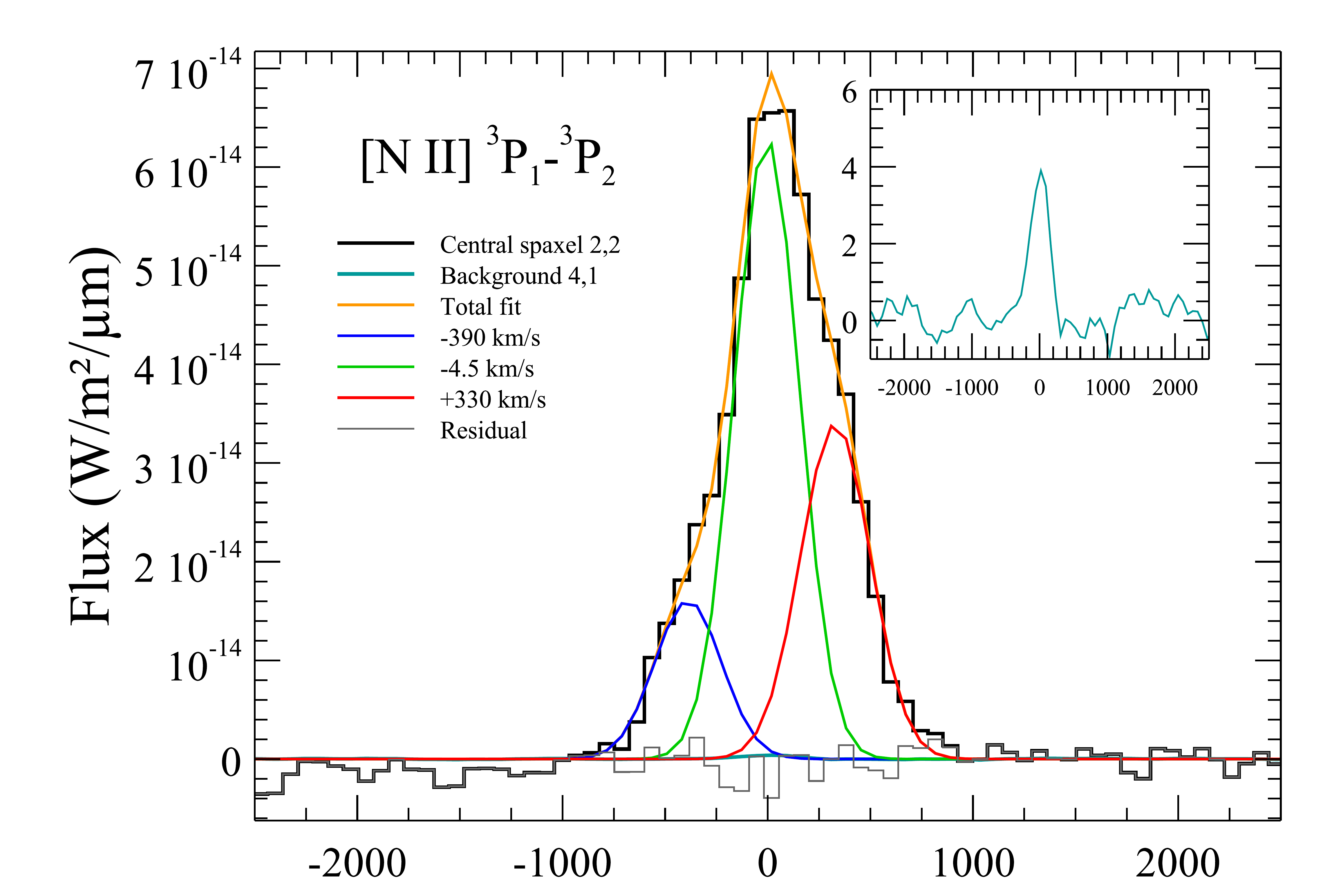}\\
\includegraphics[angle=0,width=7.8cm]{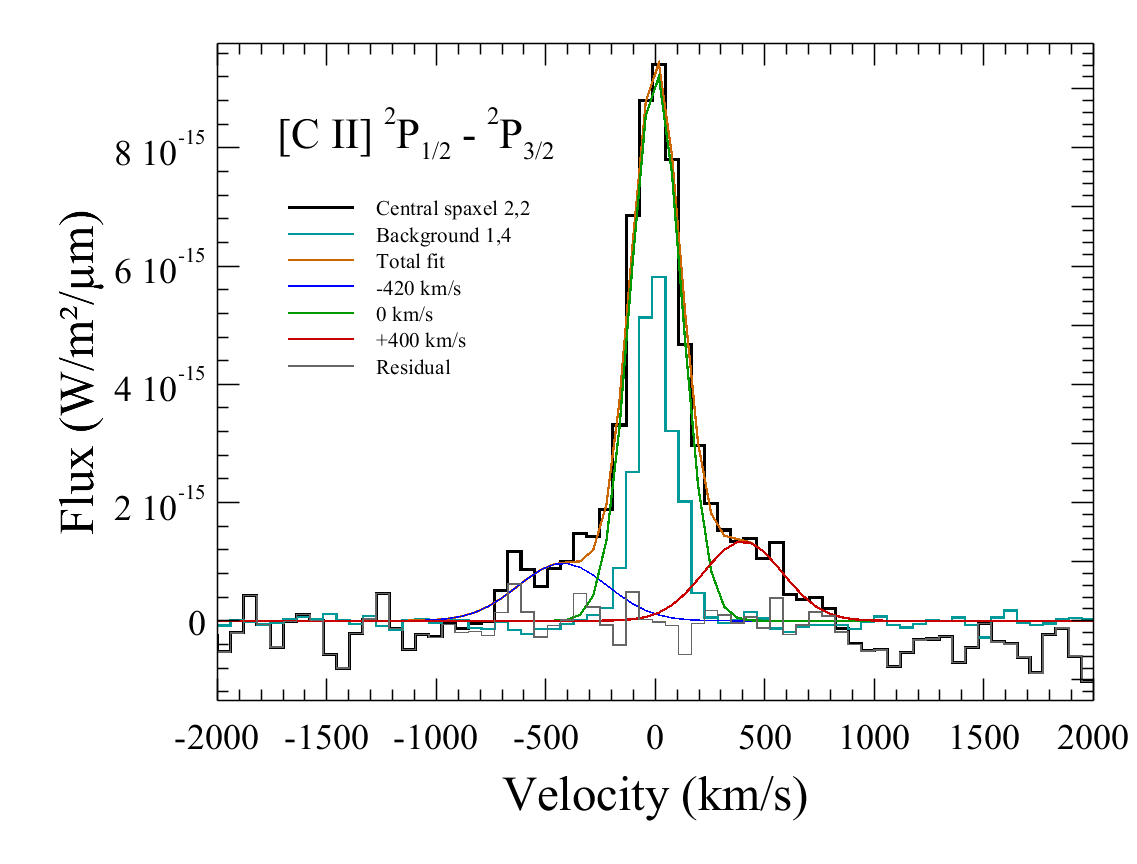}
\caption{[\ion{O}{i}], [\ion{N}{ii}], and [\ion{C}{ii}] line profiles on \ec. The observed spectra in the central spaxel (2,2), continuum subtracted, are shown with gaussian profile decomposition, demonstrating the similarities in velocity structure with broad red and blue wings and a strong spectrally-resolved central component (see Table~\ref{PACS-T}).  Extended background emission from outer spaxels for each line is shown for comparison by the cyan colored spectra, on scales of 10$^{-14}$ and 10$^{-16}$ W m$^{-2} \ \mu$m$^{-1}$ for [\ion{O}{i}] and [\ion{N}{ii}], respectively.  Only background [\ion{C}{ii}] contributes significantly to the central spectrum, $\approx$40\% to the total central component.  The background [\ion{O}{I}] and [\ion{C}{II}] lines are unresolved, while [\ion{N}{II}] is broadened by an extended shell around \ec\ (see text).
\label{CIIFITS}}
\end{figure}

[\ion{O}{i}] is detected in \ec, and in contrast to [\ion{O}{iii}], the profile on the central source is resolved and surprisingly structured with at least four velocity peaks; see Figure~\ref{CIIFITS}, where we present the [\ion{O}{i}], [\ion{N}{ii}], and [\ion{C}{ii}] spectra measured on the central spaxel and from the background in the insets.  We discuss the extended structure of [\ion{N}{ii}] and [\ion{C}{ii}] below.  Previously, only very weak [\ion{O}{i}] was identified in the spectrum of the Weigelt condensations \citep{Zethson12}, within $\sim$0.$''$5 of the central source.  [\ion{C}{ii}] has been observed with HIFI \citep{Morris17} but rendered uncertain in profile properties by the off-source chopping of the clumpy background.   Here all three profiles share similar velocity structure, with a resolved central component originating from the central region including the Butterfly Nebula, and broad blue- and red-shifted components originating from the expanding lobes and possibly the equatorial skirt. 

Only in the [\ion{C}{ii}] line is there a significant contribution from the background, $\sim$40\% of the total line flux.  The LSR velocities match expectations for the Carina Nebula:  [\ion{O}{i}] and [\ion{C}{ii}] background lines have measured velocities $v_{\rm{LSR}}$ = $-$20  and 0 \kms, respectively, in good agreement with SOFIA/GREAT high-resolution observations of the cometary globule G287.84-0.82 \citep{Mookerjea19}, roughly 20$'$ to the south of \ec.  Interestingly, the background [\ion{O}{i}] line flux in our observations, 4.0 $\times$ 10$^{-16}$ W m$^{-2}$ in the 9\farcs4 $\times$\ 9\farcs4 spaxel, is a factor 5 above the average measured over a $40' \times 20'$ region including the Car~I and Car~II \ion{H}{ii} regions (\ec\ lies just to the SE of Car II) with \iso/ Long Wave spectrometer (LWS) \citep{Oberst11}, taking the spaxel and aperture areas into account. 

\subsubsection{H recombination}

Five \ion{H}{i} n$\alpha$ lines and one $\beta$ line are identified, originating from a point-like, centrally-located source compared to the \pacs\ PSF (cf. \citealt{Abraham14}).  An example, from the central spaxel (2,2), is displayed in Figure \ref{H12} comparing the line identified as \ion{H}{i} 13-12 (88.795 $\mu$m) to the background [\ion{O}{iii}] (88.36$\mu$m).  These far-IR Rydberg states detected in our \hso\ scans and the \ion{H}{I} spectrum in the mid-IR observed with \iso\ are discussed in Sec.~\ref{ions} and \ref{swsH}.

\subsubsection{OH 84.5 and 119.3 $\mu$m}

Four weak features of rotational doublets of the hydroxyl radical OH at 84.415/84.615 $\mu$m and 119.245/119.416 $\mu$m  have been detected in the central spectrum.  After H$_2$, OH and CH are the first molecules detected in \ec, through their UV electronic transitions and moving in gas at $-$513 \kms, using the \hst/STIS \citep{Verner05b}.  Detection of the rotational lines is a significant confirmation of this key participant in the creation and regulation of water, which has been reported in \hso/HIFI observations by \cite{Morris20}, and may represent a primary sink of free oxygen in reaction with H$_2$, i.e., O + H$_2$ $\rightarrow$ OH + H.

\subsubsection{NH$_2$ 152.8/159.8 $\mu$m and NH$_3$ 127.1 $\mu$m} 

In addition to the NH transitions which are discussed below, two weak line complexes of NH$_2$ (152.80 and 159.75 $\mu$m) and one weak line complex of NH$_3$ (127.11$\mu$m) are detected (see Table \ref{PACS-T}).  Both molecules have {\it{ortho}} and {\it{para}} spin symmetries whose relative abundances are sensitive to the {\it{ortho}}-to-{\it{para}} ratio of H$_2$.  This subject is addressed in a more detailed study of the nitrogen hydrides observed at higher spectral and angular resolutions in \ec\ (P. Morris et al., in preparation).  The NH$_3$ 23.87 GHz inversion line has been detected by \cite{Smith06e}.  The NH$_2$ is a first detection in \ec.

\begin{figure}
\includegraphics[angle=0, width=8cm] {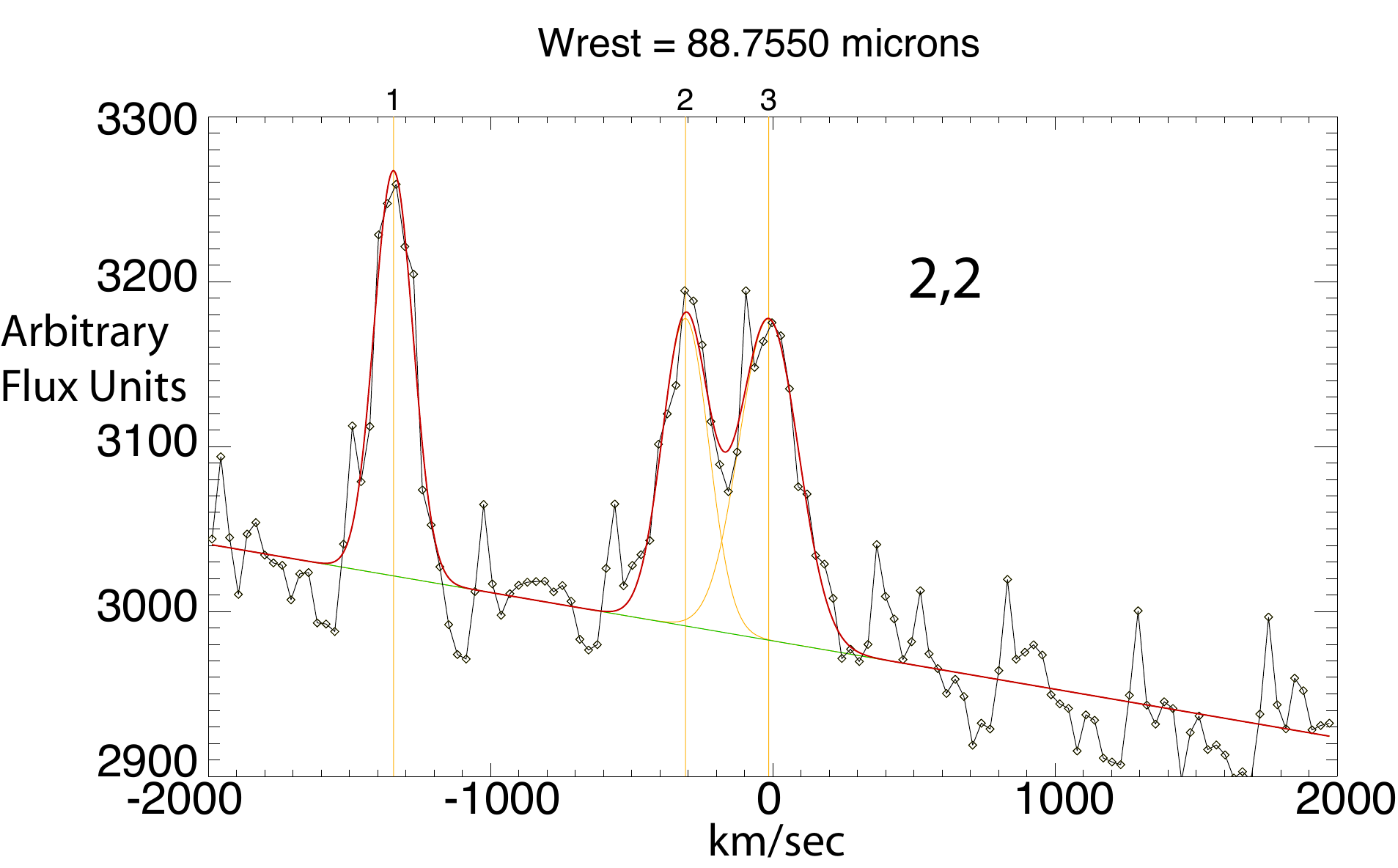}
\caption{ \ion{H}{i} 13-12 (88.755$\mu$m) line profile measured by \pacs, centered on \ec.  The line labeled '1' corresponds to background [\ion{O}{iii}] 88.36$\mu$m.  The \ion{H}{i} emission feature labeled '3'  is blended with an unidentified transition, labeled '2'. LSR velocities are referenced to the \ion{H}{i} 13-12 rest wavelength. The line is detectable within only four spaxels, strong on spaxel 2,2, and weak in three nearby spaxels which is consistent with a nearly point-like source compared to the \pacs\ spatial response. The width of the [\ion{O}{iii}] emission line is close to the instrumental profile, with a FWHM measuring 150 \kms. \label{H12}}
\end{figure}

\subsubsection {CH$_3$OH  at 69.0 $\mu$m}

Emission from CH$_3$OH 7$_4^+$--8$_5^+$ E1  $v_t=1-0$  is tentatively identified at 69.00 $\mu$m.  The plausibility of the identification is supported by a preliminary excitation model, including the effects of IR pumping and collisions at a somewhat high molecular density (1 $\times 10^8$ cm$^{-3}$), and an excitation temperature of 200 K which is adopted in our modeling of the SPIRE spectrum below. The 4346 GHz (68.9805 $\mu$m) line of CH$_3$OH is indeed one of the stronger ones in this part of the far-IR spectrum. While this model is just for guidance and not optimized across more than this isolated line, the model with $\mathcal{N}$(CH$_3$OH) = 1 $\times$ 10$^{17}$ cm$^{-2}$ over a line width of 188 \kms\, gives an integrated line flux that approximately agrees with the observed line flux (Table~\ref{PACS-T}).  A relatively high column density would be expected for a compact source of emission; the ALMA 231.28 GHz feature proposed as CH$_3$OH 10$_2$--9$_3$ A$^- \; v_t = 0$ by \cite{Morris20} originates from a compact source, $<$ 2 arcsec in diameter.  

\subsubsection{ The extended [\ion{N}{ii}] structure}\label{NIIS}
\begin{figure*}
\includegraphics[angle=0, width=17.7cm]{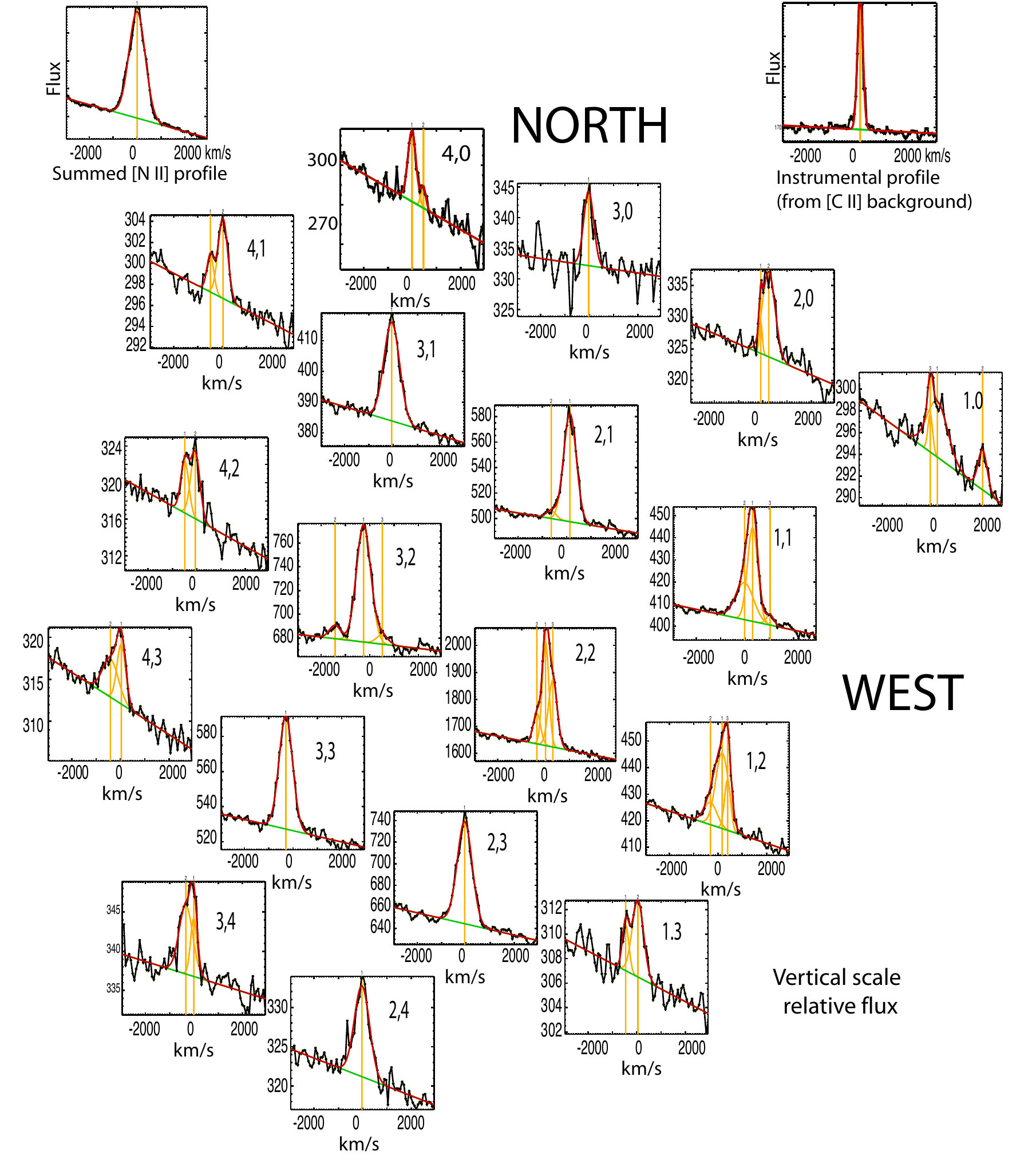}
\caption{[\ion{N}{ii}] line profiles on the Homuculus. In the top left corner is shown the [\ion{N}{ii}] profile summed over the entire \pacs\ array.  The top right shows the \pacs\ spectrometer profile of the unresolved background line of [\ion{C}{ii}] 157.736 $\mu$m.   The spectral resolution at 121 $\mu$m is $\sim$30\% lower.  In the centre is shown the mosaic of [\ion{N}{ii}] velocity profiles. The profiles, labeled with spaxel number, are arranged to reflect the spaxel positions shown in Figure~\ref{Cont}.  The summed profile, centered on 3 \kms\ has a FWHM of 804 \kms. However the peak flux, velocity FWHM and central velocity change from spaxel to spaxel. Profile variations are consistent with  [\ion{N}{ii}] emission from the Homunculus and the external, fast-moving ejecta. 
  \label{NII}}
\end{figure*}

The [\ion{N}{ii}] (121.918 $\mu$m) emission from the Homunculus is bright -- two to three times above the background levels -- and variable with position. From spaxel to spaxel, the flux and the profile wings change considerably.   These extensions are associated with the central core plus the outer, fast-moving ejecta that extends from the Southeast to Northwest, well beyond the Homunculus \citep{Weis99, smith08b, Kiminki16}.   The map provides an example of a spatially-extended source measured by the 2,2 spaxel to be centered at $-$49~\kms, but the FWHM is nearly 700 \kms, much broader than the \pacs\ spectral response. The centroid velocity shifts to $-$206 \kms\ to the southeast and to $+$80 \kms\ to the northwest, which is significantly larger than the measured instrumental resolution of the \pacs\ ($\approx$ 150 to 190 \kms. See Figure \ref{H12}). These quoted velocities ignore possible pointing-induced shifts in the wavelength calibration mentioned above, applying to compact sources.

The $[$\ion{N}{ii}] $^2$P$_{3/2}$--$^2$P$_{1/2}$ fine structure transition actually consists of six hyperfine components, which collect into three  potentially resolvable lines at 121.894, 121.907, and 121.912 $\mu$m. With respect to the weighted mean wavelength, 121.898 $\mu$m, these have offsets in Doppler velocity of $-$22, $+$8, and $+$36 km/s, respectively (see \cite{Brown94}). These splittings, are significantly smaller than the \pacs\ resolution (at 121 $\mu$m, ${\Delta}v$=150 \kms). Thus the velocity gradients and line widths observed here are dominated by kinematics. 

The six brightest spectral profiles extend up to $\pm$800 \kms\ on each side of the centroid velocity, indicating  large bulk motions of gas to the southeast and the northwest. The general behavior of the spatial and velocity distributions follow those noted by  \cite{Currie02} in their description of the so-called ``ghost nebula'', which was found to be expanding at $-$500 to $+$875 \kms\ and located exterior to the Homunculus. Additional studies of this outer, rapidly moving structure were done  by \cite{Kiminki16}, who measured the proper motion of expanding structures external to the Homunculus.  \cite{Mehner16} used the \vlt/MUSE instrument to map velocity/spatial structures external to the Homunculus. The resulting 3-dimensional model defines the large outer structure of the very fast moving, numerous 'bullets' external to the Homunculus moving at spatial velocities of several thousand \kms \citep{Smith08}.

\subsubsection{The [\ion{C}{ii}] extended  structure} \label {CIIS}

The [\ion{C}{ii}] (157.736 $\mu$m) emission mapping (Figure \ref{CIImap}) {exhibits both bright, narrow-line background emission from the Carina Nebula, and an underlying broad-line, extended emission component from the Homunculus.

The extended narrow line component is measured to be centered at $v_{\rm{LSR}}$ = 0 $\pm$10 \kms\ across the \pacs\ array with a nearly constant flux and instrument-limited velocity width.  For the central and six adjacent spaxels, weaker, broad-emission components  contribute (see Figure \ref{CIImap}, lower right for this discussion). To the northwest and west, a red-shifted component is present. To the east and south, a blue-shifted component is detected. The [\ion{C}{ii}] broad component extends over the region of the Homunculus, including the extended skirt structure to the north-northeast (NNE) and to the west-southwest (WSW), but is not external to the Homunculus as is the [\ion{N}{ii}]. The [\ion{C}{ii}]  emission is not associated with the fast moving ejecta as mapped in [\ion{N}{ii}] in the visible \citep{Mehner16}. At spaxel 3,2 (NE of 2,2) and 3,2 (SE), the broad component is shifted to the blue, consistent with material flowing towards the observer. The broad component is approximately centered in velocity in spaxel 2,3 (S) and shifted to the red on the central spaxel (2,2) and 2,1 (N). The broad component shifts to the red for both spaxels 1,1 (NW) and 1,2 (W). These velocities are consistent with the central core at small velocities, the foreground lobe (SE) moving towards the observer and the background lobe receding. The blue-shifted components seen to the east correlate with the forward moving portion of the extended skirt and the red-shifted components seen to the NW and W with a receding portion of the skirt. 

Visible/ultraviolet spectral studies of the partially ionized ejecta were accomplished with the \hst/\stis. In line of sight,  \cite{Nielsen05a,Gull06} found multiple shells of partially-ionized gas at velocities between $-$380 and $-$530 \kms. The most noticeable shell at $-$512 \kms\ included nearly a thousand NUV absorption lines from singly-ionized iron-peak elements and approximately a similar number of far-UV absorption lines from H$_2$, along with ground-state absorptions of CH and OH \citep{Verner05b}. \cite{Zethson01a, Hartman04} discovered the Strontium Filament, located a few arcseconds to the north of \ec\ in the expanding skirt, which proved to be a metal-ionized intermediate region between the classical \ion{H}{ii} and neutral, molecular regions. The [\ion{C}{ii}] emission, given that the ionization potential (I{) of  neutral carbon is 11.3 eV, below the IP of hydrogen (13.6 eV) and that of nitrogen (14.5 eV)  traces the metal-ionized and neutral regions. Hence, molecules might be present in and near the regions of [\ion{C}{ii}]. Given that the two structures semi-overlap, velocities and positions should be similar.

\begin{figure*}
\includegraphics[angle=0, width=17cm]{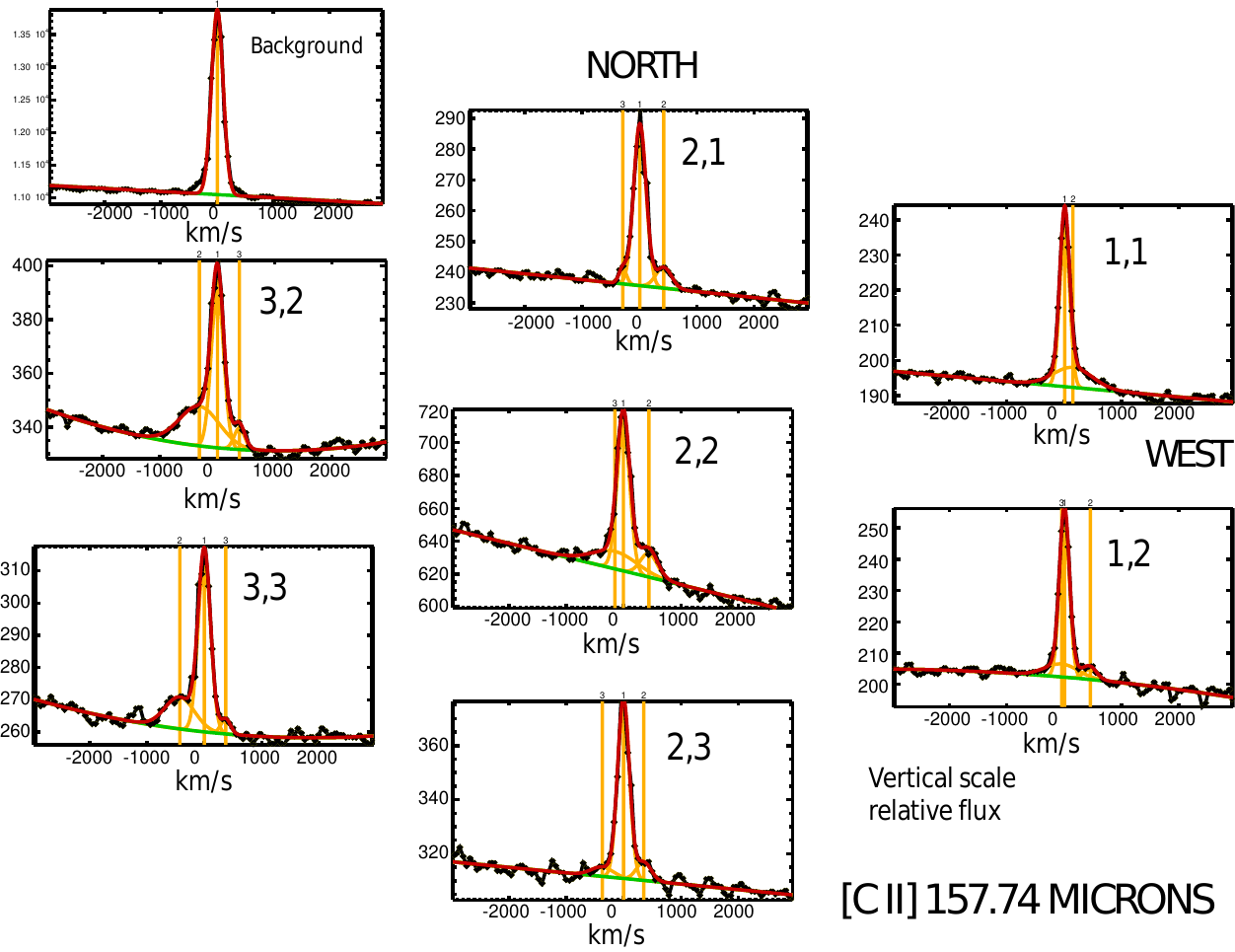}
\caption{PACS mapping of [\ion{C}{ii}].  Profiles all have at least two major components:  a constant, partially-resolved central component and a weaker, broadened component from the Homunculus that follows the large-scale SE to NW velocity shifts of the lobes and equatorial skirt.  In some profiles as in (2,2) above, the broad feature is better fit by blue- and red-shifted wings around the central component, consistent with lobe expansion velocities. 
 \label{CIImap}}
\end{figure*}

\begin{figure}
\includegraphics[angle=0, width=8.5cm]{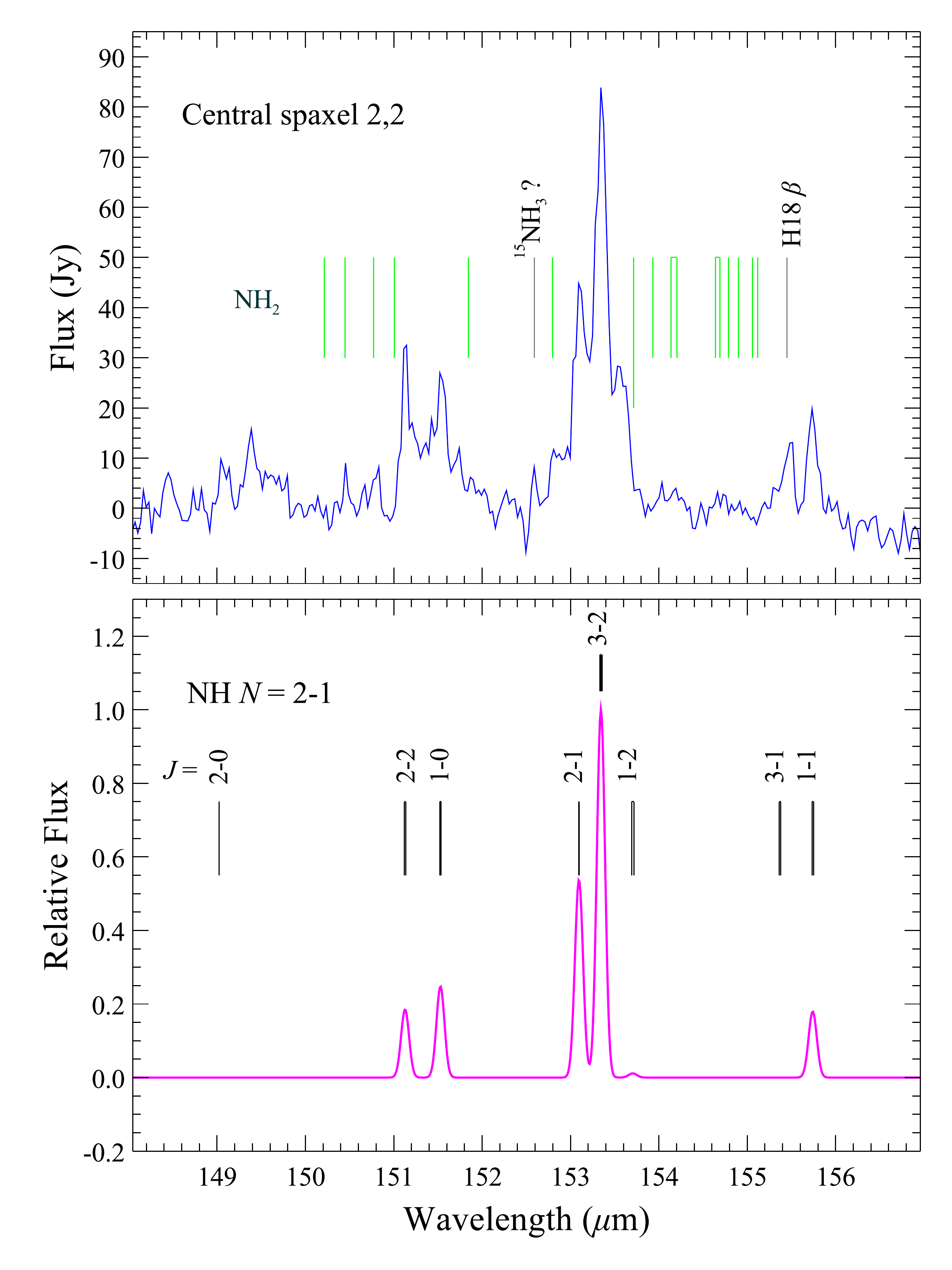}
\caption{NH $N$ = 2--1 complex between 151 to 156 $\mu$m. Top: continuum-subtracted spectrum centered on \ec\ with vacuum wavelengths shown for H18$\beta$, a weak, low significance identification of $^{15}$NH$_3$, and NH$_2$ transitions all with upper level energies below 1000 K.  The spatial extent of NH  coincides with the Homunculus lobes and skirt as demonstrated in Figure~\ref{NHFIT}. Bottom: A reference non-LTE model of NH over the full range of full range of spin-rotation-hyperfine components, computed at 100 K and zero LSR velocity, convolved with the \pacs\ resolution of 0.05 $\mu$m at 150 $\mu$m.  The width of fiducial lines at each $J$ state indicates the range of hyperfine transitions.  Note that three of the eight spin rotation components contribute practically no measurable emission, consistent with their low transition probabilities. \label{NH}}
\end{figure}

\subsubsection{The extended NH structure}\label{NHS}

The NH $N$ = 2--1 transition is absent in the background, but detected by the strongest $J$ = 3--2 spin-rotational component at 153.56 $\mu$m in the central region (Table~\ref{PACS-T}) and the surrounding Homunculus.   This transition is complex, consisting of 4 spin-rotational components, each with hyperfine structure owing to nuclear-spin interactions with both  $^{14}$N and $^1$H nuclei; see Figure~\ref{NH}.

Three of the spin-rotation components, J=1-2, 2-0, and 3-1, have small line strengths and together contribute less than 0.5\%\ of the total N=2-1 emission. However, significant emission occurs at 153.55 $\mu$m in the observed spectrum, which cannot be attributed to either the $J$ = 1--2 transition or possible NH$_2$.  This is a strong indicator of more than one velocity component to the NH emission.  

Using the same theoretical methods described below in Section~\ref{molecules} to model the \spire\ spectral survey, we computed synthetic profiles for this NH $N$ = 2--1 transition, including including all spin-rotation and hyperfine substates, to provide a clearer picture of the velocity characteristics.  The upper level energies of these transitions run from 140.37~K to 141.97~K, approximately 92 to 96 K above the lower energy levels.  The results are shown in Figure~\ref{NHFIT}, where we compare synthetic and observed spectra in each spaxel, schematically laid out consistent with the \pacs\ mapping in Fig.~\ref{Cont}.  Not shown are the flat spectra from the outer spaxels sampling the background Carina Nebula over this range.  The fitting has been weighted to the two strongest rotational states $J$ = 3--2 and 2--1.  Only in the central spaxel do we adopt an interacting  continuum of 1200 Jy (integrated thermal emission) based on the extended SED shown by \cite{Morris17}; continuum emission is negligible in the surrounding spaxels.   

We find that every spectrum can be well-approximated by a 2-component model, consisting of a fast component to around $\pm$(400 -- 500) \kms, and a slower component with $v_{\rm{LSR}}$ within $\pm$100 \kms.  The slower component generally dominates the emission.  While the fitting is weighted on the two strongest $J$ states, the agreement with the weaker lines is qualitatively acceptable, where differences can be attributed to either data quality, or an indication of additional, weaker velocity components.  As a test, we added such a third component to the spaxel (2,3) spectrum covering the cap of the SE lobe and backside of the skirt (in projection), which provided marginal improvement in the comparison between observed and total model spectra.  

For each model component, the best-fit LSR velocities, FWHM, excitation temperature, and NH column density are given in the plot legends. Best-fit temperatures of 50 to 60 K are derived for the fast moving material, and 150 to 200 K for the component with $v_{\rm{LSR}} < 100$ \kms.  Because of the calibration issues in these \pacs\ observations explained in Section~\ref{PACS_obs}, we stress that the modeling here has been intended mainly to distinguish the principle NH velocities.  That said, the total column density for both components in the central spectrum, $N_{2,2}$(NH) = 1.67 $\times$ 10$^{15}$ cm$^{-2}$ assuming line widths of 275 and 290 \kms\ for the slower and faster components respectively, is only slightly lower than determined from analysis of the \spire\ spectrum (presented below in Sec.~\ref{molecules}).  Summing all spaxels results in excellent agreement.  

The velocity structure of NH shown in Figure~\ref{NHFIT} presents something of a challenge to interpret.  We can immediately exclude detection of NH in the fast outer ejecta or the ghost nebula observed in H Balmer and certain forbidden lines by \cite{Currie02}.  Being judiciously wary about potential stray light contamination of the spaxels around the central region, the overall structure of red- and blue-shifted components is nonetheless consistent with the expanding lobes and skirt as projected by the 45$^\circ$ tilt of the polar axis onto the sky.  We are tempted to associate the fast NH with the thin walls of the expanding lobes.  The wider spatial extent and 400 to 500 \kms velocities would not be inconsistent with an origin of a portion of the NH gas in the equatorial skirt, however, which is more aptly described as break-out material,  possibly from more than one of the lesser eruptions that followed the so-called Great Eruption that formed the bipolar lobes.   The measured line strengths and column density estimates of the fast component favor association with the bipolar lobes.

Likewise, the slower NH component behaves as expected on the scale of the Homunculus, from approaching to receding in the SW to NE direction, and may point to more wide-scale formation in the Homunculus than, e.g., CO, which is restricted to layers of the dusty central torus over the inner 5$''$ -- 7$''$ diameter of the structure (see Fig.~\ref{nebular}).  While the CO profile is asymmetric and broadly flat-topped due to a number of partially overlapping velocity components in the dust rings that make up the disrupted torus \citep{Morris17},  the total velocity width of the integrated profile is only roughly consistent with the widths of the NH model profiles, 190 \kms $\leq$ FWHM $\leq$ 300 \kms with corrections for the instrument profile.  It is therefore possible that most NH is formed in regions distinct from CO.  A summary of our kinematic interpretations of the emission is given in Section~\ref{concl}. High angular resolution observations of NH $N$ = 1--0 are needed to further elucidate on the distribution. 

\begin{figure*}   
\includegraphics[angle=0, width=17.5cm] {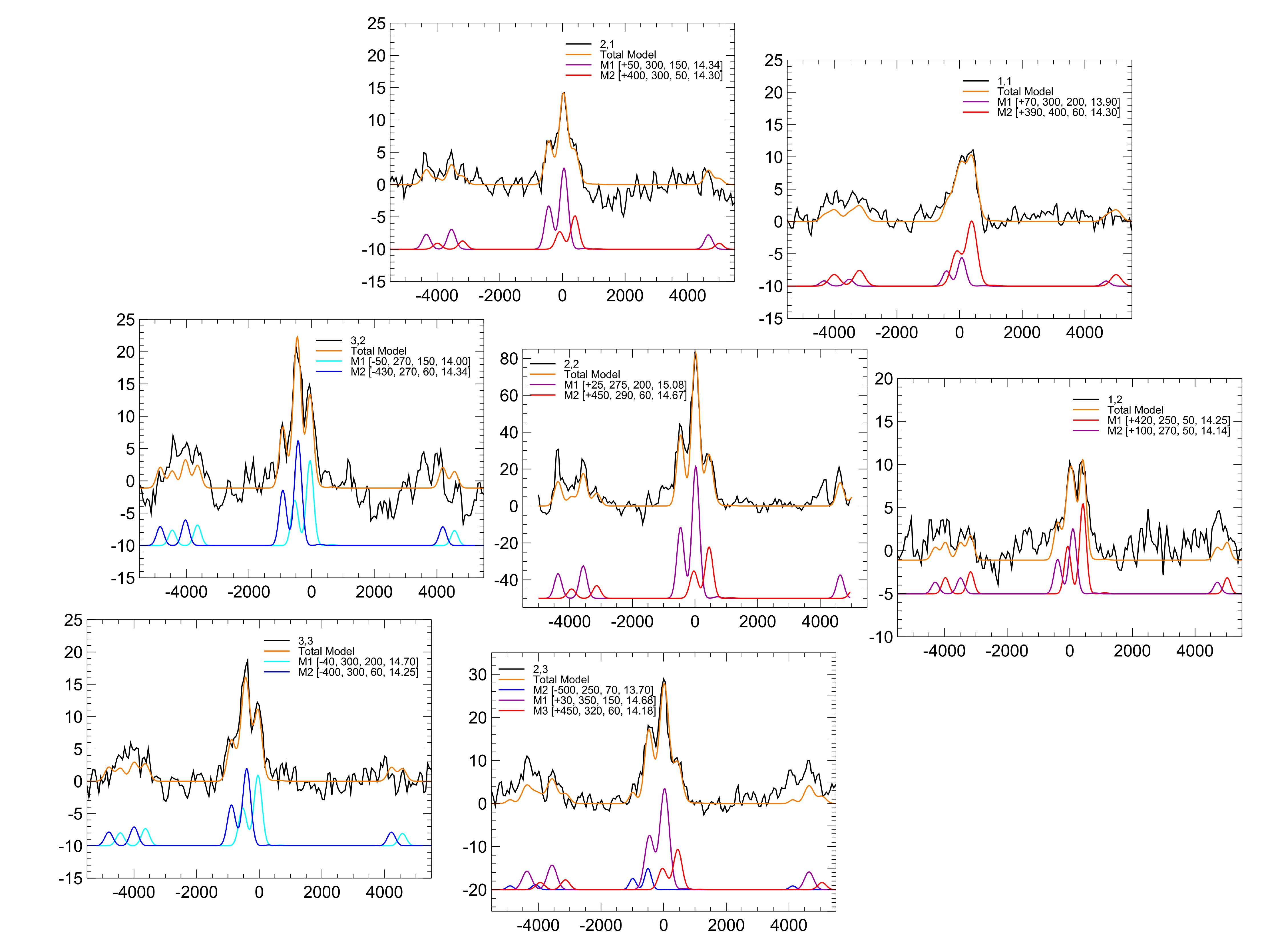}
\caption{NH $N$ = 2--1 profiles observed at seven positions on the Homunculus. with synthetic spectra comparisons. Each plot shows LSR velocity in \kms\ (referenced to the vacuum wavelength 153.450 $\mu$m of the highest transition probability $J$= 3 --2, F$_1$=7/2 -- 5/2 line) vs. line flux in Jy and are arranged to correspond schematically to the spaxel positions in Figure~\ref{Cont}.  All observed profiles are fitted with two velocity components, except for a 3-component model for spaxel (2,3).   Total synthetic fits are shown in orange.  Individual model components are offset for clarity and colored to indicate LSR velocity shift ranges of $<$ $-$100 \kms (blue), $-$100 to 0 \kms (cyan), 0 to $+$100 \kms\ (violet), and $>$ $+$100 \kms\ (red).  For each model component, the LSR velocity (in \kms), FWHM (\kms), excitation temperature (K), and log column density (cm$^{-2}$) are given in the legend.  Details on the modeling are given in the text. Features not fitted near $-$1000 \kms\ and $+$4000 \kms\ are due to NH$_2$ and H18 $\beta$, respectively.
\label{NHFIT}}
\end{figure*}

\subsection{The \spire\ Data Analysis}

\subsubsection{Non-equilibrium models of the molecular line spectra}
\label{molecules}

In order to aid identifications and provide a quantitative analysis of the observed line intensities, we have constructed 
simple one-zone models of the excitation and line formation for many simple molecules and several atoms and atomic 
ions. These models treat atomic and molecular processes in some detail, but greatly simplify the radiative transfer. 
Because the continuum radiation in the infrared is so intense near \ec, the induced radiative processes
of absorption and stimulated emission are likely to be important in molecular excitation. As a result,
the molecular excitation might depart from LTE.  Moreover,  the amount of continuum radiation absorbed by different species is greatly influenced by the  shielding by dust in complex structures around the central source in relation to the distribution of the gas. This is poorly known except for the few species observed at high angular resolution with \alma, or inferred by velocity-resolved observations with the \hso/HIFI and \APEX\ receivers. 

We have retained as much detail as possible about atomic and molecular spectra and processes with an enhanced version of the {\tt RADEX} program to compute non-LTE models of the excitation
of a large sample of atoms and molecules. The computed line list is then used to generate a synthetic
spectrum, which is finally convolved to the resolution of the \spire\ instrument for comparison 
with the observed spectrum. The radiative coupling is described with reference to a fixed, empirical
model of a passive continuum; that is, the radiative transfer in the lines is not treated self-consistently
with the continuum. The {\tt RADEX} program was described in detail by \cite{vanderTak07}. [NB:  The online version and downloadable source code have not yet included the routines for a passive continuum and explicit terms for formation and destruction processes
in the rate equations, even though these were present in the equations published by \cite{vanderTak07}.]

To summarize, statistical steady state 
requires a complete balance among all the rates that populate
and de-populate each state $i$ of a molecule. This can be described by a coupled set of 
rate equations for the state densities $n_i$ subject to the condition that all the densities are constant
in time,
\begin{equation}
 {\frac{dn_i}{dt}} = 0 
\end{equation}
\noindent so that loss equals gain for each state $i$ and
\begin{equation}
  \begin{aligned}[b]
     & n_i = \Bigl[D_i +  \sum_{j<i} \bigl( A_{ij} + \Jbar_{\nu} B_{ij} + \sum_p n_p q_{ij,p}(T_k) \bigr) + {} \\
     & \indent \sum_{k>i} \bigl(\Jbar_{\nu} B_{ik} + \sum_p n_p q_{ik,p}(T_k) \bigr)\Bigr]
  \end{aligned}
\end{equation}
\noindent and the right-hand side...
\begin{equation}
  \begin{aligned}[b]
   &  = F_i + \sum_{m<i} n_m \bigl(\Jbar_{\nu} B_{mi} + \sum_p n_p q_{mi,p}(T_k)  \bigr) + {} \\ 
   & \indent \sum_{n>i} n_n \bigl(A_{ni} + \Jbar_{\nu} B_{ni} + \sum_p n_p q_{ni,p}(T_k)  \bigr) . 
  \end{aligned}
\end{equation}
This system of equations must be solved simultaneously with the transfer
equation so that the internal, integrated, mean intensity $\Jbar$ can be evaluated. Here

\begin{equation}
\Jbar_{nu} = \int_0^{\infty} J_{\nu} \phi(\nu) d\nu /  \int_0^{\infty} \phi(\nu) d\nu  
\end{equation}

\noindent is the profile-weighted integral of the internal intensity integrated over solid angle and averaged over all
directions

\begin{equation}
J_{\nu} = {\frac{1}{4\pi}} \int I_{\nu} d\Omega . 
\end{equation}

In the rate equations, the notation $j<i$ means all states $j$ for which the energy $E_j$ is lower than the energy $E_i$ of the higher state $i$, and so on. All possible collision partners are denoted by $p$, and $n_p q_p$ are the number of collisional transitions, downward for $j < i$ and upward for $k > i$,  per unit volume and time [cm$^{-3}$ s$^{-1}$].    $D_i$ is the rate [s$^{-1}$] at which a molecule in state $i$ is destroyed (removed), and $F_i$ is the rate per unit volume [cm$^{-3}$ s$^{-1}$] at which molecules are formed (added) in state $i$. 

The mean intensity at each transition frequency is obtained through a solution to a transfer equation
based on escape probabilities that are averaged over a single uniform emitting zone and over an
assumed Gaussian line profile. \cite{Morris17} modeled the extended infrared/sub-millimeter source to be a partial annulus 5\arcsec\ in diameter. For this idealized model, we assumed the thermal continuum source to be distributed in a sphere 5\arcsec\ in diameter.  The mean intensity of 
the internal continuum at the source is then simply

\begin{equation}
 J_{\nu}^{cont} = f_{\nu}/\Omega_s  
\end{equation}

\noindent where $f_{\nu}$ is the observed continuum flux and $\Omega_s$ is the solid angle corresponding
to an angular diameter of 5\arcsec. The continuum model is shown in two forms in Figure \ref{flux}.
\begin{figure}
\includegraphics[scale=0.39]{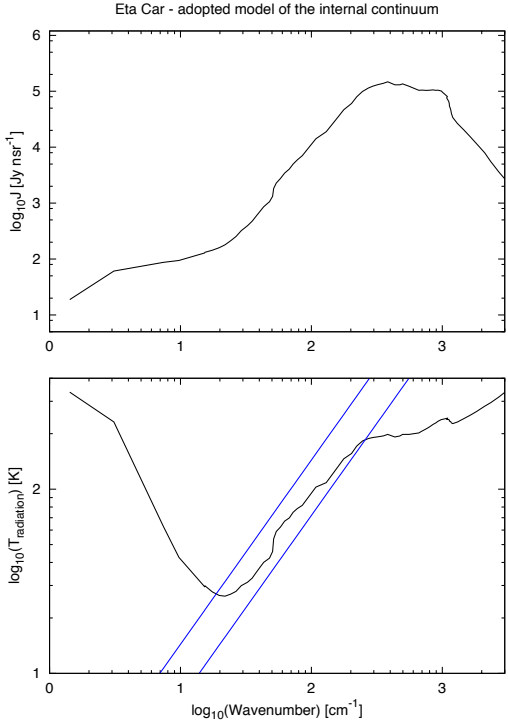}
\caption{The upper panel shows the internal mean intensity in units of Jy nsr$^{-1}$. The lower panel
shows the same intensity as a Planckian brightness temperature. The straight lines in the lower panel
show the photon energy $h\nu/k$ and $h\nu/2k$ in temperature units.\label{flux} }
\end{figure}
The lower panel of the figure illustrates that the brightness temperature of the internal continuum
is greater than half the photon energy throughout the sub-millimeter and far-IR part of the spectrum
for $\nubar < 250$ cm$^{-1}$ (wavelengths greater than 40 $\mu$m). 
The induced radiative transition rates are related to
\begin{equation}
 \rho = {\frac{A}{\exp(hc\nubar/kT_{\rm radiation}) - 1}}  \;\;\; {\rm s}^{-1} 
\end{equation}

\noindent where $A$ is the spontaneous transition probability, $k$ is Boltzmann's constant, and $T_{\rm radiation}$
is the brightness temperature at which the Planck function is equal to the mean intensity, $\Jbar = 
B_{\nu}(T_{\rm radiation})$. In short, for many common molecules the induced radiative rates in
this environment readily compete with inelastic collisions that would otherwise thermalize the 
excitation, even at densities of the order of $10^8$ cm$^{-3}$ or higher.

The enhanced version of {\tt RADEX} requires both the physical parameters of a model and extensive
molecular data files. The basic parameters are the number densities of collision partners (H$_2$, H, He,
$e^-$, and H$^+$), the kinetic temperature $T_k$, and a specification of the continuum radiation. For 
each atom or molecule, a column density and line width must be specified. Although the formation
and destruction rates are formally included, these appear to be important only for the Rydberg transitions
of nebular atoms (see section \ref{ions} below). For the molecules of the neutral emitting region, nominal
values of the formation and destruction rates have been assumed in order to ensure rapid convergence,
but these are small enough to have no effect on the computed spectra.  With reference to the low-resolution \spire\ spectra, we assume that
all species have the same intrinsic line width, 188  \kms\ full-width at half-maximum (FWHM), which is 
large enough to ensure that all of the observed line emission is optically thin. The much higher velocity 
resolution of \hso/HIFI and ground-based molecular observations reveals complex profiles of
lines of CO and other species, with spatially resolved line widths as low as $\sim$40 \kms.  In the unresolved SPIRE date, the integrated
line intensities scale with molecular column density, ${\mathcal N}$, which refers to an average over
the assumed size of the source, 5\arcsec\ in diameter, corresponding to a linear scale of $L=1.72\times
10^{17}$ cm based upon the accepted distance, 2.35 kpc \citep{Smith06}. The result for each molecular transition is an excitation temperature and a line-centre
optical depth, from which the line intensity in excess of the continuum 
is computed as a function of frequency
\begin{equation}
I_{\nu} = B_{\nu}(T_{\rm ex}) \bigl( 1 - \exp(-\tau(\nu)\bigr)  \;\;\;. 
\end{equation}
In the present case, where the optical depths are small, the total spectrum is a simple sum of the
contribution of all of the lines. For comparison with the observed spectrum, the computed spectrum 
is further convolved to the instrumental resolution.  The
computed spectrum includes both blends of lines of different molecules as well as blends due to 
unresolved fine- and hyperfine-structure components. 

Models were computed for a range of temperatures and hydrogen densities, but with the continuum
model fixed. For many of the common molecules with large dipole moments, the radiative coupling
is strong enough that the results are not very sensitive to density. The main exception is CO,
because  its permanent dipole moment is anomalously small and its fundamental vibration
occurs at a high enough frequency ($>2000$ cm$^{-1}$) that infrared pumping is negligible. There
are ten rotational transitions within the ground vibrational state of CO within the frequency range
of the \spire\ spectra.  Models of CO and CN were used in the first instance to place approximate
constraints on density and temperature. It was found that a total density $N({\rm H}_2) \approx 10^8$ 
cm$^{-3}$ at $T_k\approx 150 - 250$ K worked rather well.

Note that many of the CO and CN features are actually blends with lines of other species; 
therefore, the rest of the spectrum was computed at the same
time for each of the test cases. The final nominal model adopts a total density of $10^8$ cm$^{-3}$ at 
$T_k=200$ K. Molecules like HCN, HNC, HCO$^+$, and N$_2$H$^+$ have low-frequency bending
vibrations with relatively large transition moments. These species potentially couple strongly to the
mid-infrared continuum through their vibrational transitions. Molecular data files were assembled
to include such transitions for these molecules. SiO and H$_2$O have bending vibrations at somewhat
higher frequencies, and these were included as well. The model of NH$_3$ takes into account 847 vibration-rotation-inversion levels with energies up to 4734 cm$^{-1}$ above the ground level and 
vibration-rotation transition to frequencies up to $\nubar \approx 2000$ cm$^{-1}$.  Many of the 
molecular data files are incomplete in the sense that they lack accurate rate coefficients for 
inelastic collisions with H$_2$. In cases where collision rates are missing, rough estimates have 
been incorporated in order to include at least a crude treatment of the collisional component of 
excitation.

The chemical formation and destruction terms in equations 2 and 3, $F_i$ and $D_i$, are constrained so that the total number density of each molecule m is  
\begin{equation} n({\rm m}) = \sum n_i = \sum F_i/D\ cm^{-1} \end{equation} 
where $D = D_i$ for all $i$. As proposed by van der Tak et al. (2007), the formation rate is assumed to behave like $F_i \propto \exp(-E_i/kT_{\rm form})$, where $E_i$ is the energy of state $i$ and the formation temperature $T_{\rm form}$ need be merely high enough ($\sim 300$ K) to ensure that the relative populations of the nuclear-spin-symmetry species in H$_2$O, NH$_2$, and NH$_3$ approach their high-temperature equilibrium values. The meaning of the column density averaged over the adopted source area in Table 3 is ${\cal N}({\rm m}) = n( {\rm m}) L$ for each molecule, where $L$ is the projected source diameter, 5 arcsec, at the adopted distance. The value of the destruction rate has no noticeable effect on the computed excitation or line intensities as long as $D\leq 10^{-5}$ s$^{-1}$, which corresponds to destruction of neutral species by reaction with an ion whose fractional abundance is $10^{-4}$ of hydrogen at a total density $n_{\rm H}\approx 10^8$ cm$^{-3}$. 
The exceptional case of CH$^+$ is mentioned in the following section.

\begin{figure*}
\includegraphics[angle=90,scale=0.31]{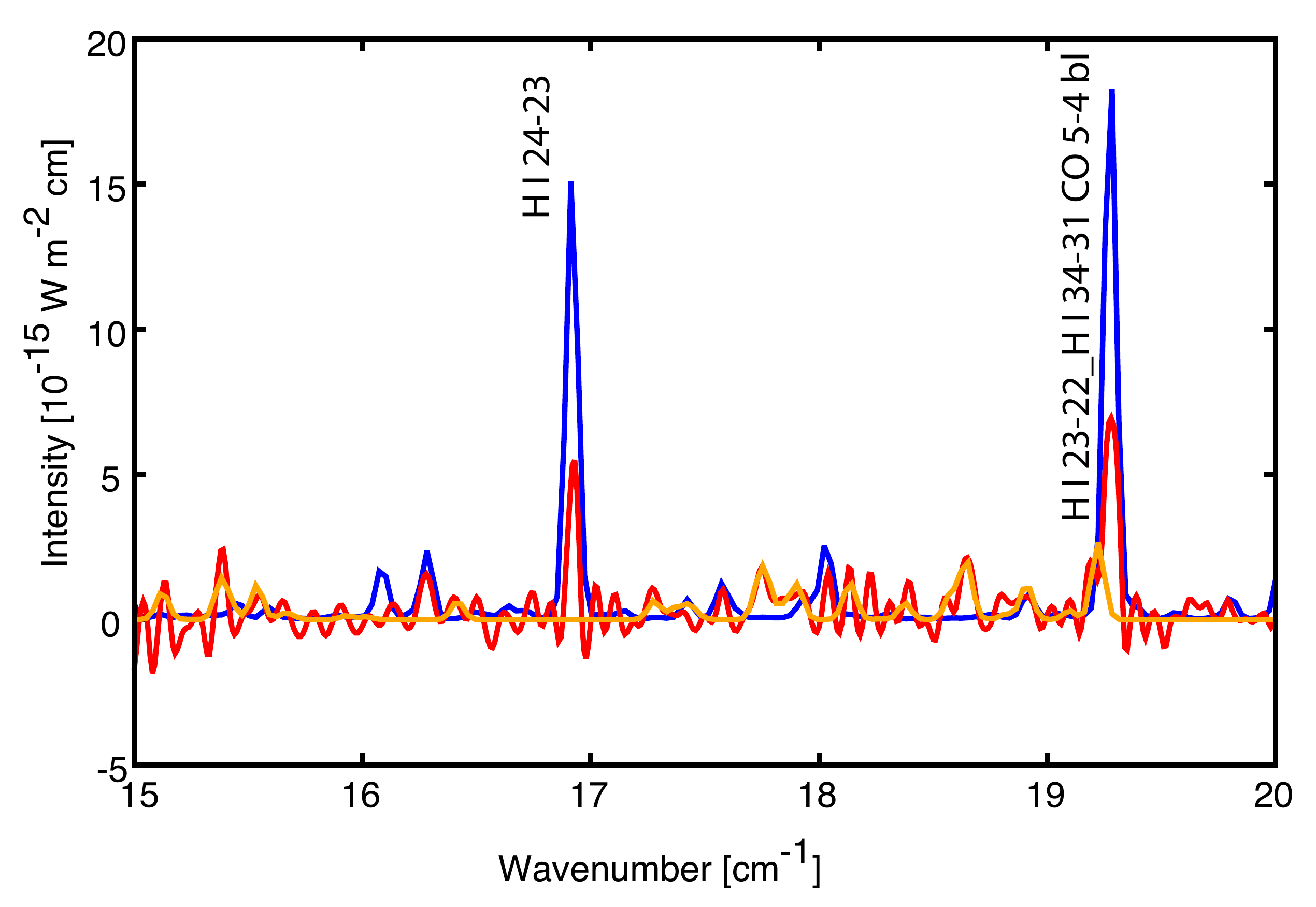}\includegraphics[angle=90,scale=0.31]{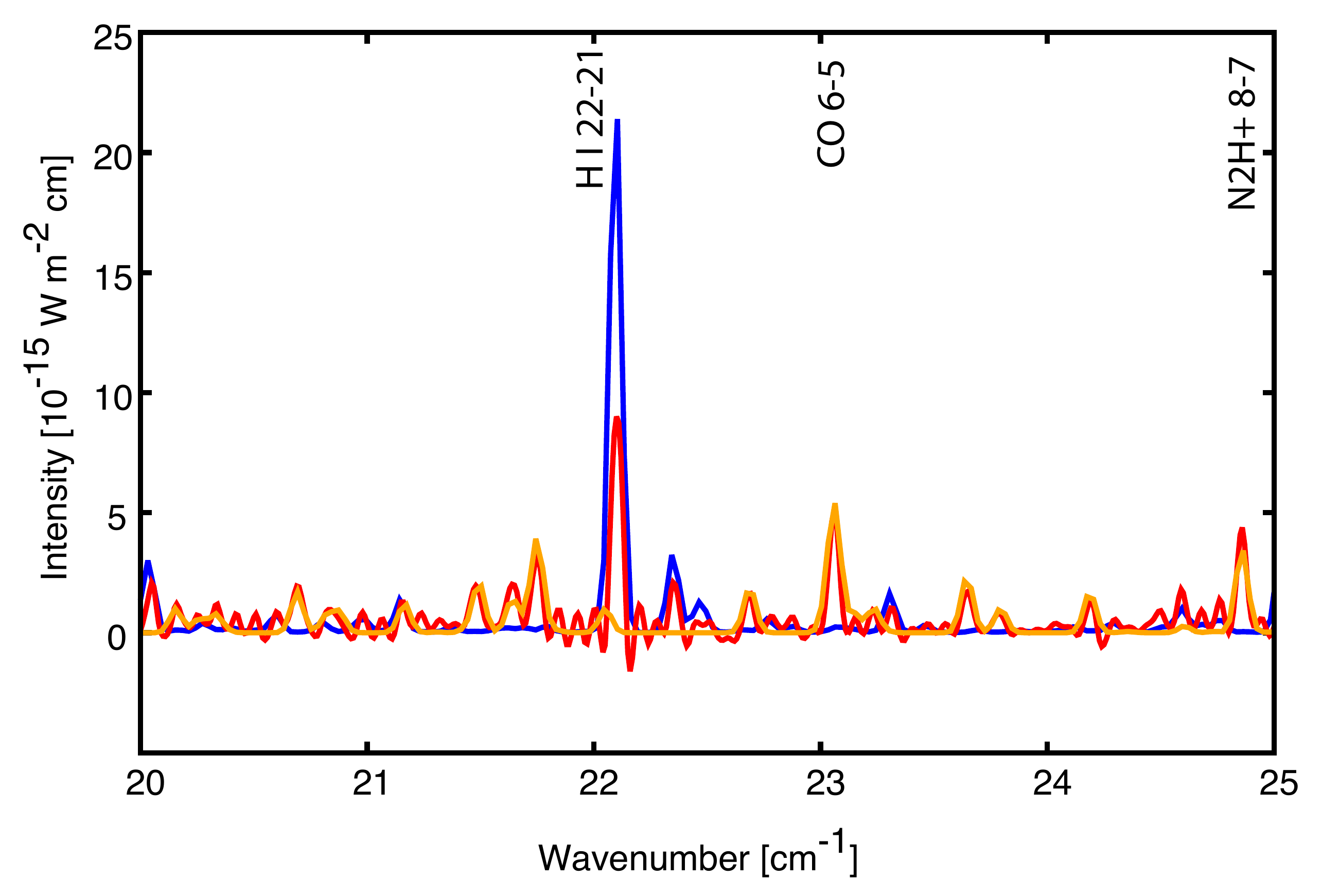}
\includegraphics[angle=90,scale=0.31]{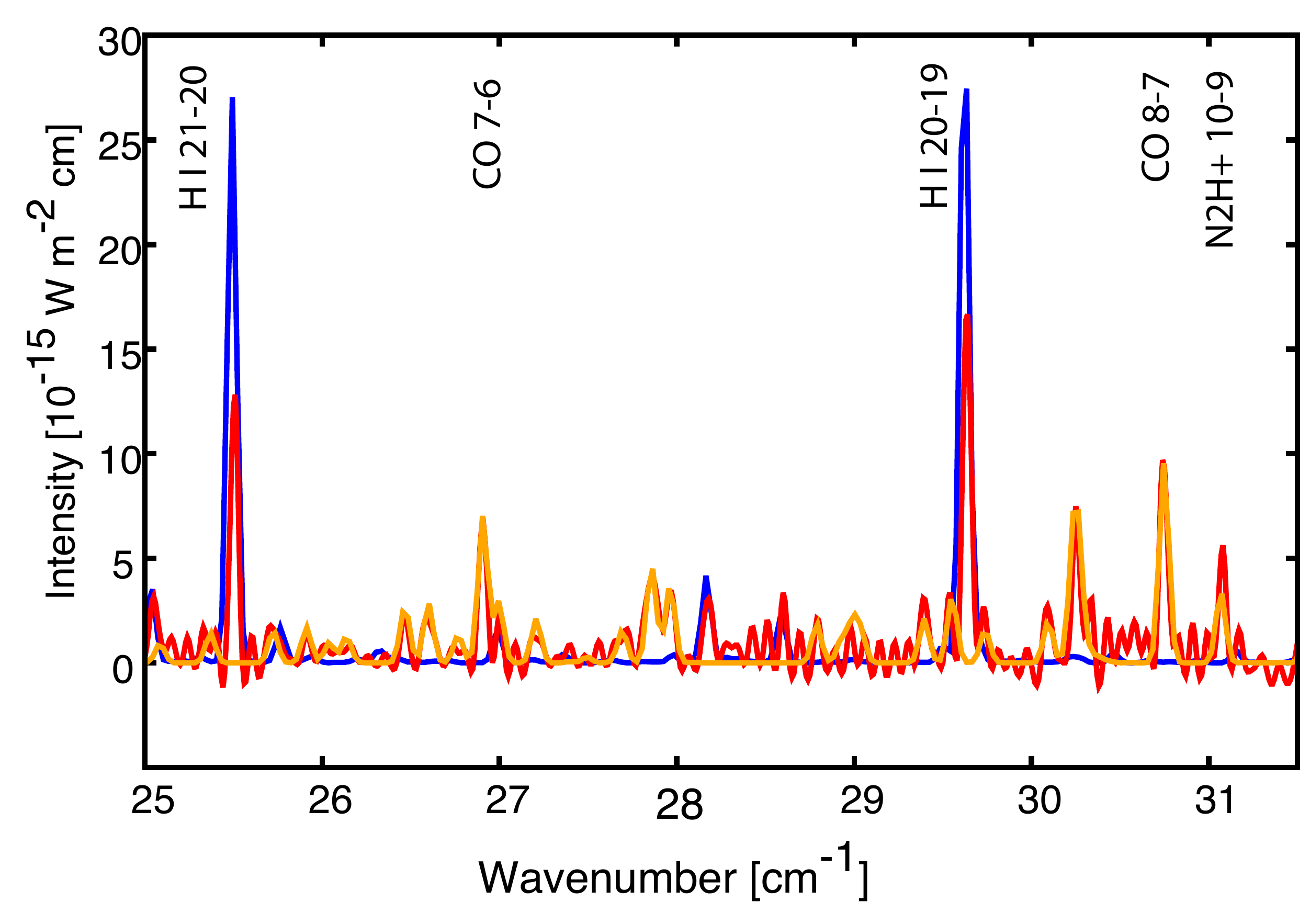}\includegraphics[angle=90,scale=0.31]{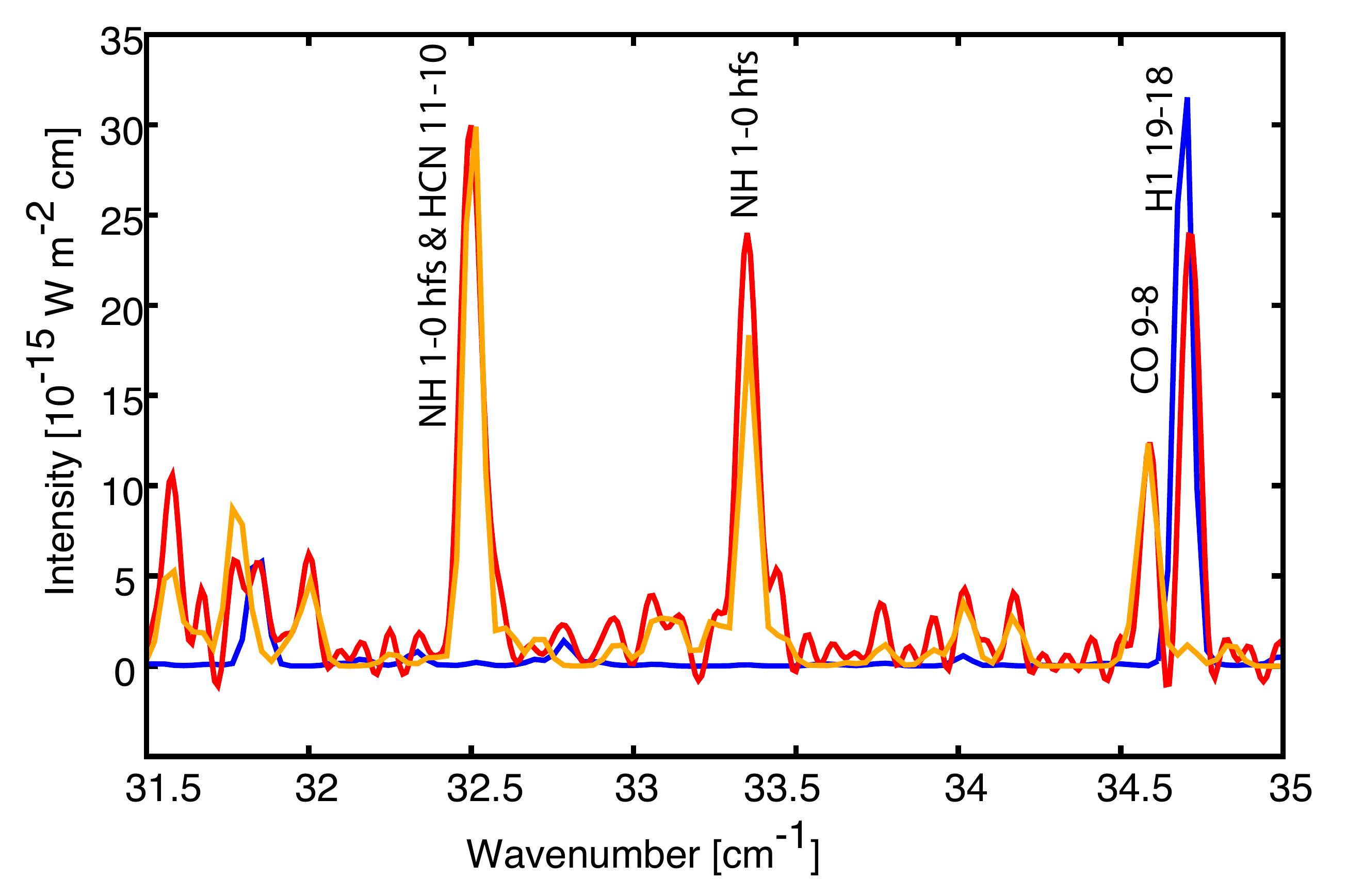}
\includegraphics[angle=90,scale=0.31]{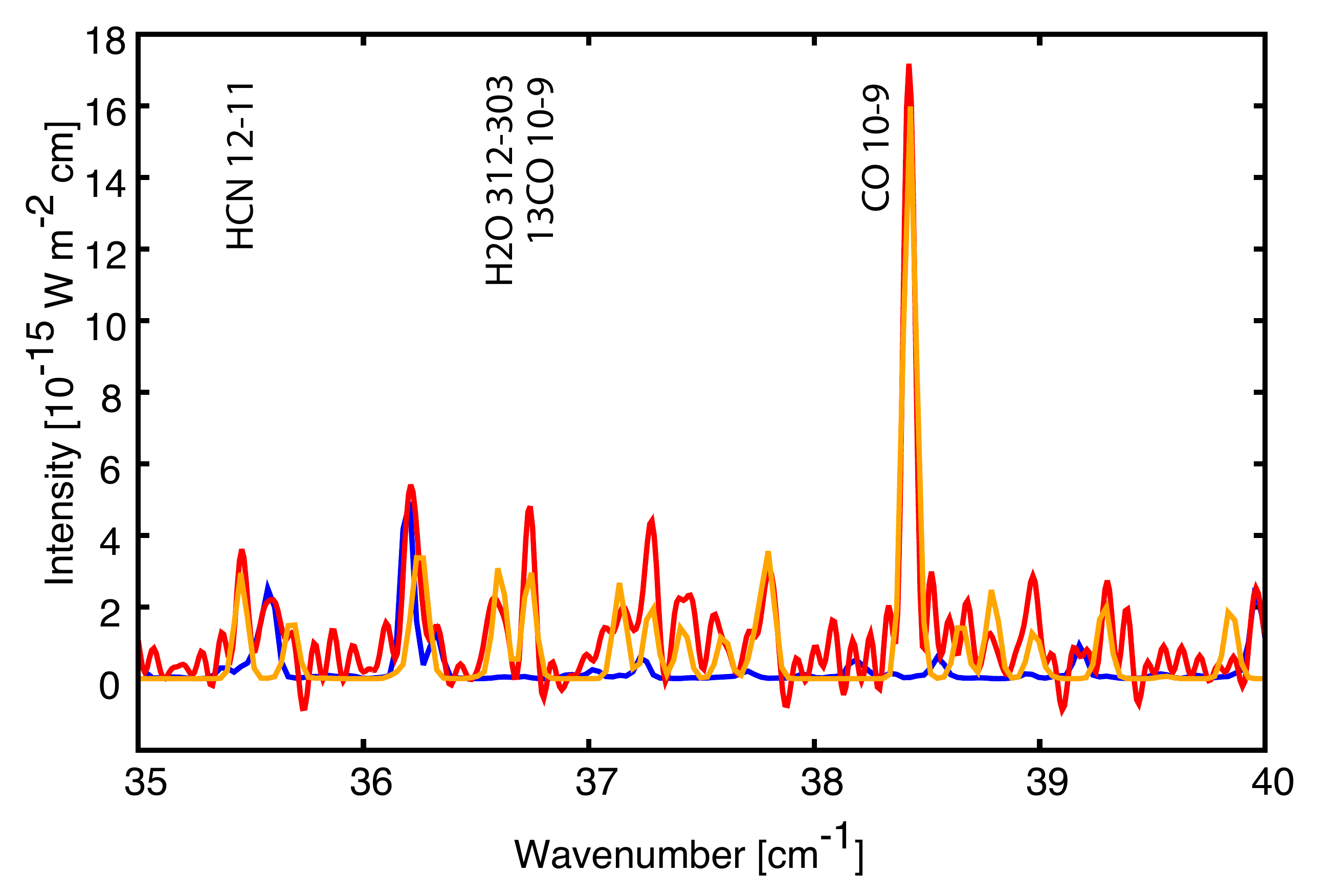}\includegraphics[angle=90,scale=0.31
]{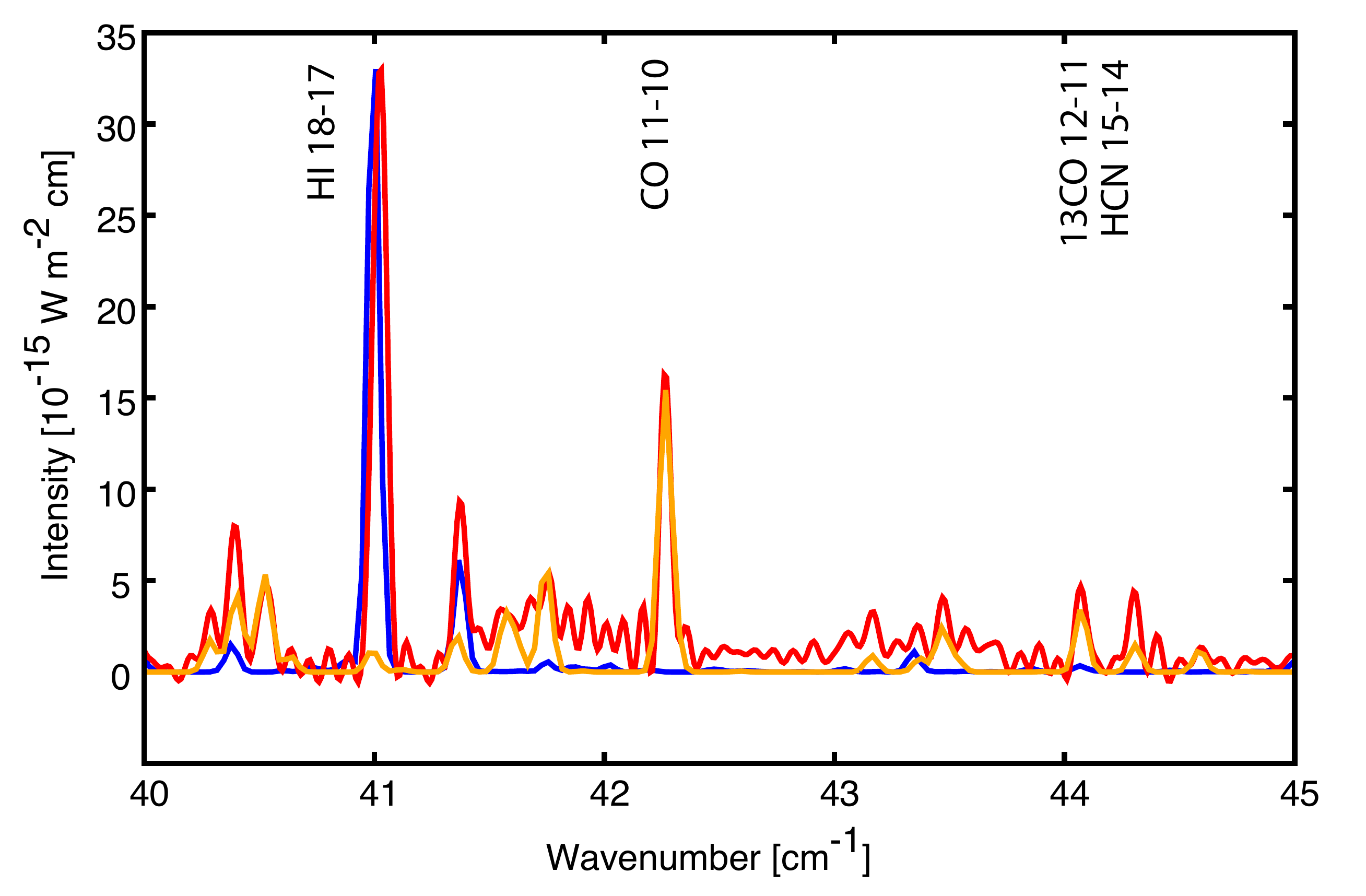}
\includegraphics[angle=90,scale=0.31]{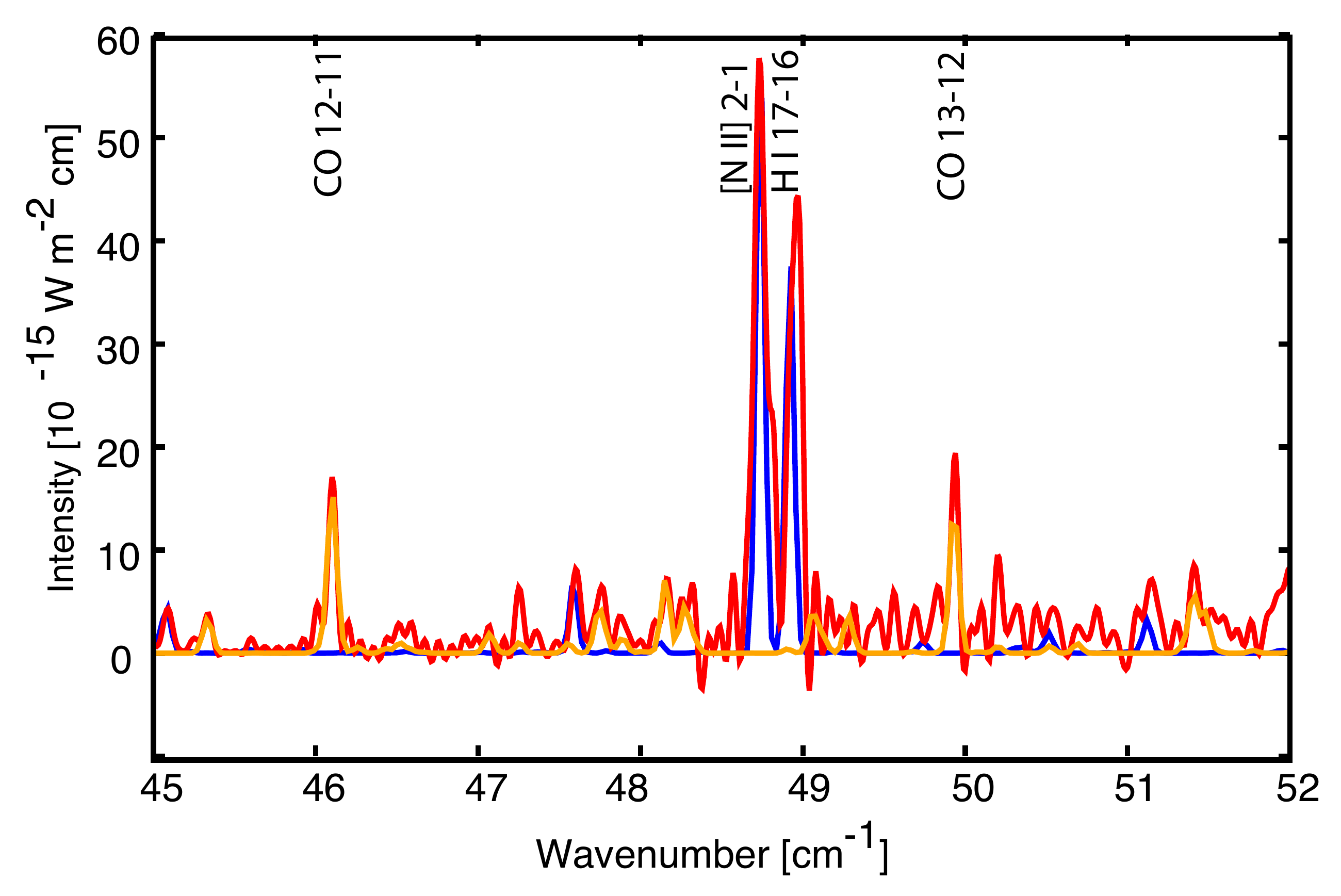}
\caption{{\bf Model fits to the \spire\ spectrum} Each subpanel includes the actual spectrum (red), the fit to the atomic recombination (blue)  and the fit to the molecular spectrum (orange). A number of brighter emission features are identified, but with this low spectral resolution, blends are abundant. Preliminary line identifications are listed in Table \ref{plotsid}1 through \ref{plotsid}4 and in more detailed plots in Figures \ref{Sfig}-1 through \ref{Sfig}-13. \label{models}}
\end{figure*}
\subsubsection{The Molecular Model Compared to the \spire\ Spectrum}\label{Smodel}
The results of the final model are displayed graphically in a series of seven panels  in Figure \ref{models}, showing the computed spectrum overlaid on the observed \spire\ spectrum.  This comparison clearly identifies 18 molecular species plus neutral
atomic carbon (see Table \ref{abundances}). At least two plausible line blends are needed for a credible identification. The 
column densities that yield the best match with observed intensities are summarized in Table \ref{abundances}.

The tabulated column densities reveal the relative abundances over the complete chemical network of species observed in the SPIRE spectrum.  The agreement in CO, $^{13}$CO, CN, HCN, H$^{13}$CN, HCO$^+$, HNC, and N$_2$H$^+$ relative abundances with previous studies of these species observed with single dish heterodyne receivers \citep{Loinard12, Morris17} is very good.  We note that the column densities tabulated by \cite{Loinard12} are more than an order of magnitude higher than our results and those of \cite{Morris17} for CO and $^{13}$CO. However this is mainly due to an assumed small source size of 1$''$ diameter, and secondarily to the exclusion of continuum interactions in the analysis by \citep{Loinard12}. CO shares a very similar velocity structure with the other molecules quoted here, and is distributed in the extended torus-like structure (see Fig.~\ref{nebular}), which we more reasonably approximate as a 5$''$ diameter sphere.  Taking these differences into account, our tabulated abundances are only slightly higher in those species by factors of 1.2 to 2.8, affording us to take confidence in chemical abundances to at least their tabulated orders of magnitude.

The ratios of $^{12}$C and $^{13}$C isotopologues require a separate, more careful assessment. The dominant $^{12}$C forms of several molecules include effects of infrared pumping through vibrational transitions; however, the minor forms of the same molecules lack
the extensive data files. As a result, the column densities of the minor isotopologues might be somewhat
overestimated, and the corresponding $^{12}$C/$^{13}$C abundance ratios underestimated. This effect 
is probably not severe for CO, suggesting that the isotope ratio might be consistent with a value of
approximately 4. Independent analysis by \cite{Morris17} of the CO ladder observed at high spectral resolution with \hifi\ and \APEX\ suggests an even lower value of [$^{12}$C/$^{13}$C$] \approx$3.}

The entry for H$_2$O in Table \ref{abundances} is very uncertain. Although several possible rotational lines of water occur over the \spire\ spectral range (Figs \ref{Sfig1} through \ref{Sfig13}) and line list (Table \ref{Tbl-SPIRE}), only one of these, the {\it{ortho}}-H$_2$O 3$_{12} - 3_{03}$ line at 36.604 cm$^{-1}$ (1097.36 GHz), is free of blending with a strong line of a different but otherwise well-identified species.  Nonetheless, H$_2$O lines have been convincingly detected with HIFI by \cite{Morris20} at 18.58 cm$^{-1}$ (556.94 GHz) and 32.96 (987.93 GHz), which are the {\it{ortho}} and {\it{para}} transitions 1$_{1,0} - 1_{0,1}$ and 2$_{0,2} - 1_{1,1}$, respectively.   They estimated column densities in the range of 2.5 - 5.0 $\times$ 10$^{13}$ cm$^{-2}$, which is about a factor of 20 to 40 lower than the upper limit set by our model of the SPIRE data.  Similarly, SiO has been included in the model, but is not detected in our \hso\ spectra or reported in ground-based observations.

The highly reactive molecular ion CH$^+$ is exceptional. CH$^+$ is destroyed on nearly every collision with the most abundant collision partners (H, H$_2$, electrons), so that "chemical pumping" can compete with inelastic collisions in its rotational excitation. CH$^+$ also couples very strongly to the intense far-infrared radiation field in eta Car. Accordingly, CH$^+$ excitation was treated with higher formation and destruction rates than the other molecules. Even with these complications, the models show that when $D({\rm CH}^+)$ is changed from 0.1 to $10^{-6}$ s$^{-1}$, the calculated intensities of the two lowest rotational transitions change by less than 30\%. Only the $J=1-0$ line is covered in the SPIRE passband, thus the entry for CH$^+$ in Table 3 is based upon a single transition.

Unfortunately, there is no simple way to estimate the total hydrogen column density for the same
neutral emitting zone that produces most of the molecular line emissions in the \spire\ spectrum.
The elemental abundances in the ejecta of $\eta$ Car are known to be unusual compared with
abundances in the interstellar medium. Both the overabundance of nitrogen and the low ratio of $^{12}$C/$^{13}$C reinforce the fact that the ejecta is heavily-processed material from a massive star.  We will return to the abundances in (Sec.~\ref{concl}).

\begin{table}
\caption{Column abundances derived from the \spire\ spectrum}
\begin{center}
\begin{tabular}{llrc}
\hline \\
 & Molecule & $\mathcal{N}^a$ & Vibration$^b$ \\
 &  & [cm$^{-2}$] & \\
 \hline
& CO &   6.1 (17) & Y \\
& $^{13}$CO &   1.5 (17) & N \\
& \ion{C}{I}   &     6.2 (18) & N \\
& HCO$^+$   &  1.0 (14) & Y \\
& H$^{13}$CO$^+$  &   6.0 (13) & N \\
& CH   &   3.0 (15) & N \\
& CH$^+$  &    7.5 (14) & Y \\
& SiO$^c$   &  <  1.0 (14)  & Y \\
& H$_2$O$^d$   &  <  1.0 (15)  & Y \\
& HCl  &    5.4 (14)  & N  \\
& NH   &    5.0 (15)  & N \\
& NH$_2$  &      2.5 (15) & N \\
& NH$_3$  &      1.0 (15) & Y \\
& N$_2$H$^+$  &     4.2 (14)  & Y \\
& CN   &   1.2 (15)  & N \\
& $^{13}$CN   &  7.5 (14) & N \\
& HCN  &    4.0 (14) & Y \\
& H$^{13}$CN  &  2.5 (14) & N \\
& HNC  &    2.7 (14) & Y \\
& HN$^{13}$C  &  1.7 (14)  & N \\
\hline
\end{tabular}
\end{center}
\noindent $^a$The column density $\mathcal{N}$ refers to an average over the adopted source area, 5 arcsec in diameter. The notation X (17) means X$\times 10^{17}$.\\
\noindent $^b$``Yes'' indicates that infrared vibration-rotation transitions are included in the excitation. \\
\noindent $^c$Not detected.  The derived abundance is an upper limit only.\\
\noindent $^d$Transitions are weak and blended, but detected in \hso/HIFI spectra \citep{Morris20}.
\label{abundances}
\end{table}

\
  
\subsubsection{Modeling the Hydrogen Rydberg transitions}}\label{ions}

The \spire\ spectrum of $\eta$ Car is dominated by strong lines of atomic hydrogen, arising in Rydberg
states with principal quantum number $n\geq 16$. These lines most likely arise in photo-ionized nebular
gas close to the star or in an ionized wind. Recent \alma\ observations of Rydberg transitions at 
longer mm wavelengths suggest that the line emission is concentrated on {sub-arcsecond angular scales
\citep{Abraham14}. Astronomical Rydberg transitions are commonly referred to as recombination lines, because
the high-$n$ levels in dilute nebulae are populated initially by the process of radiative recombination
$$ {\rm H}^+ + e^- \to {\rm H}(n) + \gamma  $$
followed by a radiative cascade $n\to n''$, with additional redistribution by inelastic collisions with 
electrons and protons. In the dense, ionized wind of a LBV like $\eta$ Car, it is possible that some of 
the excitation derives directly from electron-impact out of the lowest levels of the neutral atom, in which
case it is not strictly appropriate to call the observed transitions recombination lines. 

The atomic lines in the \spire\ spectrum are unresolved, and the projected entrance aperture is much
larger than the extent of the dense nebula and/or ionized wind. Unlike a typical idealized nebula, 
the ionized gas around $\eta$ Car is exposed to an extremely intense radiation field at far-infrared and 
sub-millimeter wavelengths. Absorption and stimulated emission in this radiation may contribute
significantly to the excitation of the observed Rydberg states in H and He. At the same time, a
realistic model must account for the intensities of the lines that are observed in excess of a strong
continuum. It is beyond the scope of the present
analysis to perform a fully self-consistent model of the ionized gas. Instead, we 
treat the atomic lines similarly to the molecular lines, striving first of all to assess the competition 
among the processes of radiative capture and cascade, inelastic collisions with electrons, and 
radiative pumping under nebular conditions.  
We apply the same expanded version of the {\tt RADEX} code discussed above for molecules
in order to compute
single-zone, non-LTE models of the excitation and line spectra of H~I, He~I, and He~II. 
Atomic data files contain levels n=1 - 500 with all allowed radiative transitions and inelastic $e$-impact rate coefficients similar to those adopted by \cite{Hummer87}  
The rate of formation of each initial state $n_i$
is computed as

\begin{equation}
 F(n_i) = \alpha_{n_i}(T_e) N(e) N({\rm ion}) \;\;\;{\rm cm}^{-3}\;{\rm s}^{-1} $$ 
  \label{eq:hformrate} 
\end{equation}
where $\alpha_{n_i}(T_e)$ is the rate coefficient of radiative recombination into level $n_i$ at 
electron temperature $T_e$, computed through use of computer routines published by 
Flower \&\ Seaton (1969). Number densities of electrons, $N(e)$, and the recombining ion, 
$N({\rm ion})$ are given upper-case symbols here to distinguish from the principal quantum number
$n$. 

The reference model is a uniform sphere of hydrogen, helium, and electrons that is assumed to 
subtend an angular diameter of $0\farcs5$. The corresponding path length is $L=1.72\times 10^{16}$ 
cm at the distance of $\eta$ Car. 
The adopted He/H abundance ratio is 0.08 by number  of nuclei. Owing to the strong effect of 
radiative pumping on Rydberg states $n=15-30$, the spectrum is not expected to be sensitive to 
temperature; therefore, we simply assume $T_e = 8000$ K. 

For guidance on the neutral fractions and formation
rates inside the nebula, we take
a very simple model of a static photo-ionized nebula illuminated by the ultraviolet light of a 37500 K 
blackbody with a bolometric luminosity of $10^6$ L$_{\odot}$. This is a poor representation of 
the marvelous complexity of $\eta$ Car, whose true ultraviolet spectrum is completely hidden from us. 
The diameter of the corresponding Str{\"o}mgren sphere is set equal to the adopted path length $L$.
Under these assumptions, the total density of hydrogen is $N_H=7.6\times 10^6$ cm$^{-3}$, and 
the column density of neutral H within the ionized zone is ${\mathcal N}({\rm H}) = 3.6\times 10^{18}$ cm$^{-2}$. These numbers set the conditions for the {\tt RADEX} models of the \ion{H}{i} spectrum, 
notably the total rates of formation (Eq.~\ref{eq:hformrate}) and destruction of neutral H within the nebular zone.  
The same model of the continuum spectrum adopted for the molecular excitation is applied here.
That spectrum was computed from the observed SED averaged over an adopted source diameter
of 5.0 arcsec.

The distributions of He, He$^+$, and He$^{++}$ within the ionized zone are sensitive to the detailed shape
of the input ionizing spectrum. In the reference model in one dimension, helium is singly ionized out
to 93\%\ of the ionized-hydrogen radius, and doubly ionized only to 8.25\%\ of the radius. The 
predicted He II spectrum from the recombination of He$^{++}$ is useful not as a quantitative prediction
but only for the sake of constructing a finding list of possible lines.  

Even though the reference model of the ionized gas is simplistic, it shows that nebular gas of 
the indicated density, temperature, and extent can provide \ion{H}{i} lines with intensities of the right
order of magnitude. Moreover, the simple model allows us to assess quantitatively the atomic 
processes that control the line emission in the environment of $\eta$ Car. In the model 
the Rydberg states of H with $n\geq 20$ follow rather closely a Boltzmann distribution but at a
temperature $T' = 3855$ K that is rather smaller than the kinetic temperature. At lower $n$, the
populations are inverted. As a result, the lowest $\alpha$ lines ($\Delta n=1$) in the \spire\ spectrum, 
H $16\alpha$ and $17\alpha$, are weak masers with optical depths $\tau > -0.1$, while H $18\alpha$ 
through $22\alpha$ have suprathermal excitation temperatures in the range $T_{\rm ex} = 10000$ to
40000 K. These features of the model are mainly due to the strong coupling to the continuum 
radiation in the far-infrared. The synthetic spectrum of the reference model is displayed as red in Figure \ref{models}.
Below in Section~\ref{swsH} we will apply our modeling to the mid-infrared H lines observed with \iso\ to examine the non-LTE effects of the coupling.

\begin{figure*}
\includegraphics[angle=0,scale=0.75]{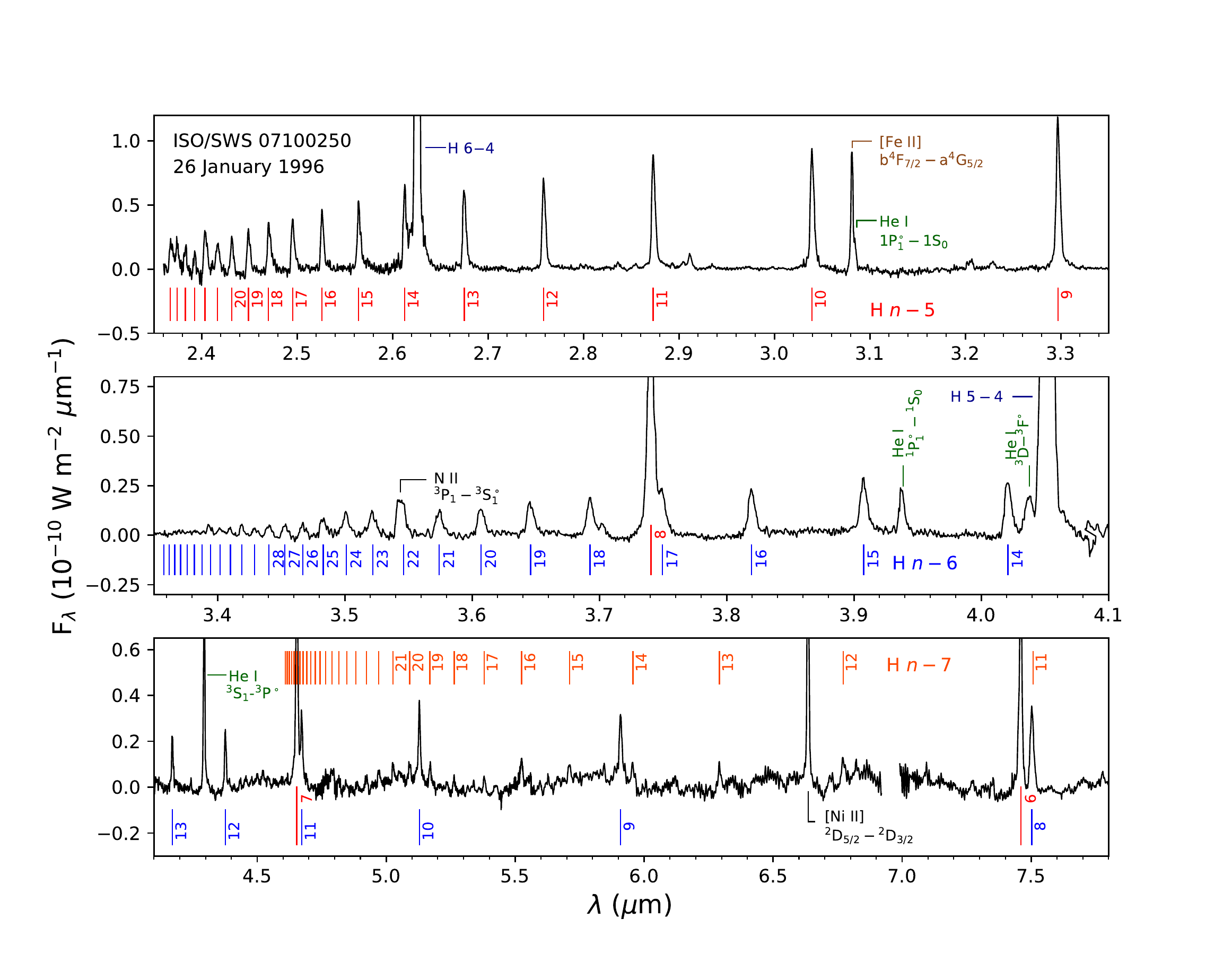}
\caption{The continuum-subtracted \iso/SWS spectrum of $\eta$ Car over the first 5 grating sub-bands of the S01 spectral scan obtained in 1996.  The continuum rises from 2.5 $\times$ 10$^{-10}$ to  6.2 $\times$ 10$^{-10}$ W m$^{-2}$ $\mu$m$^{-1}$ (500 to 2050 Jy) over this range. The \ion{H}{I} lower quantum state $n_{l}$ = 5 (Pfund), 6 (Humphreys), and 7 series are indicated, along with the Brackett $\alpha$ ($5-4$) line and identifications for leading lines of other atomic species.  The spectral resolution $R$ = 1500-2000, or 200 \kms\ at 3.0 $\mu$m.
 \label{swslines}}
\end{figure*}

The frequency interval 400 to 1600 GHz contains 2478 lines of \ion{H}{I}, \ion{He}{I}, and \ion{He}{II} with line-centre
optical depths $\tau>10^{-4}$. 
The peak intensities of lines in the model agree well with the observations at the lower frequencies,
but at $\nubar > 25$ cm$^{-1}$ ($\lambda$ < 400 $\mu$m), the $\alpha$ lines $\Delta n=1$ are approximately $1/2$ as strong 
as observed, while the weaker, higher-order $\beta$ and $\gamma$ lines ($\Delta n=2$, 3) are 
somewhat closer to the observations. The intensities of the $\alpha$ transitions are most sensitive
to the non-LTE effects at \spire\ frequencies. These effects are thought to be  closely tied 
to the balance between density-dependent recombination and collisional excitation on the one hand
and radiative pumping in the ambient continuum on the other hand. As a result, no attempt has been made to fine-tune the model parameters to achieve a better fit.

Even though the He I lines are $\sim 0.1$ times as strong 
as the corresponding \ion{H}{i} lines, it is important to include them, because they shift centroids of the 
unresolved blends into better agreement with the peak positions in the observed spectrum.  

The intensity of [\ion{N}{ii}] line at $\nubar = 48.738$ cm$^{-1}$ (1461 GHz) requires a column density of 
ionized nitrogen $\mathcal{N}({\rm N}^+) = 3.14\times 10^{21}$ cm$^{-2}$ in the nebular model,
corresponding to an abundance ratio N$^+$/H$^+$ = 0.024 in the ionized gas. 

Finally, the total mass
of gas (hydrogen and helium) in the ionized region is only 0.022 M$_{\odot}$, trivial in comparison with
the mass estimated for the neutral and dusty ejecta of $\eta$ Car.

\subsubsection{A new analysis of the mid-infrared \ion{H}{I} line spectrum}
\label{swsH}

An immediate and lasting impression of $\eta$ Car's infrared SED is the enormous amount of thermal energy produced by heated dust grains, peaking in flux density at 70,000 Jy near 25 $\mu$m; see Figure 7 in \cite{Morris17}.  Despite the steeply rising continuum in the $2-25 \; \mu$m range, a rich spectrum of \ion{H}{I} lines from the ionized gas of the wind and surrounding nebula is observed at high S/N in the \iso/SWS spectrum, from the instrument's lower wavelength cutoff of 2.4 $\mu$m to around 9 $\mu$m.  The continuum-subtracted spectrum is shown in Figure~\ref{swslines}.  An analysis of these lines complements the preceding interpretation of the far-IR Rydberg transitions, providing insight into the role of the strong continuum on the spectral energy budget of these lines over more than a factor of 100 in wavelength. 

All of the lines identified in this part of the spectrum can be attributed to the ionized gas of the wind or compact nebula, not necessarily at the same velocities. This is nicely illustrated in higher spectral resolution observations of several \ion{H}{I} $\alpha$ transitions, shown in Figure~\ref{highresH} where the SWS spectrum of H Br $\alpha$ was obtained in the line scan (S02) mode at 90 \kms\ resolution, and the HIFI spectrum of the sub-millimeter H22 $\alpha$ line is shown with $\approx$10 \kms\ smoothed resolution.  The HIFI Rydberg lines are weakly structured and quite narrow compared to the Br $\alpha$ line, which is composed of a central core, best fit by a Lorentz profile with a FWHM of 180 \kms\ (corrected for the SWS instrument profile), and extended wings of emission to which we have fit Gaussian profiles centered near $-$220 and $+$280 \kms.  If there is any mid-infrared \ion{H}{I} emission from the gas component that gives rise to the sub-millimeter line emission at $v_{\rm{LSR}} = -50$ \kms, it is completely buried in the large beam of the SWS, and too weak to influence the profile fitting.  This conclusion is consistent with our model calculations, discussed below.

\begin{figure}
\includegraphics[angle=0,scale=0.55]{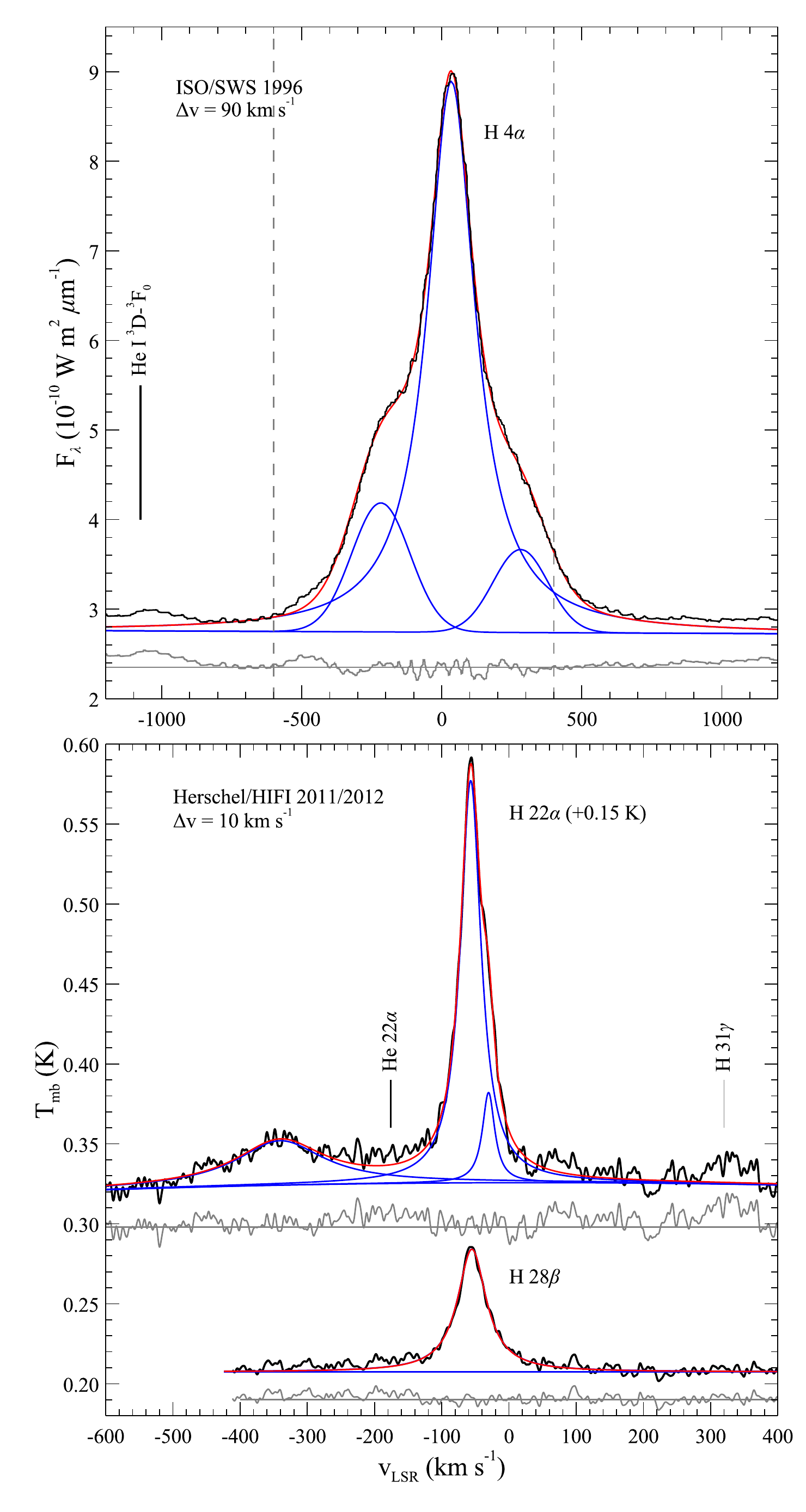}
\caption{\iso/SWS and \hso/HIFI observations of \ion{H}{I} $\alpha$ lines.  The SWS spectrum of Br $\alpha$ (top) was obtained in the highest grating resolution mode for line scanning, at $R$ = 3500.  Vertical dashed gray lines indicate the plotted range of the H22 $\alpha$ line shown in the lower panel, to emphasize the difference in profile widths.  The broad, weak feature near $-$350 \kms\ in the lower plot is unidentified.}
 \label{highresH}
\end{figure}

The continuum subtraction in the SWS spectrum at $\lambda$ > 4.5 $\mu$m is affected by the presence of broad dust bands, leading to a somewhat choppy residual baseline in the lower panel of Figure~\ref{swslines}; however there are no gas-phase molecular lines evident in the SWS spectrum. For example, vibration-rotation first overtone and fundamental lines of CO at 2.3 and 4.6 $\mu$m, NH$_2$ near 6.7 $\mu$m, and NH$_3$ near 10 $\mu$m (out of the plotted range) might be expected, although the detectability of such lines would depend both on the excitation temperatures relative to the continuum brightness, on the fraction of the continuum covered by the molecular gas, and on the instrumental resolution. The non-LTE molecular model of the \hso\ spectra can be used to predict the mid-infrared spectrum: no lines are predicted to be detectable against the intense continuum at the resolution of the SWS spectra.

The rich mid-IR \ion{H}{I} spectrum complements the interpretation of the far-IR Rydberg transitions, as illustrated in Figure~\ref{excitation} giving a view on  excitation in which the measured line fluxes (listed in Table~\ref{table_H}) are displayed as violet crosses with statistical error bars versus wavelength of the transition. The green circles indicate the intensities expected in LTE at a temperature $T_e$ = 10$^4$ K and reddened by $A_V$ = 5 mag of visual extinction using a standard Galactic extinction law (e.g., \citealt{Indebetouw05}). The numeral inside each green circle is the principal quantum number of the upper state $n_{u}$ of the transition. The red circles indicate the intensities predicted by the same {\it{non-LTE}} model that was applied to the SPIRE spectra, but scaled to the observed flux of the H~Br~$\alpha$ (5-4) line at 4.08 $\mu$m. 

\begin{table*}
\caption{Measured \iso/SWS 2.4$-$8.8 $\mu$m line fluxes$^*$}
\begin{center}
\begin{tabular}{ccrrrl}
\hline \\
Observed & Vacuum & \multicolumn{1}{c}{Line Flux}    & $n_{l}$ & $n_{u}$ & \multicolumn{1}{l}{Identification}\\ 
$\mu$m  & $\mu$m &  \multicolumn{1}{c}{10$^{-14}$ W m$^{-2}$}  &   &    &  \\
\hline
2.49577 (0.00008)  & 2.49526  &  10.77 (0.79)  & 5  & 17  & \ion{H}{I}	unless indicated	\\		
2.52643 (0.00006)  & 2.52610  &  14.36 (0.75)  & 5  & 16  & \\					
2.56477 (0.00006)  & 2.56433  &  19.04 (0.79)  & 5  & 15  & \\					
2.61397 (0.00008)  & 2.61266  &  35.01 (1.19)  & 5  & 14  & \\					
2.62580 (0.00001)  & 2.62588  & 250.14 (0.94)  & 4  & 6   & \\					
2.67551 (0.00005)  & 2.67514  &  27.95 (0.91)  & 5  & 13  & \\					
2.75864 (0.00004)  & 2.75828  &  32.37 (0.86)  & 5  & 12  & \\					
2.87344 (0.00003)  & 2.87300  &  39.65 (0.79)  & 5  & 11  & \\					
2.91110 (0.00030)  & 2.91144  &   6.12 (0.89)  & \dotfill &  \dotfill & [\ion{Ni}{II}] $^4$F$_{5/2}-^2$F$_{7/2}$ \\					
3.03961 (0.00003)  & 3.03921  &  43.93 (0.75)  & 5  & 10  & \\					
3.08159 (0.00003)  & 3.08095  &  28.00 (0.61)  &  \dotfill &  \dotfill & [\ion{Fe}{II}] b$^4$F$_{7/2}-$a$^2$G$_{5/2}$ \\					
3.29728 (0.00002)  & 3.29700  &  57.71 (0.63)  & 5  & 9   & \\					
3.41937 (0.00031)  & 3.41911  &   2.53 (0.53)  & 6  & 30  & \\					
3.44072 (0.00034)  & 3.44032  &   2.18 (0.53)  & 6  & 28  & \\					
3.45309 (0.00034)  & 3.45286  &   1.97 (0.51)  & 6  & 27  & \\					
3.46701 (0.00033)  & 3.46698  &   2.39 (0.53)  & 6  & 26  & \\					
3.48349 (0.00027)  & 3.48297  &   4.94 (0.63)  & 6  & 25  & \\					
3.50163 (0.00023)  & 3.50117  &   7.73 (0.69)  & 6  & 24  & \\					
3.52203 (0.00023)  & 3.52203  &   7.18 (0.74)  & 6  & 23  & \\					
3.54403 (0.00014)  & 3.54611  &  14.99 (0.82)  & 6  & 22  & + \ion{N}{II} $^3$P$_1-^3$S$^\circ_1$ \\		
3.57467 (0.00022)  & 3.57411  &   9.35 (0.81)  & 6  & 21  & \\					
3.60744 (0.00018)  & 3.60698  &  10.03 (0.77)  & 6  & 20  & \\					
3.64639 (0.00016)  & 3.64593  &  12.55 (0.78)  & 6  & 19  & \\					
3.69277 (0.00018)  & 3.69264  &  13.68 (0.92)  & 6  & 18  & \\					
3.74046 (0.00004)  & 3.74057  &  76.39 (0.95)  & 5  & 8   & \\					
3.74976 (0.00021)  & 3.74940  &  20.81 (1.17)  & 6  & 17  & \\					
3.81991 (0.00010)  & 3.81946  &  16.36 (0.69)  & 6  & 16  & \\					
3.90783 (0.00009)  & 3.90756  &  14.51 (0.62)  & 6  & 15  & \\					
4.02111 (0.00008)  & 4.02088  &  15.49 (0.57)  & 6  & 14  & \\					
4.04313 (0.00026)  & 4.04429  &  25.71 (1.29)  &  \dotfill &  \dotfill & \ion{He}{I} $^3$D	$-^3$F$^\circ$ triplet\\			
4.05211 (0.00000)  & 4.05228  & 361.88 (0.68)  & 4  & 5   & \\					
4.17157 (0.00013)  & 4.17080  &  28.72 (1.06)  & 6  & 13  & \\					
4.29471 (0.00002)  & 4.29596  &  72.61 (0.80)  &  \dotfill &  \dotfill & \ion{He}{I} $^3$S$_1-^3$P$^\circ$	\\			
4.37693 (0.00009)  & 4.37646  &  18.58 (0.76)  & 6  & 12  & \\					
4.65386 (0.00002)  & 4.65379  & 105.82 (0.79)  & 5  & 7   & \\					
4.67250 (0.00009)  & 4.67252  &  32.79 (0.87)  & 6  & 11  & \\					
4.97177 (0.00053)  & 4.97089  &   2.05 (0.60)  & 7  & 22  & \\					
5.02635 (0.00032)  & 5.02610  &   3.91 (0.57)  & 7  & 21  & \\					
5.12883 (0.00007)  & 5.12867  &  30.16 (0.66)  & 6  & 10  & \\					
5.17013 (0.00024)  & 5.16929  &   8.58 (0.66)  & 7  & 19  & \\					
5.26336 (0.00028)  & 5.26369  &   3.71 (0.52)  & 7  & 18  & \\					
5.52351 (0.00054)  & 5.52520  &  16.32 (1.77)  & 7  & 16  & \\					
5.70929 (0.00072)  & 5.71147  &   9.02 (1.52)  & 7  & 15  & \\					
5.90880 (0.00019)  & 5.90823  &  30.71 (1.40)  & 6  & 9   & \\					
5.95737 (0.00046)  & 5.95685  &   9.17 (1.24)  & 7  & 14  & \\					
6.29316 (0.00048)  & 6.29193  &  11.42 (1.26)  & 7  & 13  & \\					
6.63551 (0.00004)  & 6.63596  &  87.70 (0.99)  &  \dotfill &  \dotfill & [\ion{Ni}{II}] $^2$D$_{5/2}-^2$D$_{3/2}$ \\			
6.77325 (0.00040)  & 6.77200  &   8.76 (0.99)  & 7  & 12  & \\					
7.45930 (0.00004)  & 7.45988  & 168.75 (1.48)  & 5  & 6   & \\					
7.50340 (0.00010)  & 7.50251  &  82.69 (1.59)  & 6  & 8   & blend 7.50812 \ion{H}{I} 7$-$11\\		
8.76050 (0.00063)  & 8.76008  &  12.45 (3.69)  & 7  & 10  & \\					
\hline
\end{tabular}
\end{center}
\begin{flushleft}
\noindent $^*$Values in parentheses are statistical (formal) measurement uncertainties.  The wavelength calibration of the SWS grating is accurate to $\lambda/\delta\lambda \sim 10000$, or 0.0003 $\mu$m at 3.0 $\mu$m \cite{Valentijn96}.  Relative line intensity calibration errors are less than 3\% from 2.4 to 4.1 $\mu$m, and around 5\% over 4.1 - 8.8 $\mu$m.
\end{flushleft}
\label{table_H}
\end{table*}

The key result of Figure~\ref{excitation} is that the lines of the \ion{H}{I} Pf, Hu, and $n_l$ = 7 series show significant deviations from LTE intensities for upper-state quantum numbers $n_{u}$ in the range 13 through 25. This is the same range of quantum numbers sampled by the $\alpha$ lines, $\Delta n$ = 1, in the SPIRE and PACS spectra. These deviations can be attributed to the radiative pumping of these Rydberg states by the intense continuum radiation of $\eta$ Car. The magnitude of these effects is quite sensitive to the details of the radiative transfer, which our single-zone escape-probability model does not completely represent, and deserves follow-up. The model does, however, reveal that the $\alpha$ and $\beta$ transitions with $n_{u}$ $\approx 13-25$ have inverted populations and negative excitation temperatures. The maser amplification factors in the nominal model are close to unity.

\begin{figure}
\includegraphics[angle=0, scale=0.65]{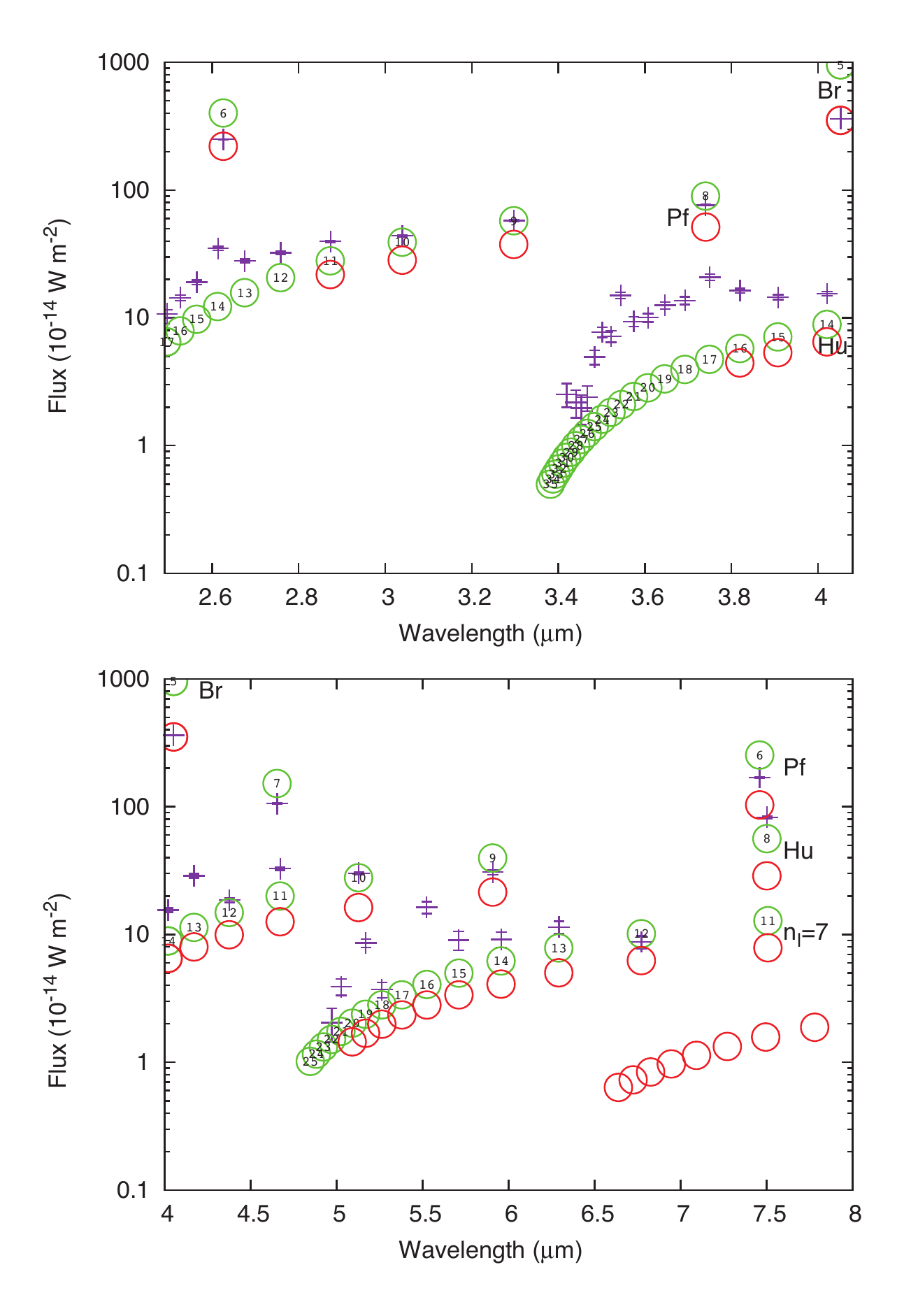}
\caption{Excitation diagram for the mid-infrared hydrogen lines observed with \iso/SWS.  The integrated fluxes of the \ion{H}{I} Br, Pf, Hu, and $n_l$ = 7 series are shown as violet crosses as a function of wavelength. The upper-state quantum numbers are indicated inside the green circles, which show fluxes expected in LTE at $T_e$ = 10$^4$ K when reddened by an interstellar extinction curve with $A_V$ = 5 mag. The red circles show the predicted fluxes of the non-LTE capture-cascade model used to analyze the far-infrared \hso\ spectra.
 \label{excitation}}
\end{figure}

\section{Summary and Conclusions}\label{concl}

\subsection{$\eta$ Car's molecular inventory}

A critical view generally held prior to the launch of \iso\ of \ec\ lacking the conditions to support a significant observable component of molecular gas other than H$_2$ is transformed almost immediately by the rich \pacs\ and \spire\ spectral scans presented in this paper, as well as subsequent ground-based observations acquired during the analysis and in followup to our \iso\ program.  Indeed, while the detected lines span tens to hundreds of \kms\ in width and are relatively weak over the thermal continuum particularly at $\lambda <$ 300 $\mu$m, the inventory of molecules detected in \ec\ stands now at 20, including the 18 molecules listed in Table~\ref{abundances} (i.e., excluding SiO which is not detected), H$_2$ \citep{Smith02}, and possibly CH$_3$OH \citep{Morris20}.   Of the species tabulated here, about half have previously been reported through transitions accessible from the ground in the sub-millimeter range (CO, $^{13}$CO, CN, HCN, H$^{13}$CN, HCO$^+$, HNC, and N$_2$H$^+$ by \citealt{Loinard12}) and in the UV from space with \hst/STIS (CH and OH by \citealt{Verner05b}).  Unambiguous detections of H$_2$O and NH with \hso/HIFI, and identification of a feature in \alma\ observations and supporting weaker features in HIFI spectra as CH$_3$OH have been published as part of our \hso\ program \citep{Morris20}.  

Our observational results also remove ambiguity about NH$_3$ in the Homunculus, based on an identification of the $J_K$ = 3$_3$ inversion line at 23.87 GHz by \cite{Smith06} but challenged by \cite{Loinard16} to be associated instead with the H81 $\beta$ recombination line.   The observations by \cite{Smith06} show that the line emission has a different velocity structure and is more extended than we would expect from H recombination as observed with \alma\ (cf. \citealt{Abraham14}).  Regardless, the NH$_3$ rotational transitions detected in our \hso\ spectra (including HIFI, which will be published elsewhere) confirm this molecule along with nitrogen-bearing NH, NH$_2$, N$_2$H$^+$, CN, HCN, HNC, and $^{13}$C isotopologues for the C-bearing nitrides.

\subsection{Distribution of the atomic and molecular gas}\label{gasdistribution}
The \pacs\ and \spire\ spectral scans are rich in lines, however at the angular resolutions of both instruments compared to the $10'' \; \times \; 18''$ nebula, we can surmise the locations where the bulk of the far-IR line emission originates only on the scales of the physical structures described in the Introduction and Figure~\ref{nebular}. Our current understanding of the distribution of the atomic and molecular species detected in these observations is illustrated in Figure~\ref{moleculemap}, and summarized as follows: 

\begin{itemize}
  \item The continuum is unresolved with \pacs\ in 9.4\arcsec $\times$ 9.4\arcsec\ spaxels over the full usable wavelength range fo the spectrum, 55 - 190 $\mu$m.  In the \spire\ observations, \cite{Morris17} has shown that the continuum at 305 $\mu$m is within  a diameter 2$''$-5$''$ across, based on a semi-extended source size analysis where the Short Wave and Long Wave sections overlap at different beam sizes. 
  \item Hydrogen recombination and most atomic lines likewise appear to originate from a compact source well within the 10\arcsec diameter region centered on \ec. Based upon the radio observations of \cite{Duncan03} and \alma\ observations by \cite{Abraham14}, the bulk of the hydrogen recombination lines originate from within $0\farcs5$ of \ec. 
  
 \item The [\ion{N}{ii}] emission, while centered on \ec\ is extended from the SE (blue-shifted) to the NE (red-shifted). While the bulk of the [\ion{N}{ii}] emission originates from the central region, a fainter component maps to the spatial extent (40 to 50\arcsec) of the fast-moving bullets described by \cite{Weis97, smith08b,Kiminki16}.
 \item NH on the other hand is not detected in the fast outer ejecta or the ghost nebula observed in H Balmer and certain forbidden lines. The overall structure of red- and blue-shifted components is consistent with emission in the expanding lobes and break-out material from the skirt.  The fast NH may be from the thin walls of the lobes, favored by the measured line strengths and estimated column densities.  Similarly, the slower NH component behaves as expected on the scale of the Homunculus, from approaching to receding in the SW to NE direction, and may point to more wide-scale formation in the Homunculus than, e.g., CO, which is restricted to layers of the dusty central torus over the inner 5$''$ -- 8$''$ diameter of the structure.
  \item The [\ion{O}{iii}] emission originates entirely in the background Carina Nebula and is relatively constant over the \pacs\ spaxels with the exception of a relative minimum centered on \ec. This may be due to the fossil winds having driven out nebular material and  having replaced it with nitrogen-enriched processed material. We have detected [\ion{O}{i}] on the central source, where it is resolved and surprisingly structured with at least four velocity peaks.	
  \item The [\ion{C}{ii}] emission is primarily background emission from the Carina Nebula, but a broad component correlates with the spatial structure of the Homunculus and the broad component shifts to the red and blue as would be expected for various components of the expanding Homunculus bipolar lobe and skirt. [\ion{C}{I}] 492 GHz is only weakly detected in the \spire\ spectrum, and may also arise in the background. 
 \item Our results indicate that $\eta$ Car's small reservoir of carbon is held in C-bearing molecules.  CO in particular mainly traces the inner layers of the expanding Butterfly Nebula associated with the massive disrupted torus (see Fig.~\ref{nebular}). No C-bearing dust species have been identified in $\eta$ Car's complex solid state spectrum, while the N-rich atomic and molecular gas appears to provide favorable conditions for the formation of AlN and Si$_3$N$_4$ dust in low abundances \citep{Morris17}.  
 \item Emission lines from most of the molecules listed in Table~\ref{abundances}  are detected in the central spaxel of \pacs\ (HCl is detected in the \spire\ spectrum), but outside of CO and NH, they are generally too weak or blended to determine their extent in the rest of the nebula.  We know from ground-based  and HIFI observations, however, that N-bearing molecules NH$_2$, N$_2$H$^+$, CN, HCN, and HNC all share a very similar velocity structure with CO, and are thus principally associated with inner layers of the dusty equatorial torus.   Certain molecules such as CH$_3$OH and possibly NH$_3$ and H$_2$O are more centrally confined, $\lesssim$ 2$''$ across, possibly tracing shock-heated gas.   
\end{itemize}

\subsubsection{Abundance patterns}

The abundance pattern resulting from our broad-brush modeling of $\eta$ Car's sub-millimeter line spectrum provides further support for the Homunculus and its interior structures having been formed in explosive events from at least one massive star (but possible two in a merger scenario) with surface abundances consistent with yields and elemental depletions resulting from CNO-cycle.  While precise abundance estimates  are hindered by certain necessary simplifying assumptions on the geometry of the line emitting regions and interactions with the thermal continuum, the scales of the abundances certainly point to a N-rich, C-deficient environment with temperatures and densities supporting the formation and survival of molecules with upper level energies up to $\sim$500~K, and no more complex than CH$_3$OH (so far).   

The derived[CO/H$_2$] and [$^{12}$C/$^{13}$C] abundance ratios of 2.0 $\times$ 10$^{-5}$ and $\sim$3-4, respectively, are both quite low compared to cosmic abundances (by factors of 10-20 each), as expected from evolutionary models for the depletion of $^{12}$C and enhancement of $^{13}$C for rotating massive starts prior to entering the core He-burning phase of Wolf-Rayet stars (e.g., \citealt{Ekstrom12}).  Our results are consistent with those of  \cite{Morris17, Morris20} who analyzed velocity-resolved HIFI and \alma\ CO observations. 

\begin{figure}
\includegraphics[angle=0, scale=0.9]{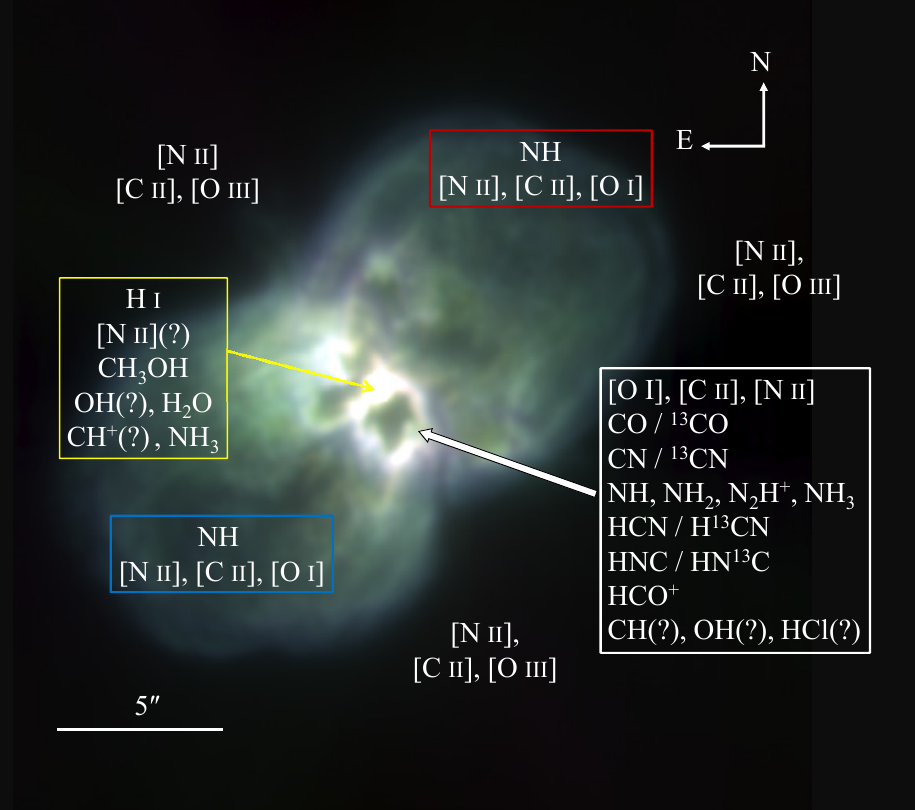}
\caption{A false color mid-infrared 9.0 $\mu$m image of $\eta$ Car with bulk locations of the atomic and molecular species detected in the \pacs\ and \spire\ spectral survey as summarized in Sec.~\ref{concl}.  The color scheme provides a good assocation of the white high contrast emission defining the Butterfly Nebula and species in the white box.  Species surmised to originate within 1$''$ of the central source (i.e., 2$''$ across) are indicated in the yellow box.  The red and blue boxes contain molecules associated with the expanding lobes (or break-out material of the equatorial skirt).  Detected background species are indicated without boxes. The image was taken with the VLT/VISIR instrument through the [\ion{Ar}{III}] filter \citep{Mehner19}, sampling the T $\geq$ 400 K continuum.
 \label{moleculemap}}
\end{figure}

The numerous transitions from N-bearing molecules in the \iso\ scans are evidence of the nitrogen-rich environment; however, our results for the abundances of the N hydrides may be several factors to an order of magnitude in error, in part because they are distributed on different spatial scales, although the Butterfly Nebula always dominates the emission.  NH is found throughout the Homunculus (Sec.~\ref{NHS}), including fast-moving material possibly located in the thin walls of the two lobes and/or the skirt.  The velocity structure of NH$_2$ and N$_2$H$^+$ from HIFI and \alma\ observations show a closer association with the expanding Butterfly Nebula, while NH$_3$ may be more closely confined to the compact central source \citep{Smith06}.  The main effect will be on the estimated column densities, decreasing $\mathcal{N}$(NH) and increasing $\mathcal{N}$(NH$_3$) such that the NH/NH$_2$/NH$_3$ abundance ratio should be qualitatively taken as  (<2.0)/1.0/(>0.4).   

Not surprisingly, the NH/NH$_2$/NH$_3$ ratio trends opposite of that estimated from absorption lines towards Sgr B2 star forming clouds \cite{Goicoechea04} and in cold envelopes around protostellar cores (e.g., \citealt{Hily-Blant10}) where collisional processes are weak at low temperatures, and NH is observed in smallest relative abundance, NH$_3$ the highest, separated by two orders or magnitude.  Surprisingly, the NH/NH$_2$/NH$_3$ ratio is {\it{very}} similar to those deduced from absorption lines in the diffuse or translucent sightlines studied by \cite{Persson12}, indeed almost {\it{equivalent}} towards the massive star forming region G10.6$-$0.4.  This is probably coincidental. The densities, column densities, and radiation environments are very different in eta Car compared with diffuse molecular gas in the Galactic plane.  

\subsubsection{Theoretical considerations}

Potentially more important to the veracity of the abundances than simplifying the geometry of the line forming gas (which we can eventually account for more accurately with, e.g., \alma\ observations) is a full accounting of the physical processes which drive excitation, including collisional excitation, radiative excitation, spontaneous radiative decay, and formation pumping. LTE will be reached through collisions at sufficiently high gas densities, populating energy levels at the gas kinetic temperature. However, as summarized by \cite{Gerin16} (see also \citealt{Roueff13}), the smaller momenta of inertia, larger rotational constants, and much higher spontaneous radiative decay rates for the rotational states of the hydrides compared to other molecules result in much higher densities needed to achieve LTE.  Below such ``critical'' densities, excitation becomes sub-thermal as total production and loss rates reach a state of quasi-equilibrium, and the excitation temperatures for most transitions are lower than the gas kinetic temperature.  We demonstrated in our analysis of the mid-IR and sub-millimeter \ion{H}{I} lines that departures from LTE occur due to radiative coupling of the Rydberg states to the strong thermal continuum.

The molecular abundances and ionization may be partly controlled by radiative processes in the Homunculus, in the presence of the X-ray and UV radiation field of the hot, luminous stellar pair at the center.  Energetic photons attenuated by complex shielding conditions in the Homunculus will dissociate and ionize molecules and atoms with ionization potentials below 13.6 eV, to an extent determined by the hardness of the photodissociation region (PDR).   \ec\ is a source of X-ray emission \citep{Pittard02}, and a diagnostic of the hardness of the radiation field incident on the gas may be found in the HNC/HCN line ratio. It has been suggested that molecular line intensity ratios, e.g. HNC/HCN, can distinguish between X-ray-dominated (XDR) and UV-photon-dominated (PDR) regions in luminous infrared galaxies \citep{Meijerink07}. Such claims are model-dependent. Further investigation of the photochemistry and ionization in eta Car might lead to a better understanding of other such complex, dust-obscured environments.  

Shielding is provided mainly by the enormous torus-like structure where the bulk of the 0.22 M$_\odot$ of dust  resides, but also in material in the poorly understood immediate vicinity of the binary, containing an unknown fraction of vibrationally-excited H$_2$ (which is a key component in energetic reactions; e.g., \citealt{Black87, Agundez10}), and dense ``coronagraphic'' structures of unknown distance from the central source \cite{Damineli19}, possibly associated with the compact source of methanol, water, and ammonia \cite{Morris20,Smith06}.  These species are sometimes considered as tracers of shock-heated gas, which has not been treated in our modeling.  

Shocks are expected to have a significant effect on the chemistry of the nebula, with the destruction of certain molecules and enhancement of others.  Many of the species we have observed but especially NH$_3$, CH$_3$OH, H$_2$O, and CH$^+$ may be useful as indicators of shock-heated gas in different parts of the Homunculus and central region.   While we cannot easily follow up with H$_2$O observations from the ground or with SOFIA instruments (except in very excited states), further observations of OH should be rewarding if at the high temperatures of the gas, most of the gas-phase oxygen is driven into OH in collisions with vibrationally excited, or simply hot, molecular hydrogen H$_2^*$, and then into water OH + H$_2^*$ $\rightarrow$ H$_2$O + H.    The balance of OH and H$_2$O in both forward and reverse reactions in the nebula will be regulated by the local UV field strength or shock velocities.   This is a topic we will address in a future study of the chemistry-driving roles of UV irradiation and shocks in $\eta$ Car, requiring additional high-angular observations to disentangle the kinematics of the bulk motion of structures in and around the central region from shock-broadened emission.
\subsection{The uniqueness of \ec}
 
Most evolved stars with mass loss have circumstellar envelopes that are rich in carbon or oxygen at the expense of nitrogen through the CNO cycle. AGB stars (initial masses, M $\approx$\ 1 to 8\Msun) and RSG stars (M $\approx$\ 8 to 30 \Msun) have circumstellar shells rich in oxygen- and carbon-bearing molecules.  In contrast, the ejecta around \ec\ contain proportionately more nitrogen-bearing molecules compared to  circumstellar envelopes around less-massive, evolved stars.  Such would be expected given the strong nitrogen presence and near-absence of oxygen and carbon in the ionized ejecta as noted previously by \cite{Davidson86a, Verner05b} as there is little evidence of abundance stratification in the ejecta. The over-abundance of nitrogen plus the detected C/O < 4 is consistent with the ejecting star having an initial mass > 60 \Msun\ \citep{Ekstrom12}.

The massive binary, \ec, has gone through a transition on historical timescales and is sufficiently close that present observatories can study the ejecta in detail. This system is providing ample information on binary systems with stellar members greatly exceeding 30 \Msun. At any given epoch, the more massive stars, like those found in \ec, occur far less frequently, their lifetimes are also much shorter, and likely contribute significantly to enrichment of the interstellar medium. What is learned from this massive binary has potential impact on our knowledge of the earliest stars in our Universe, the massive members of which evolved on rapid timescales quickly enriching the ISM and may be revealing information on Type IIn or Ib/c progenitors.

\section*{Acknowledgments}
The content of this paper is based on observations with the \herschel, which is an ESA space observatory with science instruments provided by European-led Principal Investigator consortia and with important participation from NASA.  We have made use software in the \hso\ Interactive Processing Environment (HIPE), which was a joint development by the \hso\ Science Ground Segment Consortium, consisting of ESA, the NASA Herschel Science Center, and the HIFI, PACS and SPIRE consortia. T.R.G.,  K.E.N. and P.M. acknowledge partial financial support from NASA grant SCEX22012D for the \hso\ program OT1\_tgull3. M.J.B. acknowledges support from European Research Council Grant SNDUST ERC-2015-AdG-694520.  T.R.G. acknowledges the support of Onsala Observatory and Max Planck Institute for RadioAstronomy during visits at those institutions. 

We thank the referee for many useful comments leading to significant improvement of this paper.}

\section*{Data Availability}
The observations presented in this paper are based on observations which are available in the public \hso\ {http://archives.esac.esa.int/hsa/whsa/} and \iso\ data archives {https://www.cosmos.esa.int/web/iso/}. Reasonable requests for advanced data products may be addressed to P.W.M.




\bibliographystyle{mnras}
\bibliography{ref.bib}

\begin{thebibliography}{}
\makeatletter
\relax
\def\mn@urlcharsother{\let\do\@makeother \do\$\do\&\do\#\do\^\do\_\do\%\do\~}
\def\mn@doi{\begingroup\mn@urlcharsother \@ifnextchar [ {\mn@doi@}
  {\mn@doi@[]}}
\def\mn@doi@[#1]#2{\def\@tempa{#1}\ifx\@tempa\@empty \href
  {http://dx.doi.org/#2} {doi:#2}\else \href {http://dx.doi.org/#2} {#1}\fi
  \endgroup}
\def\mn@eprint#1#2{\mn@eprint@#1:#2::\@nil}
\def\mn@eprint@arXiv#1{\href {http://arxiv.org/abs/#1} {{\tt arXiv:#1}}}
\def\mn@eprint@dblp#1{\href {http://dblp.uni-trier.de/rec/bibtex/#1.xml}
  {dblp:#1}}
\def\mn@eprint@#1:#2:#3:#4\@nil{\def\@tempa {#1}\def\@tempb {#2}\def\@tempc
  {#3}\ifx \@tempc \@empty \let \@tempc \@tempb \let \@tempb \@tempa \fi \ifx
  \@tempb \@empty \def\@tempb {arXiv}\fi \@ifundefined
  {mn@eprint@\@tempb}{\@tempb:\@tempc}{\expandafter \expandafter \csname
  mn@eprint@\@tempb\endcsname \expandafter{\@tempc}}}

\bibitem[\protect\citeauthoryear{{Abraham}, {Falceta-Gon{\c c}alves}  \&
  {Beaklini}}{{Abraham} et~al.}{2014}]{Abraham14}
{Abraham} Z.,  {Falceta-Gon{\c c}alves} D.,   {Beaklini} P.~P.~B.,  2014,
  \mn@doi [\apj] {10.1088/0004-637X/791/2/95}, \href
  {http://adsabs.harvard.edu/abs/2014ApJ...791...95A} {791, 95}

\bibitem[\protect\citeauthoryear{{Ag{\'u}ndez}, {Goicoechea}, {Cernicharo},
  {Faure}  \& {Roueff}}{{Ag{\'u}ndez} et~al.}{2010}]{Agundez10}
{Ag{\'u}ndez} M.,  {Goicoechea} J.~R.,  {Cernicharo} J.,  {Faure} A.,
  {Roueff} E.,  2010, \mn@doi [\apj] {10.1088/0004-637X/713/1/662}, \href
  {https://ui.adsabs.harvard.edu/abs/2010ApJ...713..662A} {713, 662}

\bibitem[\protect\citeauthoryear{{Artigau}, {Martin}, {Humphreys}, {Davidson},
  {Chesneau}  \& {Smith}}{{Artigau} et~al.}{2011}]{Artigau11}
{Artigau} {\'E}.,  {Martin} J.~C.,  {Humphreys} R.~M.,  {Davidson} K.,
  {Chesneau} O.,   {Smith} N.,  2011, \mn@doi [\aj]
  {10.1088/0004-6256/141/6/202}, \href
  {http://adsabs.harvard.edu/abs/2011AJ....141..202A} {141, 202}

\bibitem[\protect\citeauthoryear{{Black} \& {van Dishoeck}}{{Black} \& {van
  Dishoeck}}{1987}]{Black87}
{Black} J.~H.,  {van Dishoeck} E.~F.,  1987, \mn@doi [\apj] {10.1086/165740},
  \href {https://ui.adsabs.harvard.edu/abs/1987ApJ...322..412B} {322, 412}

\bibitem[\protect\citeauthoryear{{Bocchio}, {Bianchi}  \& {Abergel}}{{Bocchio}
  et~al.}{2016}]{Bocchio16}
{Bocchio} M.,  {Bianchi} S.,   {Abergel} A.,  2016, \mn@doi [\aap]
  {10.1051/0004-6361/201628665}, \href
  {http://adsabs.harvard.edu/abs/2016A%26A...591A.117B} {591, A117}

\bibitem[\protect\citeauthoryear{{Brown}, {Varberg}, {Evenson}  \&
  {Cooksy}}{{Brown} et~al.}{1994}]{Brown94}
{Brown} J.~M.,  {Varberg} T.~D.,  {Evenson} K.~M.,   {Cooksy} A.~L.,  1994,
  \mn@doi [\apjl] {10.1086/187387}, \href
  {https://ui.adsabs.harvard.edu/abs/1994ApJ...428L..37B} {428, L37}

\bibitem[\protect\citeauthoryear{{Chesneau} et~al.,}{{Chesneau}
  et~al.}{2005}]{Chesneau05a}
{Chesneau} O.,  et~al., 2005, \mn@doi [\aap] {10.1051/0004-6361:20041395},
  \href {http://adsabs.harvard.edu/abs/2005A%26A...435.1043C} {435, 1043}

\bibitem[\protect\citeauthoryear{{Cox}, {Mezger}, {Sievers}, {Najarro},
  {Bronfman}, {Kreysa}  \& {Haslam}}{{Cox} et~al.}{1995}]{Cox95}
{Cox} P.,  {Mezger} P.~G.,  {Sievers} A.,  {Najarro} F.,  {Bronfman} L.,
  {Kreysa} E.,   {Haslam} G.,  1995, \aap, \href
  {http://adsabs.harvard.edu/abs/1995A%26A...297..168C} {297, 168}

\bibitem[\protect\citeauthoryear{{Currie}, {Dorland}  \& {Kaufer}}{{Currie}
  et~al.}{2002}]{Currie02}
{Currie} D.~G.,  {Dorland} B.~N.,   {Kaufer} A.,  2002, A\&A, \href
  {http://adsabs.harvard.edu/cgi-bin/nph-bib_query?bibcode=2002A%26A...389L..65C&amp;db_key=AST}
  {389, L65}

\bibitem[\protect\citeauthoryear{{Damineli}}{{Damineli}}{1996}]{Damineli96}
{Damineli} A.,  1996, \mn@doi [\apjl] {10.1086/309961}, \href
  {http://adsabs.harvard.edu/cgi-bin/nph-bib_query?bibcode=1996ApJ...460L..49D&db_key=AST}
  {460, L49}

\bibitem[\protect\citeauthoryear{{Damineli} et~al.,}{{Damineli}
  et~al.}{2019}]{Damineli19}
{Damineli} A.,  et~al., 2019, \mn@doi [\mnras] {10.1093/mnras/stz067}, \href
  {http://adsabs.harvard.edu/abs/2019MNRAS.484.1325D} {484, 1325}

\bibitem[\protect\citeauthoryear{{Davidson}}{{Davidson}}{1997}]{Davidson97a}
{Davidson} K.,  1997, \mn@doi [New Astronomy] {10.1016/S1384-1076(97)00028-6},
  \href
  {http://adsabs.harvard.edu/cgi-bin/nph-bib_query?bibcode=1997NewA....2..387D&db_key=AST}
  {2, 387}

\bibitem[\protect\citeauthoryear{{Davidson}, {Walborn}  \& {Gull}}{{Davidson}
  et~al.}{1982}]{Davidson82a}
{Davidson} K.,  {Walborn} N.~R.,   {Gull} T.~R.,  1982, \mn@doi [ApJ]
  {10.1086/183754}, \href
  {http://adsabs.harvard.edu/cgi-bin/nph-bib_query?bibcode=1982ApJ...254L..47D&db_key=AST}
  {254, L47}

\bibitem[\protect\citeauthoryear{{Davidson}, {Dufour}, {Walborn}  \&
  {Gull}}{{Davidson} et~al.}{1986}]{Davidson86a}
{Davidson} K.,  {Dufour} R.~J.,  {Walborn} N.~R.,   {Gull} T.~R.,  1986, ApJ,
  \href
  {http://adsabs.harvard.edu/cgi-bin/nph-bib_query?bibcode=1986ApJ...305..867D&amp;db_key=AST}
  {305, 867}

\bibitem[\protect\citeauthoryear{{Dorland}}{{Dorland}}{2007}]{Dorland07}
{Dorland} B.~N.,  2007, PhD thesis, University of Maryland, College Park

\bibitem[\protect\citeauthoryear{{Duncan} \& {White}}{{Duncan} \&
  {White}}{2003}]{Duncan03}
{Duncan} R.~A.,  {White} S.~M.,  2003, \mn@doi [MNRAS]
  {10.1046/j.1365-8711.2003.06287.x}, \href
  {http://adsabs.harvard.edu/abs/2003MNRAS.338..425D} {338, 425}

\bibitem[\protect\citeauthoryear{{Ekstr{\"o}m} et~al.,}{{Ekstr{\"o}m}
  et~al.}{2012}]{Ekstrom12}
{Ekstr{\"o}m} S.,  et~al., 2012, \mn@doi [\aap] {10.1051/0004-6361/201117751},
  \href {http://adsabs.harvard.edu/abs/2012A%26A...537A.146E} {537, A146}

\bibitem[\protect\citeauthoryear{{Frew}}{{Frew}}{2004}]{Frew04}
{Frew} D.~J.,  2004, Journal of Astronomical Data, \href
  {http://adsabs.harvard.edu/abs/2004JAD....10....6F} {10, 6}

\bibitem[\protect\citeauthoryear{{Gerin}, {Neufeld}  \& {Goicoechea}}{{Gerin}
  et~al.}{2016}]{Gerin16}
{Gerin} M.,  {Neufeld} D.~A.,   {Goicoechea} J.~R.,  2016, \mn@doi [\araa]
  {10.1146/annurev-astro-081915-023409}, \href
  {https://ui.adsabs.harvard.edu/abs/2016ARA&A..54..181G} {54, 181}

\bibitem[\protect\citeauthoryear{{Goicoechea}, {Rodr{\'\i}guez-Fern{\'a}ndez}
  \& {Cernicharo}}{{Goicoechea} et~al.}{2004}]{Goicoechea04}
{Goicoechea} J.~R.,  {Rodr{\'\i}guez-Fern{\'a}ndez} N.~J.,   {Cernicharo} J.,
  2004, \mn@doi [\apj] {10.1086/379704}, \href
  {https://ui.adsabs.harvard.edu/abs/2004ApJ...600..214G} {600, 214}

\bibitem[\protect\citeauthoryear{{Grant}, {Blundell}  \& {Matthews}}{{Grant}
  et~al.}{2020}]{Grant20}
{Grant} D.,  {Blundell} K.,   {Matthews} J.,  2020, \mn@doi [\mnras]
  {10.1093/mnras/staa669}, \href
  {https://ui.adsabs.harvard.edu/abs/2020MNRAS.494...17G} {494, 17}

\bibitem[\protect\citeauthoryear{{Griffin} et~al.,}{{Griffin}
  et~al.}{2010}]{Griffin10}
{Griffin} M.~J.,  et~al., 2010, \mn@doi [\aap] {10.1051/0004-6361/201014519},
  \href {http://adsabs.harvard.edu/abs/2010A%26A...518L...3G} {518, L3}

\bibitem[\protect\citeauthoryear{{Groh}, {Hillier}, {Madura}  \&
  {Weigelt}}{{Groh} et~al.}{2012}]{Groh12}
{Groh} J.~H.,  {Hillier} D.~J.,  {Madura} T.~I.,   {Weigelt} G.,  2012, \mn@doi
  [\mnras] {10.1111/j.1365-2966.2012.20984.x}, \href
  {http://adsabs.harvard.edu/abs/2012MNRAS.tmp.3024G} {423, 3024}

\bibitem[\protect\citeauthoryear{{Groh}, {Meynet}, {Ekstrom}  \&
  {Georgy}}{{Groh} et~al.}{2014}]{Groh14}
{Groh} J.,  {Meynet} G.,  {Ekstrom} S.,   {Georgy} C.,  2014, preprint, \href
  {http://adsabs.harvard.edu/abs/2014arXiv1401.7322G} {} (\mn@eprint {arXiv}
  {1401.7322})

\bibitem[\protect\citeauthoryear{{Gull}, {Kober}  \& {Nielsen}}{{Gull}
  et~al.}{2006}]{Gull06}
{Gull} T.~R.,  {Kober} G.~V.,   {Nielsen} K.~E.,  2006, \mn@doi [ApJS]
  {10.1086/500113}, \href
  {http://adsabs.harvard.edu/cgi-bin/nph-bib_query?bibcode=2006ApJS..163..173G&db_key=AST}
  {163, 173}

\bibitem[\protect\citeauthoryear{{Gull} et~al.,}{{Gull} et~al.}{2009}]{Gull09}
{Gull} T.~R.,  et~al., 2009, \mn@doi [\mnras]
  {10.1111/j.1365-2966.2009.14854.x}, \href
  {http://adsabs.harvard.edu/abs/2009MNRAS.396.1308G} {396, 1308}

\bibitem[\protect\citeauthoryear{{Gull} et~al.,}{{Gull} et~al.}{2016}]{Gull16}
{Gull} T.~R.,  et~al., 2016, \mn@doi [\mnras] {10.1093/mnras/stw1829}, \href
  {http://adsabs.harvard.edu/abs/2016MNRAS.462.3196G} {462, 3196}

\bibitem[\protect\citeauthoryear{{Hartman}, {Gull}, {Johansson}, {Smith}  \&
  {HST Eta Carinae Treasury Project Team}}{{Hartman} et~al.}{2004}]{Hartman04}
{Hartman} H.,  {Gull} T.,  {Johansson} S.,  {Smith} N.,   {HST Eta Carinae
  Treasury Project Team} 2004, A\&A, \href
  {http://adsabs.harvard.edu/cgi-bin/nph-bib_query?bibcode=2004A%26A...419..215H&db_key=AST}
  {419, 215}

\bibitem[\protect\citeauthoryear{{Hillier}, {Davidson}, {Ishibashi}  \&
  {Gull}}{{Hillier} et~al.}{2001}]{Hillier01}
{Hillier} D.~J.,  {Davidson} K.,  {Ishibashi} K.,   {Gull} T.,  2001, \mn@doi
  [\apj] {10.1086/320948}, \href
  {http://adsabs.harvard.edu/cgi-bin/nph-bib_query?bibcode=2001ApJ...553..837H&db_key=AST}
  {553, 837}

\bibitem[\protect\citeauthoryear{{Hily-Blant} et~al.,}{{Hily-Blant}
  et~al.}{2010}]{Hily-Blant10}
{Hily-Blant} P.,  et~al., 2010, \mn@doi [\aap] {10.1051/0004-6361/201015253},
  \href {https://ui.adsabs.harvard.edu/abs/2010A&A...521L..52H} {521, L52}

\bibitem[\protect\citeauthoryear{{Hopwood} et~al.,}{{Hopwood}
  et~al.}{2015}]{Hopwood15}
{Hopwood} R.,  et~al., 2015, \mn@doi [\mnras] {10.1093/mnras/stv353}, \href
  {http://adsabs.harvard.edu/abs/2015MNRAS.449.2274H} {449, 2274}

\bibitem[\protect\citeauthoryear{{Hummer} \& {Storey}}{{Hummer} \&
  {Storey}}{1987}]{Hummer87}
{Hummer} D.~G.,  {Storey} P.~J.,  1987, \mn@doi [\mnras]
  {10.1093/mnras/224.3.801}, \href
  {https://ui.adsabs.harvard.edu/abs/1987MNRAS.224..801H} {224, 801}

\bibitem[\protect\citeauthoryear{{Indebetouw} et~al.,}{{Indebetouw}
  et~al.}{2005}]{Indebetouw05}
{Indebetouw} R.,  et~al., 2005, \mn@doi [\apj] {10.1086/426679}, \href
  {https://ui.adsabs.harvard.edu/abs/2005ApJ...619..931I} {619, 931}

\bibitem[\protect\citeauthoryear{{Ishibashi} et~al.,}{{Ishibashi}
  et~al.}{2003}]{Ishibashi03}
{Ishibashi} K.,  et~al., 2003, \mn@doi [\aj] {10.1086/375306}, \href
  {http://adsabs.harvard.edu/cgi-bin/nph-bib_query?bibcode=2003AJ....125.3222I&db_key=AST}
  {125, 3222}

\bibitem[\protect\citeauthoryear{{Kessler} et~al.,}{{Kessler}
  et~al.}{1996}]{Kessler96}
{Kessler} M.~F.,  et~al., 1996, \aap, \href
  {https://ui.adsabs.harvard.edu/abs/1996A&A...315L..27K} {500, 493}

\bibitem[\protect\citeauthoryear{{Kiminki}, {Reiter}  \& {Smith}}{{Kiminki}
  et~al.}{2016}]{Kiminki16}
{Kiminki} M.~M.,  {Reiter} M.,   {Smith} N.,  2016, \mn@doi [\mnras]
  {10.1093/mnras/stw2019}, \href
  {http://adsabs.harvard.edu/abs/2016MNRAS.463..845K} {463, 845}

\bibitem[\protect\citeauthoryear{{Loinard}, {Menten}, {G{\"u}sten}, {Zapata}
  \& {Rodr{\'{\i}}guez}}{{Loinard} et~al.}{2012}]{Loinard12}
{Loinard} L.,  {Menten} K.~M.,  {G{\"u}sten} R.,  {Zapata} L.~A.,
  {Rodr{\'{\i}}guez} L.~F.,  2012, \mn@doi [\apjl]
  {10.1088/2041-8205/749/1/L4}, \href
  {http://adsabs.harvard.edu/abs/2012ApJ...749L...4L} {749, L4}

\bibitem[\protect\citeauthoryear{{Loinard}, {Kami{\'n}ski}, {Serra}, {Menten},
  {Zapata}  \& {Rodr{\'\i}guez}}{{Loinard} et~al.}{2016}]{Loinard16}
{Loinard} L.,  {Kami{\'n}ski} T.,  {Serra} P.,  {Menten} K.~M.,  {Zapata}
  L.~A.,   {Rodr{\'\i}guez} L.~F.,  2016, \mn@doi [\apj]
  {10.3847/1538-4357/833/1/48}, \href
  {https://ui.adsabs.harvard.edu/abs/2016ApJ...833...48L} {833, 48}

\bibitem[\protect\citeauthoryear{{Madura} et~al.,}{{Madura}
  et~al.}{2013}]{Madura13}
{Madura} T.~I.,  et~al., 2013, \mn@doi [\mnras] {10.1093/mnras/stt1871}, \href
  {http://adsabs.harvard.edu/abs/2013MNRAS.436.3820M} {436, 3820}

\bibitem[\protect\citeauthoryear{{Mauerhan} et~al.,}{{Mauerhan}
  et~al.}{2013}]{Mauerhan13}
{Mauerhan} J.~C.,  et~al., 2013, \mn@doi [\mnras] {10.1093/mnras/stt009}, \href
  {http://adsabs.harvard.edu/abs/2013MNRAS.430.1801M} {430, 1801}

\bibitem[\protect\citeauthoryear{{Mehner}, {Davidson}, {Ferland}  \&
  {Humphreys}}{{Mehner} et~al.}{2010}]{Mehner10}
{Mehner} A.,  {Davidson} K.,  {Ferland} G.~J.,   {Humphreys} R.~M.,  2010,
  \mn@doi [\apj] {10.1088/0004-637X/710/1/729}, \href
  {http://adsabs.harvard.edu/abs/2010ApJ...710..729M} {710, 729}

\bibitem[\protect\citeauthoryear{{Mehner} et~al.,}{{Mehner}
  et~al.}{2016}]{Mehner16}
{Mehner} A.,  et~al., 2016, \mn@doi [\aap] {10.1051/0004-6361/201628770}, \href
  {http://adsabs.harvard.edu/abs/2016A%26A...595A.120M} {595, A120}

\bibitem[\protect\citeauthoryear{{Mehner} et~al.,}{{Mehner}
  et~al.}{2019}]{Mehner19}
{Mehner} A.,  et~al., 2019, \mn@doi [\aap] {10.1051/0004-6361/201936277}, \href
  {https://ui.adsabs.harvard.edu/abs/2019A&A...630L...6M} {630, L6}

\bibitem[\protect\citeauthoryear{{Meijerink}, {Spaans}  \&
  {Israel}}{{Meijerink} et~al.}{2007}]{Meijerink07}
{Meijerink} R.,  {Spaans} M.,   {Israel} F.~P.,  2007, \mn@doi [\aap]
  {10.1051/0004-6361:20066130}, \href
  {https://ui.adsabs.harvard.edu/abs/2007A&A...461..793M} {461, 793}

\bibitem[\protect\citeauthoryear{{Mookerjea}, {Sandell}, {G{\"u}sten},
  {Riquelme}, {Wiesemeyer}  \& {Chambers}}{{Mookerjea}
  et~al.}{2019}]{Mookerjea19}
{Mookerjea} B.,  {Sandell} G.,  {G{\"u}sten} R.,  {Riquelme} D.,  {Wiesemeyer}
  H.,   {Chambers} E.,  2019, \mn@doi [\aap] {10.1051/0004-6361/201935482},
  \href {https://ui.adsabs.harvard.edu/abs/2019A&A...626A.131M} {626, A131}

\bibitem[\protect\citeauthoryear{{Morris} et~al.,}{{Morris}
  et~al.}{1999}]{Morris99}
{Morris} P.~W.,  et~al., 1999, \mn@doi [\nat] {10.1038/990048}, \href
  {http://adsabs.harvard.edu/abs/1999Natur.402..502M} {402, 502}

\bibitem[\protect\citeauthoryear{{Morris}, {Gull}, {Hillier}, {Barlow},
  {Royer}, {Nielsen}, {Black}  \& {Swinyard}}{{Morris} et~al.}{2017}]{Morris17}
{Morris} P.~W.,  {Gull} T.~R.,  {Hillier} D.~J.,  {Barlow} M.~J.,  {Royer} P.,
  {Nielsen} K.,  {Black} J.,   {Swinyard} B.,  2017, \mn@doi [\apj]
  {10.3847/1538-4357/aa71b3}, \href
  {http://adsabs.harvard.edu/abs/2017ApJ...842...79M} {842, 79}

\bibitem[\protect\citeauthoryear{{Morris} et~al.,}{{Morris}
  et~al.}{2020}]{Morris20}
{Morris} P.~W.,  et~al., 2020, \mn@doi [\apjl] {10.3847/2041-8213/ab784a},
  \href {https://ui.adsabs.harvard.edu/abs/2020ApJ...892L..23M} {892, L23}

\bibitem[\protect\citeauthoryear{{M{\"u}ller}, {Schl{\"o}der}, {Stutzki}  \&
  {Winnewisser}}{{M{\"u}ller} et~al.}{2005}]{Mueller05}
{M{\"u}ller} H. S.~P.,  {Schl{\"o}der} F.,  {Stutzki} J.,   {Winnewisser} G.,
  2005, \mn@doi [Journal of Molecular Structure]
  {10.1016/j.molstruc.2005.01.027}, \href
  {https://ui.adsabs.harvard.edu/abs/2005JMoSt.742..215M} {742, 215}

\bibitem[\protect\citeauthoryear{{Nielsen}, {Gull}  \& {Vieira
  Kober}}{{Nielsen} et~al.}{2005}]{Nielsen05a}
{Nielsen} K.~E.,  {Gull} T.~R.,   {Vieira Kober} G.,  2005, \mn@doi [ApJS]
  {10.1086/427437}, \href {http://adsabs.harvard.edu/abs/2005ApJS..157..138N}
  {157, 138}

\bibitem[\protect\citeauthoryear{{Oberst}, {Parshley}, {Nikola}, {Stacey},
  {L{\"o}hr}, {Lane}, {Stark}  \& {Kamenetzky}}{{Oberst}
  et~al.}{2011}]{Oberst11}
{Oberst} T.~E.,  {Parshley} S.~C.,  {Nikola} T.,  {Stacey} G.~J.,  {L{\"o}hr}
  A.,  {Lane} A.~P.,  {Stark} A.~A.,   {Kamenetzky} J.,  2011, \mn@doi [\apj]
  {10.1088/0004-637X/739/2/100}, \href
  {https://ui.adsabs.harvard.edu/abs/2011ApJ...739..100O} {739, 100}

\bibitem[\protect\citeauthoryear{{Persson} et~al.,}{{Persson}
  et~al.}{2012}]{Persson12}
{Persson} C.~M.,  et~al., 2012, \mn@doi [\aap] {10.1051/0004-6361/201118686},
  \href {https://ui.adsabs.harvard.edu/abs/2012A&A...543A.145P} {543, A145}

\bibitem[\protect\citeauthoryear{{Pickett}, {Poynter}, {Cohen}, {Delitsky},
  {Pearson}  \& {M{\"u}ller}}{{Pickett} et~al.}{1998}]{Pickett98}
{Pickett} H.~M.,  {Poynter} R.~L.,  {Cohen} E.~A.,  {Delitsky} M.~L.,
  {Pearson} J.~C.,   {M{\"u}ller} H.~S.~P.,  1998, \mn@doi [\jqsrt]
  {10.1016/S0022-4073(98)00091-0}, \href
  {https://ui.adsabs.harvard.edu/abs/1998JQSRT..60..883P} {60, 883}

\bibitem[\protect\citeauthoryear{{Pilbratt} et~al.,}{{Pilbratt}
  et~al.}{2010}]{Pilbratt10}
{Pilbratt} G.~L.,  et~al., 2010, \mn@doi [\aap] {10.1051/0004-6361/201014759},
  \href {http://adsabs.harvard.edu/abs/2010A%26A...518L...1P} {518, L1}

\bibitem[\protect\citeauthoryear{{Pittard} \& {Corcoran}}{{Pittard} \&
  {Corcoran}}{2002}]{Pittard02}
{Pittard} J.~M.,  {Corcoran} M.~F.,  2002, \mn@doi [A\&A]
  {10.1051/0004-6361:20020025}, \href
  {http://adsabs.harvard.edu/cgi-bin/nph-bib_query?bibcode=2002A%26A...383..636P&db_key=AST}
  {383, 636}

\bibitem[\protect\citeauthoryear{{Poglitsch} et~al.,}{{Poglitsch}
  et~al.}{2010}]{Poglitsch10}
{Poglitsch} A.,  et~al., 2010, \mn@doi [\aap] {10.1051/0004-6361/201014535},
  \href {http://adsabs.harvard.edu/abs/2010A%26A...518L...2P} {518, L2}

\bibitem[\protect\citeauthoryear{{Polomski}, {Telesco}, {Pi{\~n}a}  \&
  {Fisher}}{{Polomski} et~al.}{1999}]{Polomski99}
{Polomski} E.~F.,  {Telesco} C.~M.,  {Pi{\~n}a} R.~K.,   {Fisher} R.~S.,  1999,
  \mn@doi [\aj] {10.1086/301093}, \href
  {http://adsabs.harvard.edu/abs/1999AJ....118.2369P} {118, 2369}

\bibitem[\protect\citeauthoryear{{Roelfsema} et~al.,}{{Roelfsema}
  et~al.}{2012}]{Roelfsema12}
{Roelfsema} P.~R.,  et~al., 2012, \mn@doi [\aap] {10.1051/0004-6361/201015120},
  \href {https://ui.adsabs.harvard.edu/abs/2012A&A...537A..17R} {537, A17}

\bibitem[\protect\citeauthoryear{{Roueff} \& {Lique}}{{Roueff} \&
  {Lique}}{2013}]{Roueff13}
{Roueff} E.,  {Lique} F.,  2013, \mn@doi [Chemical Reviews]
  {10.1021/cr400145a}, \href
  {https://ui.adsabs.harvard.edu/abs/2013ChRv..113.8906R} {113, 8906}

\bibitem[\protect\citeauthoryear{{Smith}}{{Smith}}{2002}]{Smith02}
{Smith} N.,  2002, MNRAS, \href
  {http://adsabs.harvard.edu/cgi-bin/nph-bib_query?bibcode=2002MNRAS.337.1252S&db_key=AST}
  {337, 1252}

\bibitem[\protect\citeauthoryear{{Smith}}{{Smith}}{2006}]{Smith06}
{Smith} N.,  2006, \mn@doi [\apj] {10.1086/503766}, \href
  {http://adsabs.harvard.edu/abs/2006ApJ...644.1151S} {644, 1151}

\bibitem[\protect\citeauthoryear{{Smith}}{{Smith}}{2008a}]{Smith08}
{Smith} N.,  2008a, in {de Koter} A.,  {Smith} L.~J.,   {Waters} L.~B.~F.~M.,
  eds,  Astronomical Society of the Pacific Conference Series Vol. 388, Mass
  Loss from Stars and the Evolution of Stellar Clusters. pp 129--+

\bibitem[\protect\citeauthoryear{{Smith}}{{Smith}}{2008b}]{smith08b}
{Smith} N.,  2008b, \mn@doi [\nat] {10.1038/nature07269}, \href
  {http://adsabs.harvard.edu/abs/2008Natur.455..201S} {455, 201}

\bibitem[\protect\citeauthoryear{{Smith} \& {Gehrz}}{{Smith} \&
  {Gehrz}}{1998}]{Smith98}
{Smith} N.,  {Gehrz} R.~D.,  1998, \mn@doi [\aj] {10.1086/300447}, \href
  {http://adsabs.harvard.edu/abs/1998AJ....116..823S} {116, 823}

\bibitem[\protect\citeauthoryear{{Smith} \& {Morse}}{{Smith} \&
  {Morse}}{2004}]{Smith04d}
{Smith} N.,  {Morse} J.~A.,  2004, \mn@doi [ApJ] {10.1086/382671}, \href
  {http://adsabs.harvard.edu/abs/2004ApJ...605..854S} {605, 854}

\bibitem[\protect\citeauthoryear{{Smith}, {Gehrz}, {Hinz}, {Hoffmann}, {Hora},
  {Mamajek}  \& {Meyer}}{{Smith} et~al.}{2003}]{Smith03a}
{Smith} N.,  {Gehrz} R.~D.,  {Hinz} P.~M.,  {Hoffmann} W.~F.,  {Hora} J.~L.,
  {Mamajek} E.~E.,   {Meyer} M.~R.,  2003, AJ, 125, 1458

\bibitem[\protect\citeauthoryear{{Smith}, {Brooks}, {Koribalski}  \&
  {Bally}}{{Smith} et~al.}{2006}]{Smith06e}
{Smith} N.,  {Brooks} K.~J.,  {Koribalski} B.~S.,   {Bally} J.,  2006, \mn@doi
  [\apjl] {10.1086/505934}, \href
  {http://adsabs.harvard.edu/abs/2006ApJ...645L..41S} {645, L41}

\bibitem[\protect\citeauthoryear{{Smith}, {Ginsburg}  \& {Bally}}{{Smith}
  et~al.}{2018a}]{Smith18a}
{Smith} N.,  {Ginsburg} A.,   {Bally} J.,  2018a, \mn@doi [\mnras]
  {10.1093/mnras/stx3050}, \href
  {http://adsabs.harvard.edu/abs/2018MNRAS.474.4988S} {474, 4988}

\bibitem[\protect\citeauthoryear{{Smith} et~al.,}{{Smith}
  et~al.}{2018b}]{Smith18c}
{Smith} N.,  et~al., 2018b, \mn@doi [\mnras] {10.1093/mnras/sty1500}, \href
  {http://adsabs.harvard.edu/abs/2018MNRAS.480.1466S} {480, 1466}

\bibitem[\protect\citeauthoryear{{Steffen} et~al.,}{{Steffen}
  et~al.}{2014}]{Steffen14}
{Steffen} W.,  et~al., 2014, \mn@doi [\mnras] {10.1093/mnras/stu1088}, \href
  {http://adsabs.harvard.edu/abs/2014MNRAS.442.3316S} {442, 3316}

\bibitem[\protect\citeauthoryear{{Swinyard} et~al.,}{{Swinyard}
  et~al.}{2010}]{Swinyard10}
{Swinyard} B.~M.,  et~al., 2010, \mn@doi [\aap] {10.1051/0004-6361/201014605},
  \href {http://adsabs.harvard.edu/abs/2010A%26A...518L...4S} {518, L4}

\bibitem[\protect\citeauthoryear{{Swinyard} et~al.,}{{Swinyard}
  et~al.}{2014}]{Swinyard14}
{Swinyard} B.~M.,  et~al., 2014, \mn@doi [\mnras] {10.1093/mnras/stu409}, \href
  {http://adsabs.harvard.edu/abs/2014MNRAS.440.3658S} {440, 3658}

\bibitem[\protect\citeauthoryear{{Teodoro}, {Madura}, {Gull}, {Corcoran}  \&
  {Hamaguchi}}{{Teodoro} et~al.}{2013}]{Teodoro13}
{Teodoro} M.,  {Madura} T.~I.,  {Gull} T.~R.,  {Corcoran} M.~F.,   {Hamaguchi}
  K.,  2013, \mn@doi [\apjl] {10.1088/2041-8205/773/1/L16}, \href
  {http://adsabs.harvard.edu/abs/2013ApJ...773L..16T} {773, L16}

\bibitem[\protect\citeauthoryear{{Teodoro} et~al.,}{{Teodoro}
  et~al.}{2016}]{Teodoro16}
{Teodoro} M.,  et~al., 2016, \mn@doi [\apj] {10.3847/0004-637X/819/2/131},
  \href {http://adsabs.harvard.edu/abs/2016ApJ...819..131T} {819, 131}

\bibitem[\protect\citeauthoryear{{Teodoro}, {Gull}, {Bautista}, {Hillier},
  {Weigelt}  \& {Corcoran}}{{Teodoro} et~al.}{2020}]{Teodoro20}
{Teodoro} M.,  {Gull} T.~R.,  {Bautista} M.~A.,  {Hillier} D.~J.,  {Weigelt}
  G.,   {Corcoran} M.~F.,  2020, \mn@doi [\mnras] {10.1093/mnras/staa1311},
  \href {https://ui.adsabs.harvard.edu/abs/2020MNRAS.495.2754T} {495, 2754}

\bibitem[\protect\citeauthoryear{{Valentijn} et~al.,}{{Valentijn}
  et~al.}{1996}]{Valentijn96}
{Valentijn} E.~A.,  et~al., 1996, \aap, \href
  {https://ui.adsabs.harvard.edu/abs/1996A&A...315L..60V} {315, L60}

\bibitem[\protect\citeauthoryear{{Valtchanov} et~al.,}{{Valtchanov}
  et~al.}{2018}]{Valtchanov18}
{Valtchanov} I.,  et~al., 2018, \mn@doi [\mnras] {10.1093/mnras/stx3178}, \href
  {http://adsabs.harvard.edu/abs/2018MNRAS.475..321V} {475, 321}

\bibitem[\protect\citeauthoryear{{Verner}, {Bruhweiler}  \& {Gull}}{{Verner}
  et~al.}{2005a}]{Verner05a}
{Verner} E.~M.,  {Bruhweiler} F.,   {Gull} T.~R.,  2005a, ApJ, \href
  {http://adsabs.harvard.edu/cgi-bin/nph-bib_query?bibcode=2005ApJ...624..973V&db_key=AST}
  {624, 973}

\bibitem[\protect\citeauthoryear{{Verner}, {Bruhweiler}, {Nielsen}, {Gull},
  {Vieira Kober}  \& {Corcoran}}{{Verner} et~al.}{2005b}]{Verner05b}
{Verner} E.,  {Bruhweiler} F.,  {Nielsen} K.~E.,  {Gull} T.~R.,  {Vieira Kober}
  G.,   {Corcoran} M.,  2005b, \mn@doi [ApJ] {10.1086/431917}, \href
  {http://adsabs.harvard.edu/abs/2005ApJ...629.1034V} {629, 1034}

\bibitem[\protect\citeauthoryear{{Walborn}, {Blanco}  \& {Thackeray}}{{Walborn}
  et~al.}{1978}]{Walborn78}
{Walborn} N.~R.,  {Blanco} B.~M.,   {Thackeray} A.~D.,  1978, \mn@doi [\apj]
  {10.1086/155806}, \href
  {https://ui.adsabs.harvard.edu/abs/1978ApJ...219..498W} {219, 498}

\bibitem[\protect\citeauthoryear{{Weigelt} \& {Ebersberger}}{{Weigelt} \&
  {Ebersberger}}{1986}]{Weigelt86}
{Weigelt} G.,  {Ebersberger} J.,  1986, A\&A, \href
  {http://adsabs.harvard.edu/cgi-bin/nph-bib_query?bibcode=1986A%26A...163L...5W&db_key=AST}
  {163, L5}

\bibitem[\protect\citeauthoryear{{Weigelt} \& {Kraus}}{{Weigelt} \&
  {Kraus}}{2012}]{Weigelt12}
{Weigelt} G.,  {Kraus} S.,  2012, in {Davidson} K.,  {Humphreys} R.~M.,  eds,
  Astrophysics and Space Science Library Vol. 384, Eta Carinae and the
  Supernova Impostors. p.~129, \mn@doi{10.1007/978-1-4614-2275-4_6}

\bibitem[\protect\citeauthoryear{{Weis}, {Duschl}, {Bomans}, {Chu}  \&
  {Joner}}{{Weis} et~al.}{1997}]{Weis97}
{Weis} K.,  {Duschl} W.~J.,  {Bomans} D.~J.,  {Chu} Y.-H.,   {Joner} M.~D.,
  1997, \aap, \href {http://adsabs.harvard.edu/abs/1997A%26A...320..568W} {320,
  568}

\bibitem[\protect\citeauthoryear{{Weis}, {Duschl}  \& {Chu}}{{Weis}
  et~al.}{1999}]{Weis99}
{Weis} K.,  {Duschl} W.~J.,   {Chu} Y.-H.,  1999, \aap, \href
  {http://adsabs.harvard.edu/abs/1999A%26A...349..467W} {349, 467}

\bibitem[\protect\citeauthoryear{{Westphal} \& {Neugebauer}}{{Westphal} \&
  {Neugebauer}}{1969}]{Westphal69}
{Westphal} J.~A.,  {Neugebauer} G.,  1969, \mn@doi [\apjl] {10.1086/180346},
  \href {http://adsabs.harvard.edu/abs/1969ApJ...156L..45W} {156, L45}

\bibitem[\protect\citeauthoryear{{Wu} et~al.,}{{Wu} et~al.}{2013}]{Wu13}
{Wu} R.,  et~al., 2013, \mn@doi [\aap] {10.1051/0004-6361/201321837}, \href
  {https://ui.adsabs.harvard.edu/abs/2013A&A...556A.116W} {556, A116}

\bibitem[\protect\citeauthoryear{{Zethson}, {Gull}, {Hartman}, {Johansson},
  {Davidson}  \& {Ishibashi}}{{Zethson} et~al.}{2001}]{Zethson01a}
{Zethson} T.,  {Gull} T.~R.,  {Hartman} H.,  {Johansson} S.,  {Davidson} K.,
  {Ishibashi} K.,  2001, AJ, \href
  {http://adsabs.harvard.edu/cgi-bin/nph-bib_query?bibcode=2001AJ....122..322Z&db_key=AST}
  {122, 322}

\bibitem[\protect\citeauthoryear{{Zethson}, {Johansson}, {Hartman}  \&
  {Gull}}{{Zethson} et~al.}{2012}]{Zethson12}
{Zethson} T.,  {Johansson} S.,  {Hartman} H.,   {Gull} T.~R.,  2012, \mn@doi
  [\aap] {10.1051/0004-6361/201116696}, \href
  {http://adsabs.harvard.edu/abs/2012A%26A...540A.133Z} {540, A133}

\bibitem[\protect\citeauthoryear{{de Graauw} et~al.,}{{de Graauw}
  et~al.}{1996}]{deGraauw96}
{de Graauw} T.,  et~al., 1996, \aap, \href
  {https://ui.adsabs.harvard.edu/abs/1996A&A...315L..49D} {315, L49}

\bibitem[\protect\citeauthoryear{{van der Tak}, {Black}, {Sch{\"o}ier},
  {Jansen}  \& {van Dishoeck}}{{van der Tak} et~al.}{2007}]{vanderTak07}
{van der Tak} F.~F.~S.,  {Black} J.~H.,  {Sch{\"o}ier} F.~L.,  {Jansen} D.~J.,
   {van Dishoeck} E.~F.,  2007, \mn@doi [\aap] {10.1051/0004-6361:20066820},
  \href {http://adsabs.harvard.edu/abs/2007A%26A...468..627V} {468, 627}

\makeatother
\end{thebibliography}


\appendix

\section{\spire\ spectrum and line identification}
\begin{figure*}
\includegraphics[angle=0, scale=.65]{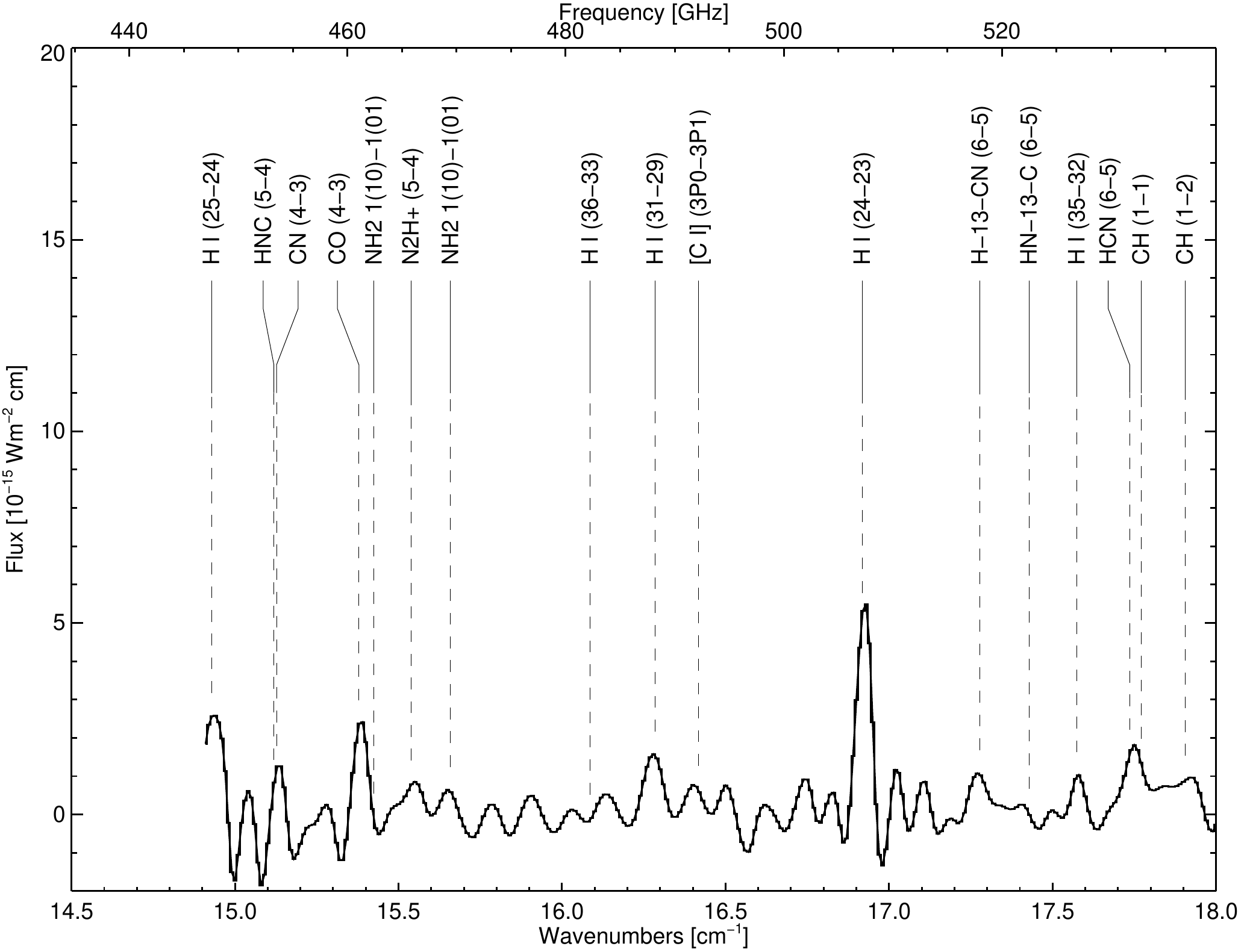}

\caption{\spire\ spectrum 14.5 to 18.0 cm$^{-1}$. As commented in section \ref{SPIRE_obs}, the  S/N exceeds 100 over most of the spectrum depicted in Figures \ref{Sfig1} through \ref{Sfig13}. Most of the features in these spectra are real and promise to reveal even more line identifications. Note that, for all of the following spectral plots, the strong continuum has been subtracted to provide an optimized baseline for line identification.}\label{Sfig1}\label{Sfig}
\end{figure*}

The complete \spire\ spectrum is  plotted in Figure \ref{Sfig} extending from 14.9 to 52.4 cm$^{-1}$. Note that the spectrum is less reliable beyond 51.5 cm$^{-1}$ with  the last line identification being at 51.4 cm$^{-1}$.

\begin{figure*}
\includegraphics[angle=0, scale=.65]{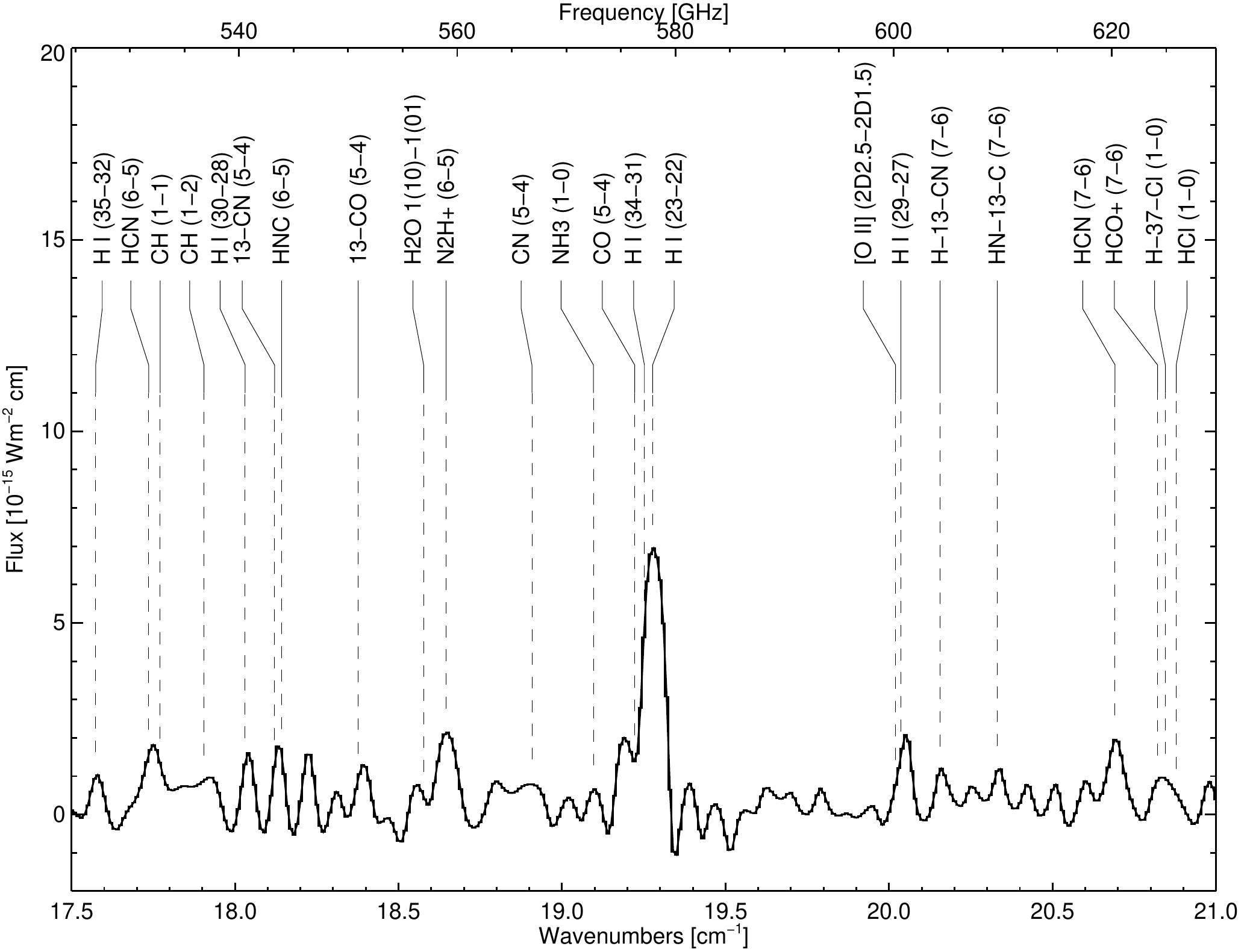}
\caption{\spire\ spectrum 17.5 to 21.0 cm$^{-1}$ \label{Sfig2}}
\end{figure*}
\begin{figure*}
\includegraphics[angle=0, scale=.65]{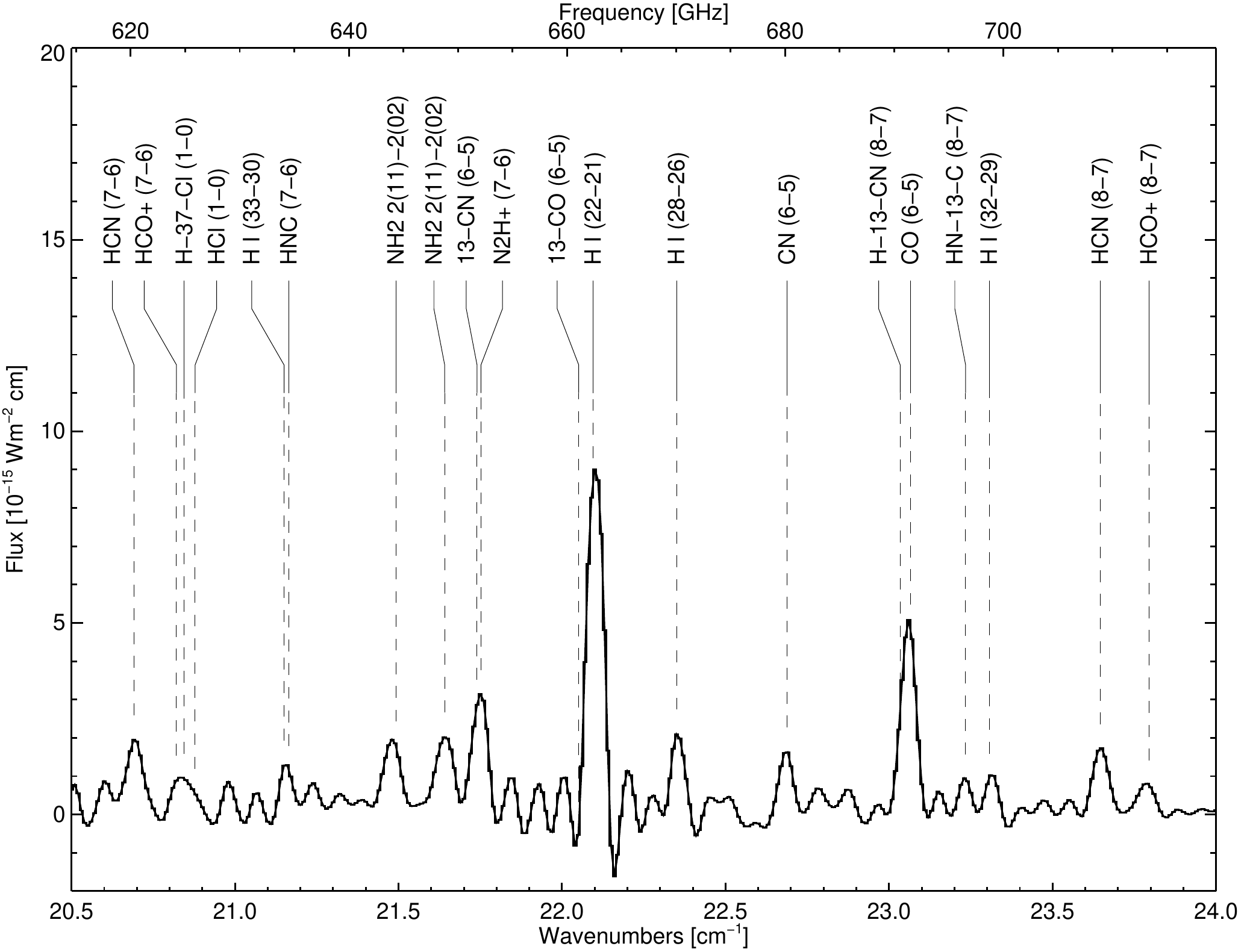}
\caption{\spire\ spectrum 20.5 to 24.0 cm$^{-1}$ \label{Sfig3}}
\end{figure*}
\begin{figure*}
\includegraphics[angle=0,scale=.65]{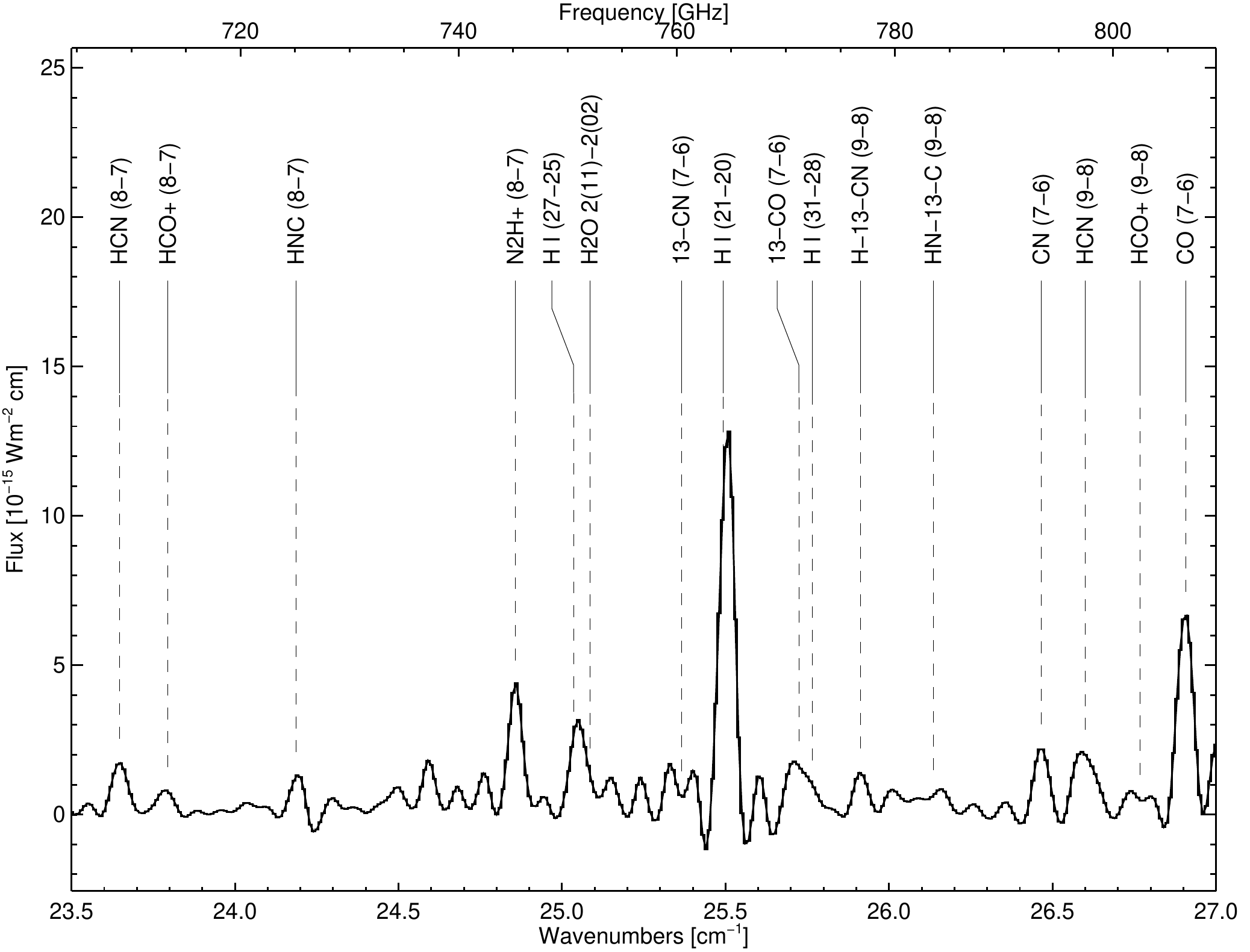}
\caption{\spire\ spectrum 23.5 to 27.0 cm$^{-1}$ \label{Sfig4}}
\end{figure*}
\begin{figure*}
\includegraphics[angle=0,scale=.65]{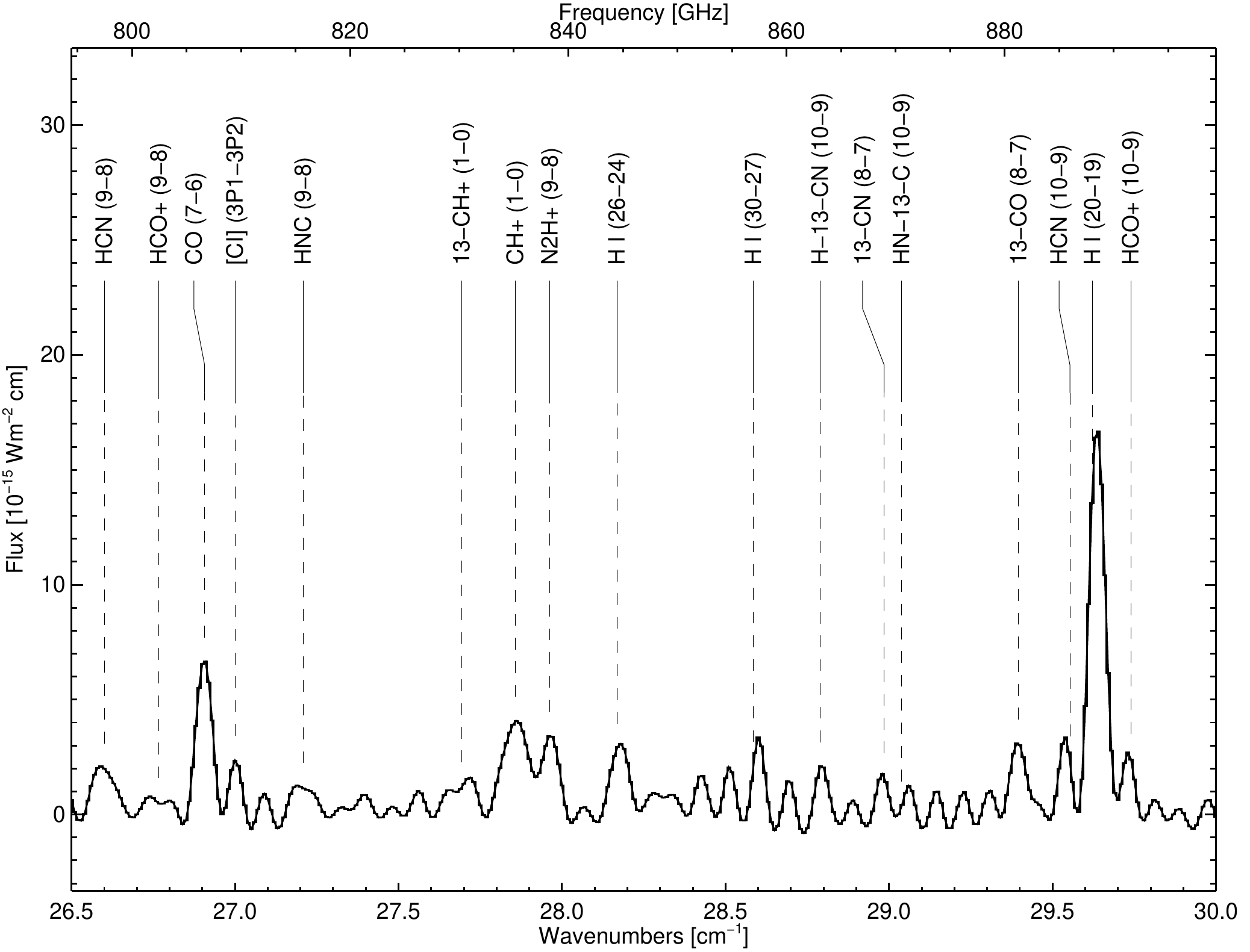}
\caption{\spire\ spectrum 26.5 to 30.0 cm$^{-1}$ \label{Sfig5}}
\end{figure*}
\begin{figure*}
\includegraphics[angle=0,scale=.65]{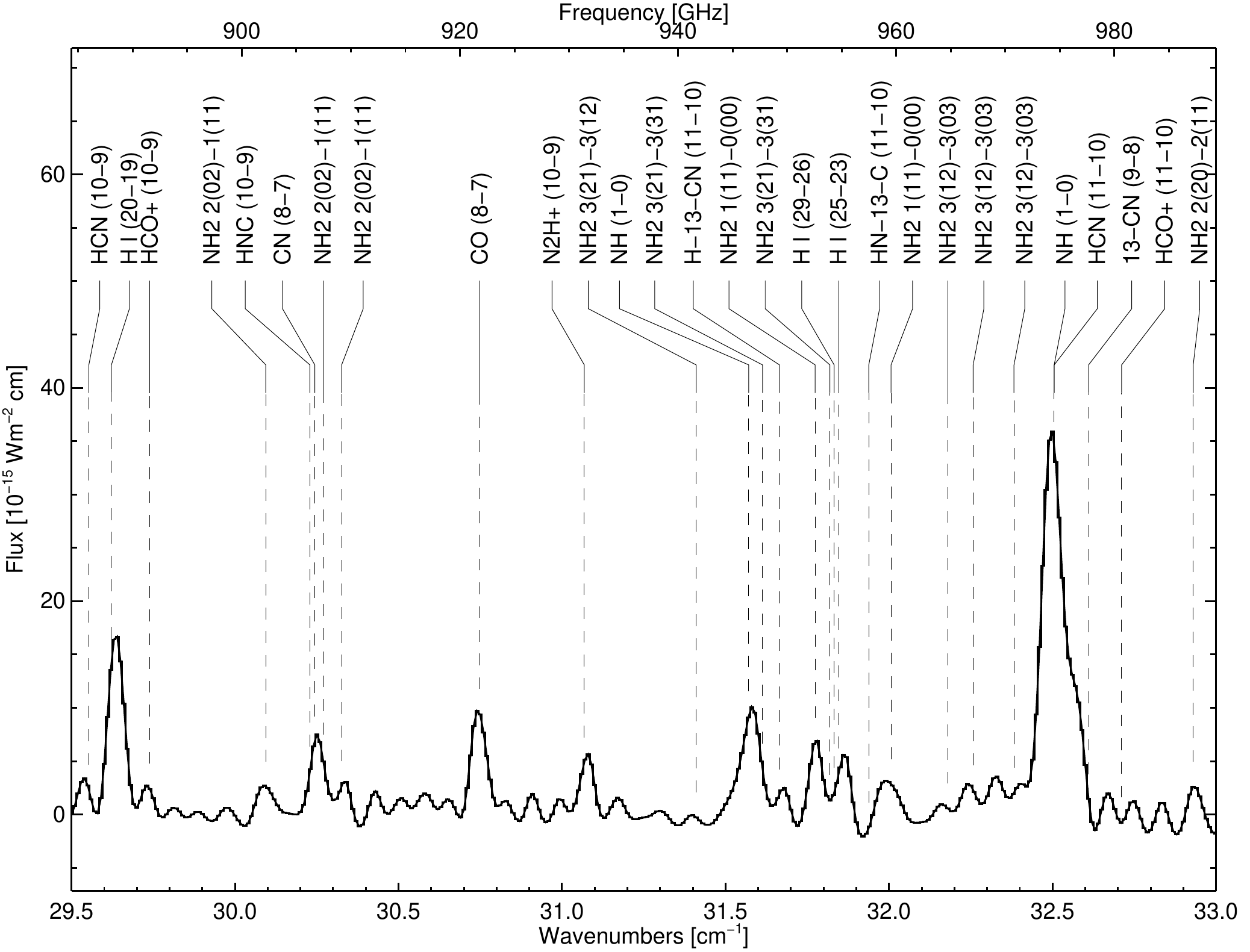}
\caption{\spire\ spectrum 29.5 to 33.0 cm$^{-1}$ \label{Sfig6}}
\end{figure*}
\begin{figure*}
\includegraphics[angle=0,scale=.65]{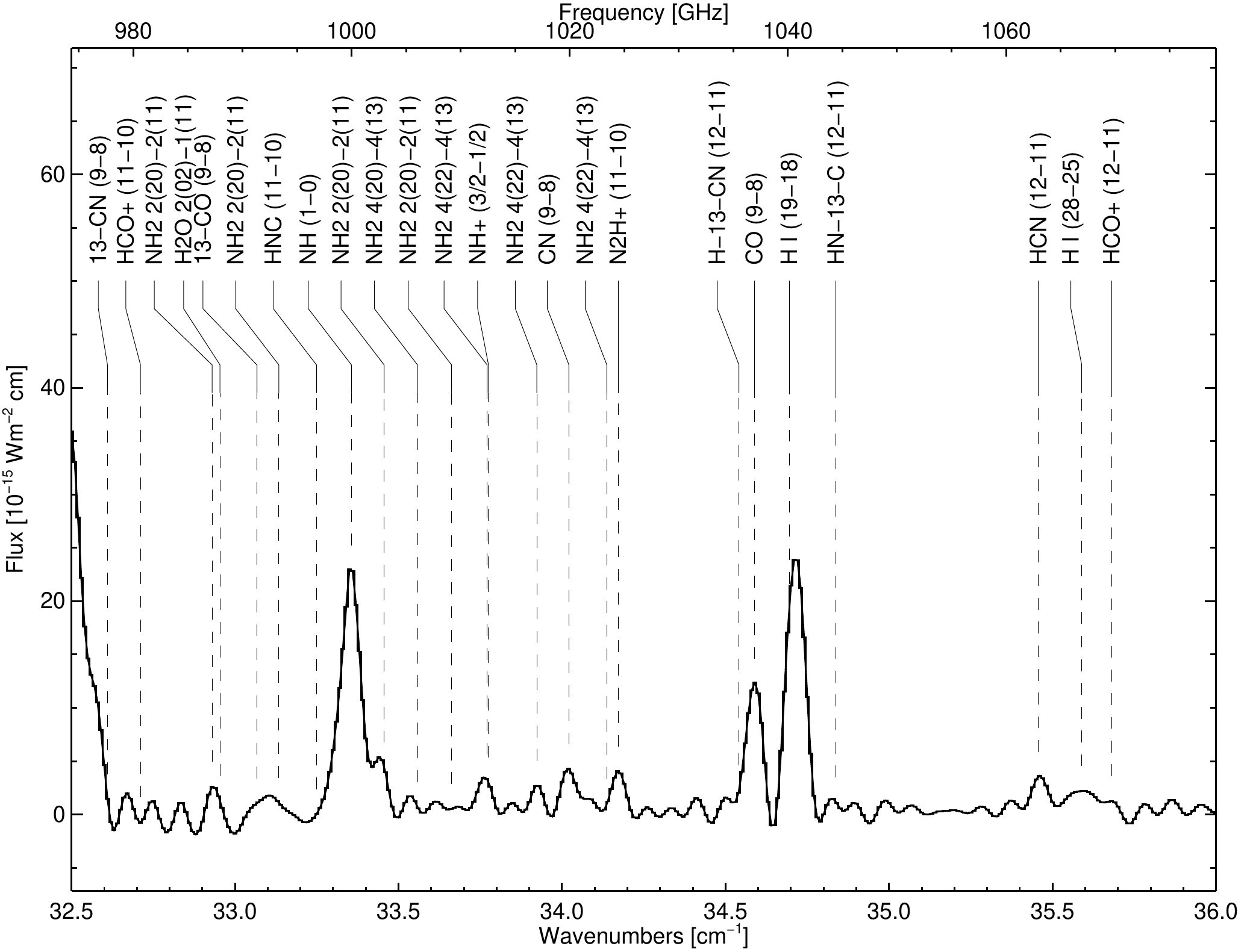}
\caption{\spire\ spectrum 32.5 to 36.0 cm$^{-1}$\label{Sfig7}}
\end{figure*}
\begin{figure*}
\includegraphics[angle=0,scale=.65]{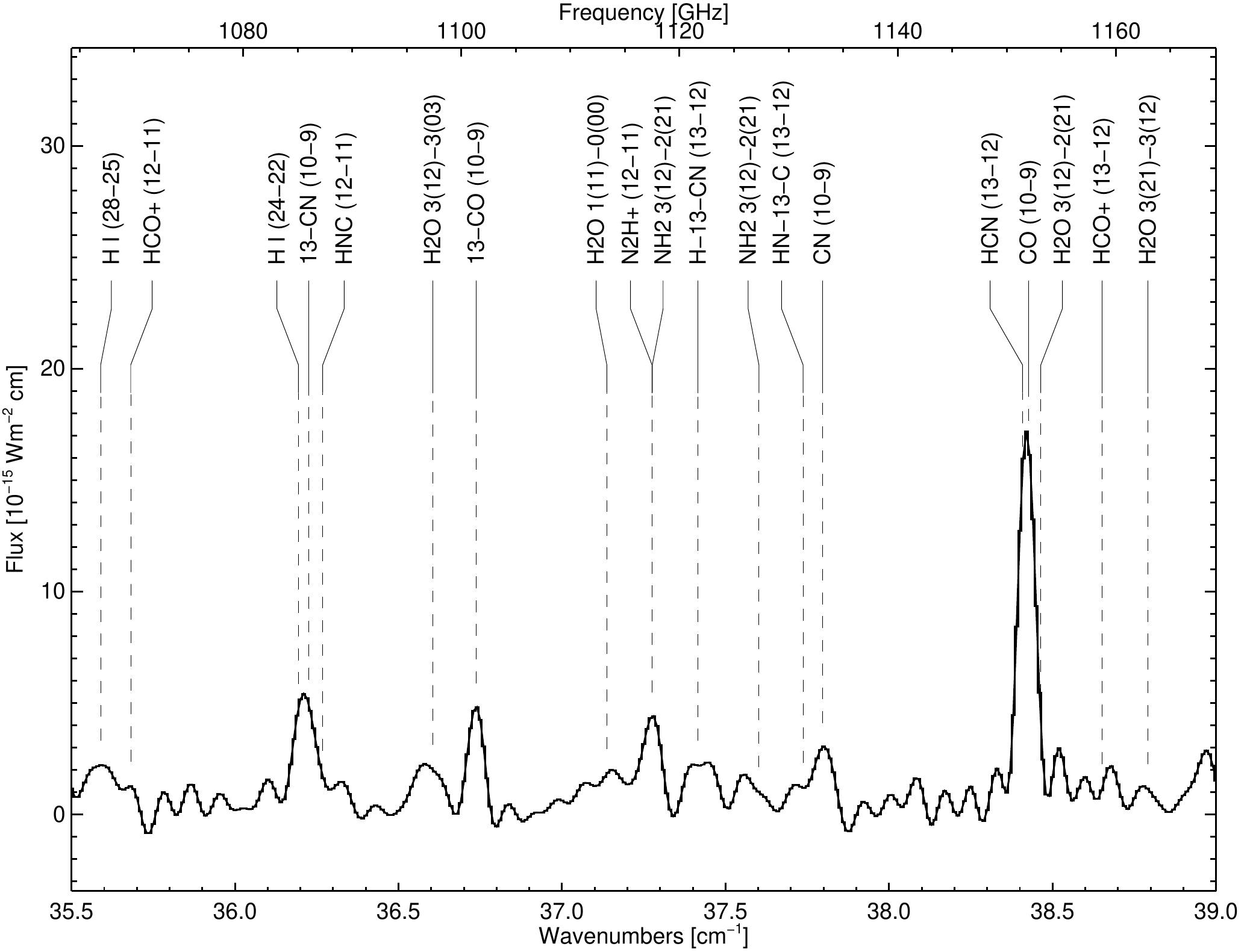}
\caption{\spire\ spectrum 35.5 to 39.0 cm$^{-1}$ \label{Sfig8}}
\end{figure*}
\begin{figure*}
\includegraphics[angle=0,scale=.65]{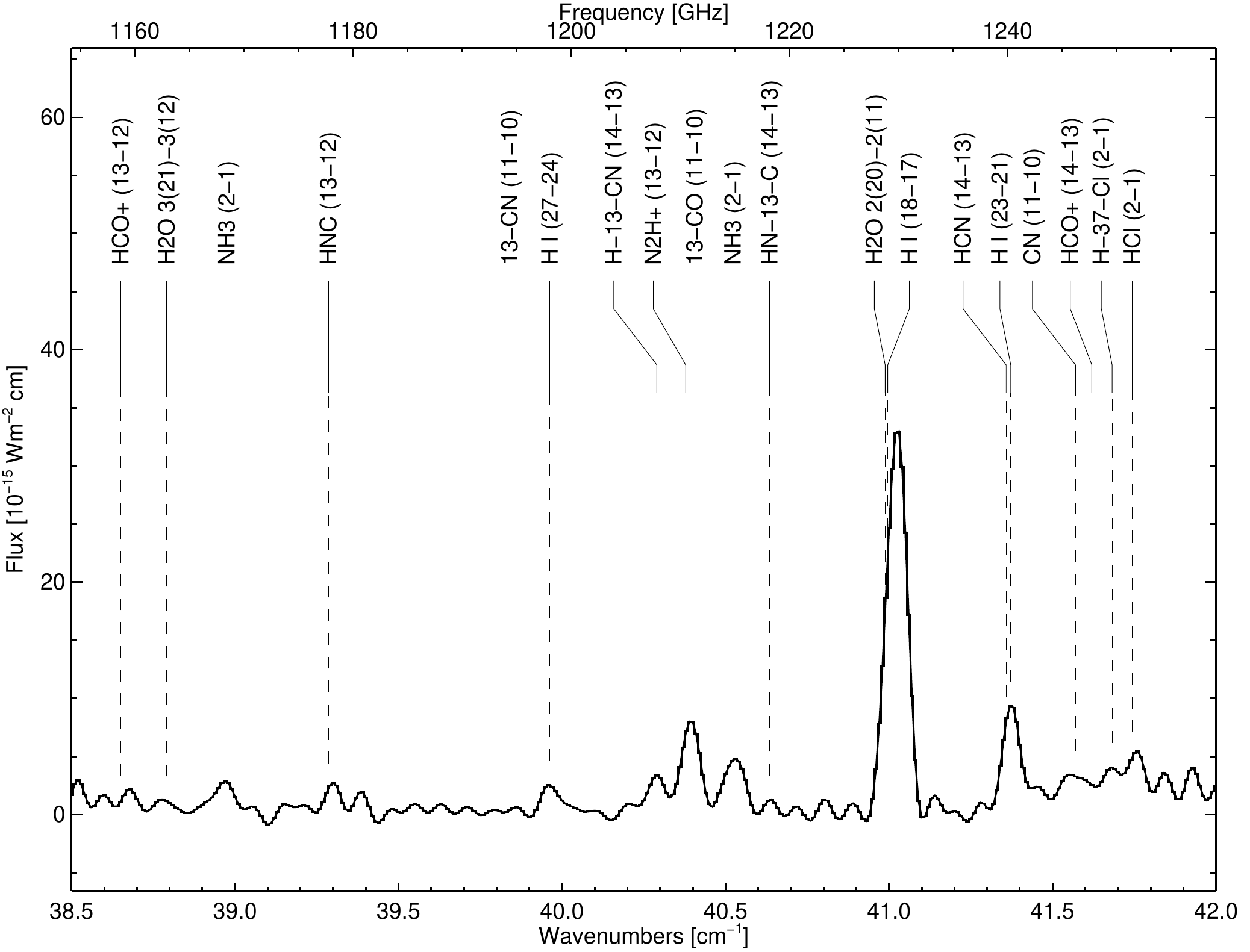}
\caption{\spire\ spectrum 38.5 to 42.0 cm$^{-1}$ \label{Sfig9}}
\end{figure*}
\begin{figure*}
\includegraphics[angle=0,scale=.65]{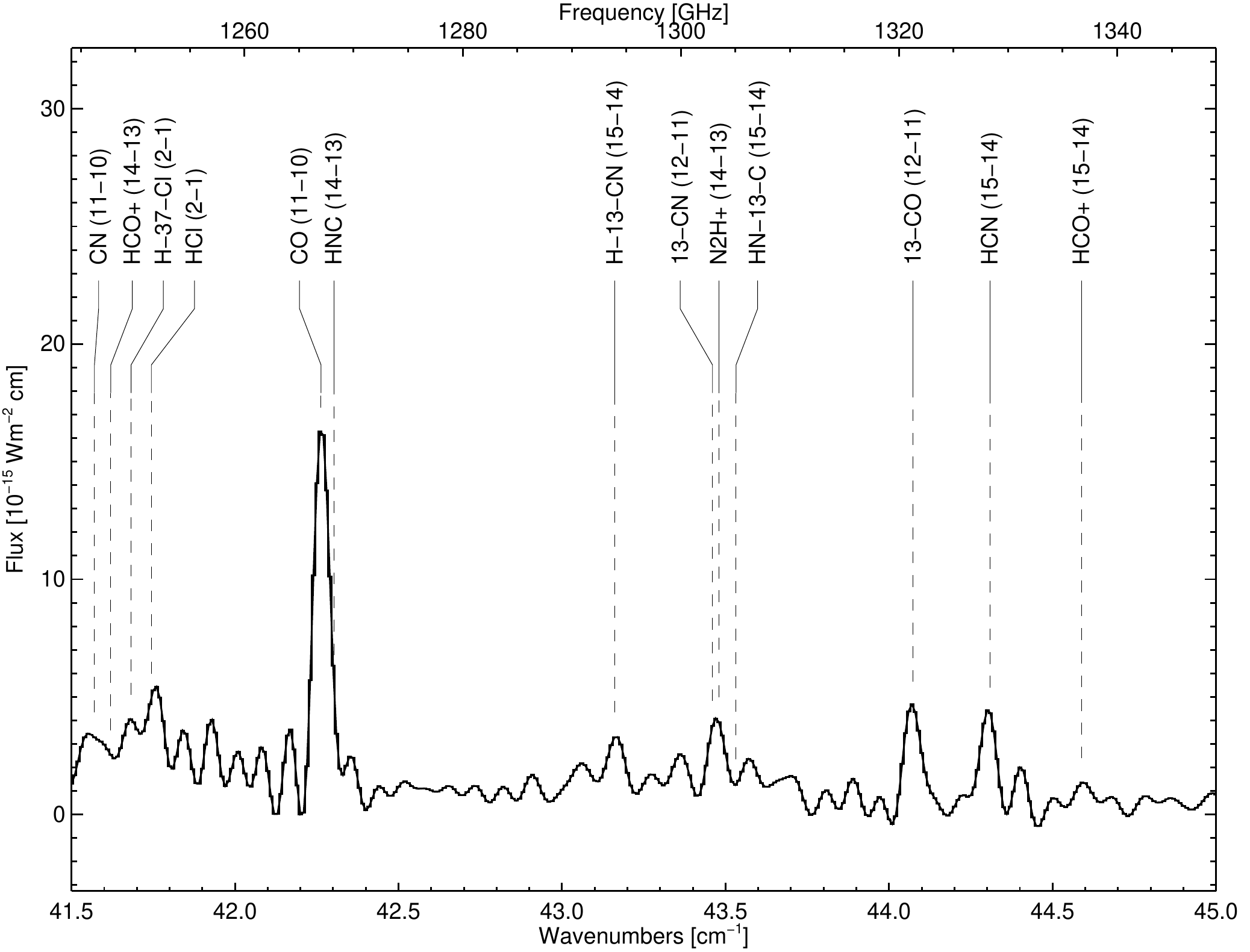}
\caption{\spire\ spectrum 41.5 to 45.0 cm$^{-1}$ \label{Sfig10}}
\end{figure*}
\begin{figure*}
\includegraphics[angle=0,scale=.65]{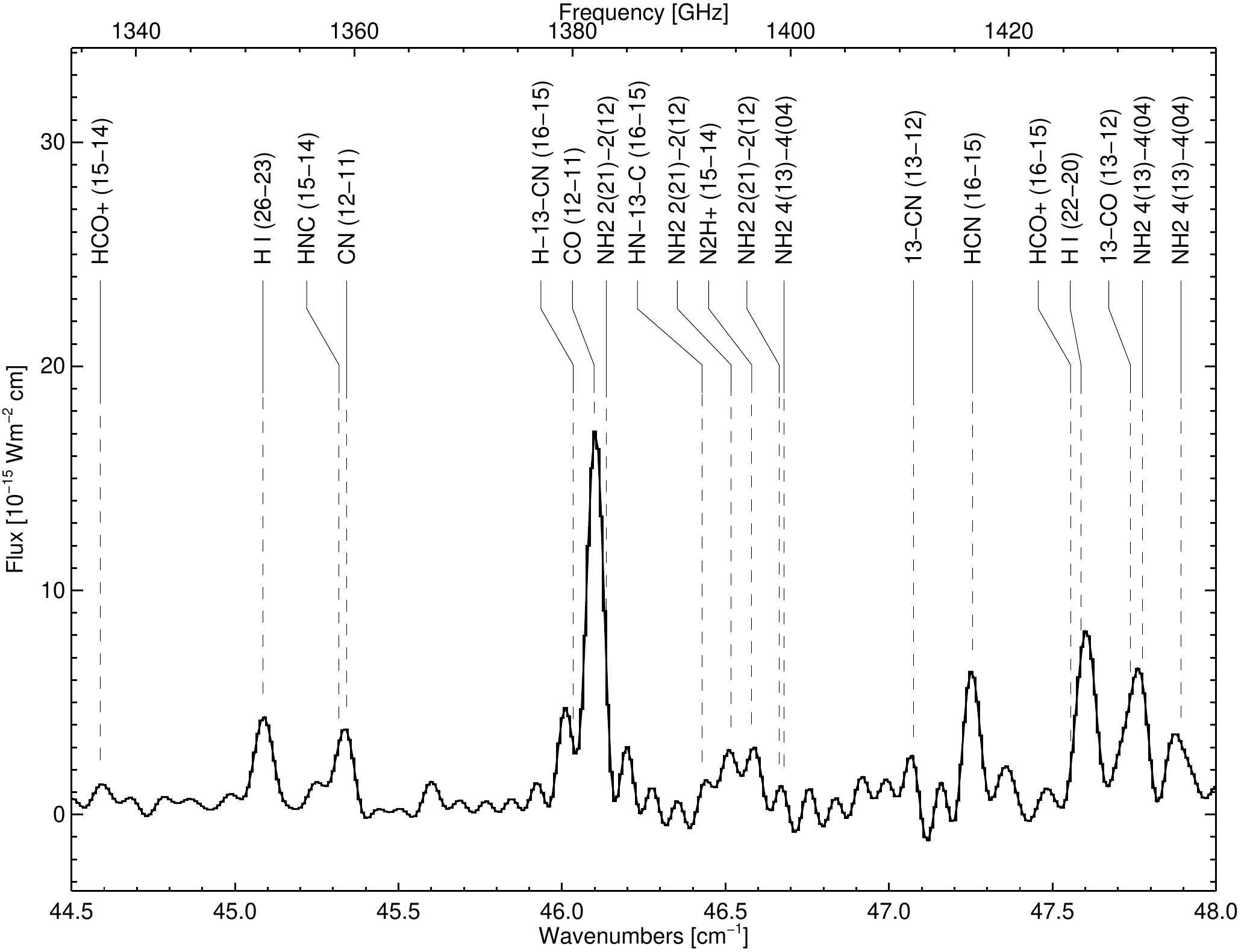}
\caption{\spire\ spectrum 44.5 to 48.0 cm$^{-1}$\label{Sfig11}}
\end{figure*}
\begin{figure*}
\includegraphics[angle=0,scale=.65]{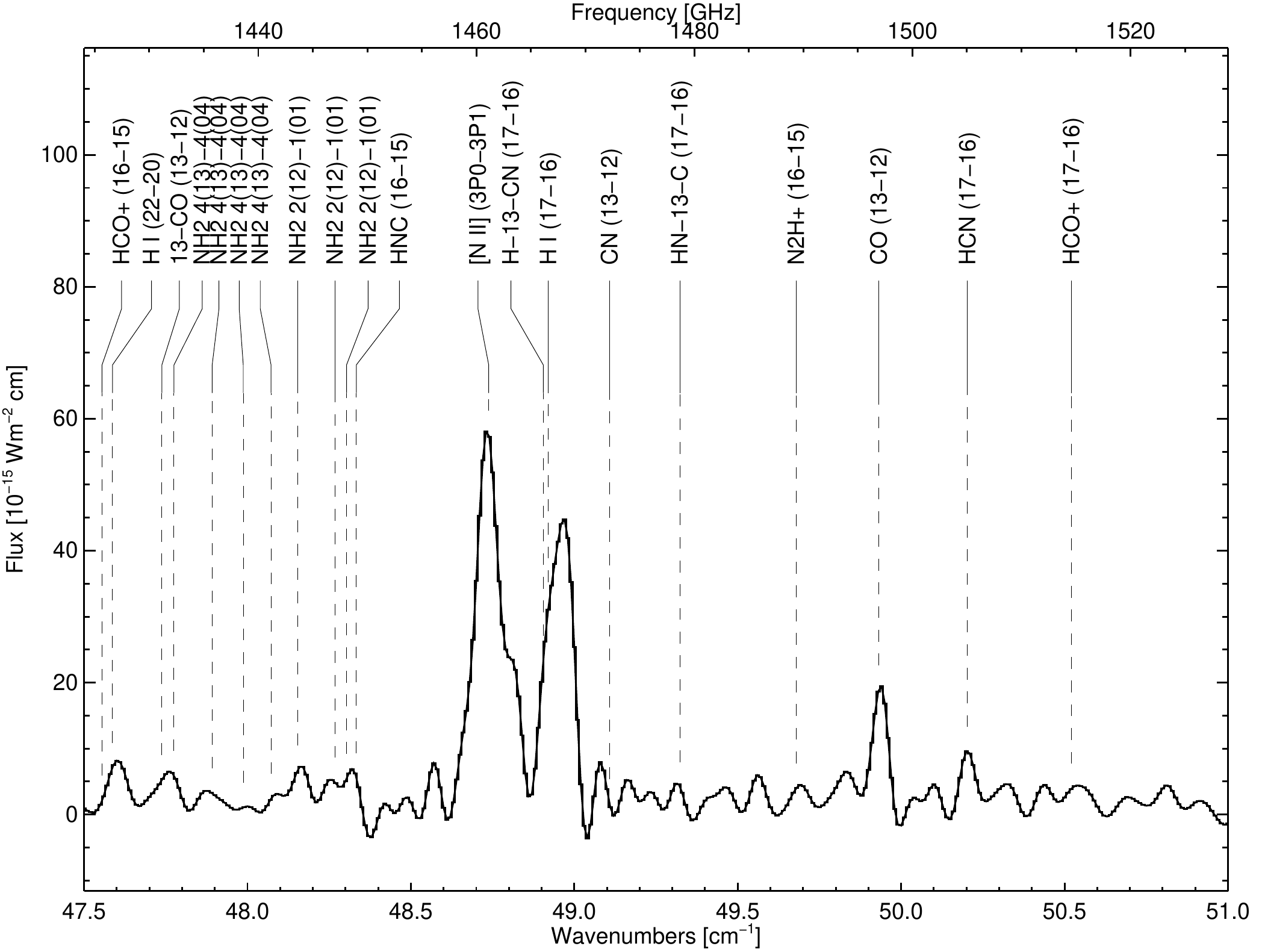}
\caption{\spire\ spectrum 47.5 to 51.0 cm$^{-1}$\label{Sfig12}}
\end{figure*}
\begin{figure*}
\includegraphics[angle=0,scale=.65]{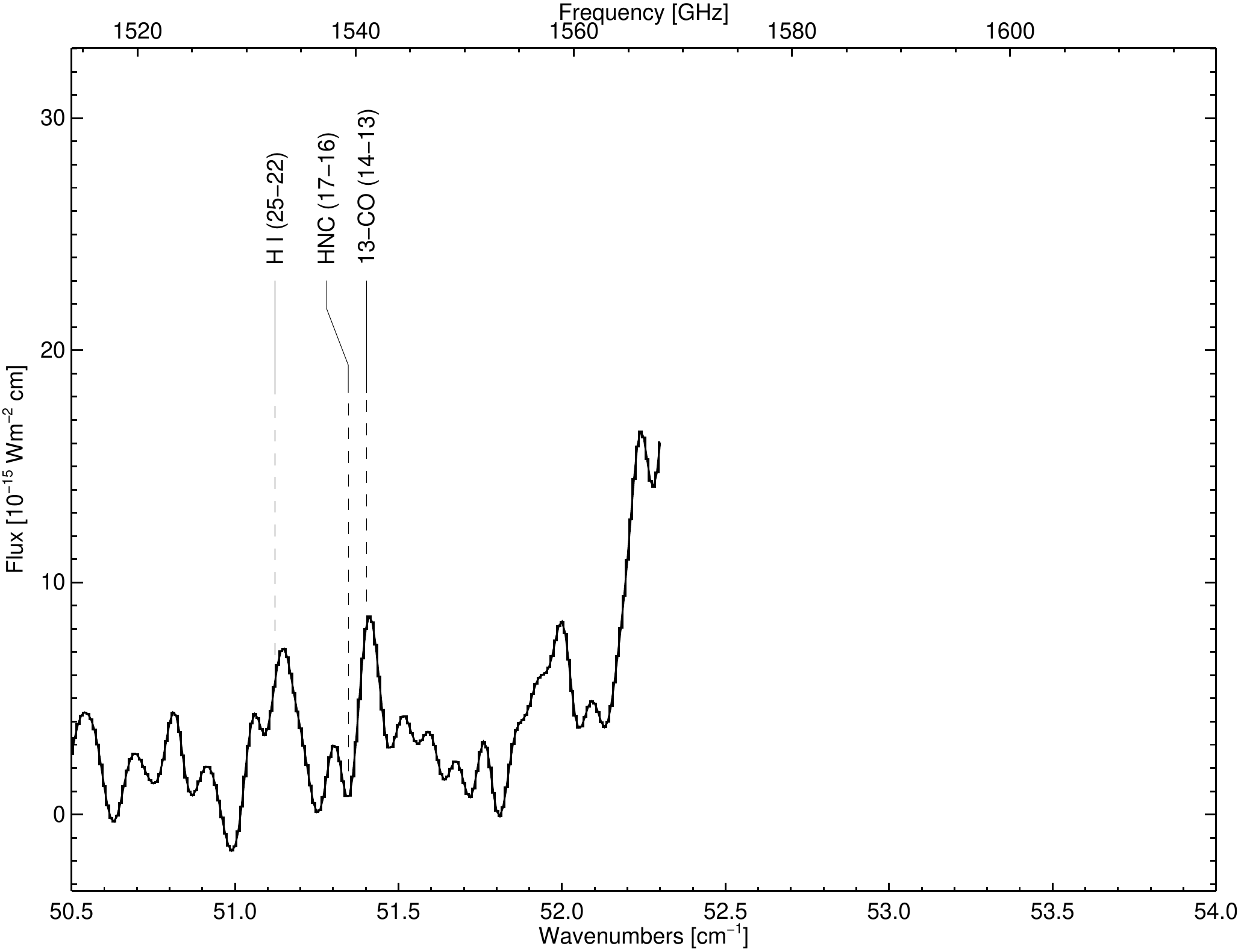}
\caption{\spire\ spectrum 50.5 to 54.0 cm$^{-1}$ Note that while the plot extends out to 52.3 cm$^{-1}$, the spectrum is not reliable beyond 51.5 cm$^{-1}$ with the last line tabulated in Table \ref{Tbl-SPIRE} at 51.4 cm$^{-1}$. \label{Sfig13}}
\end{figure*}

Table \ref{Tbl-SPIRE} lists line identifications and measures for the more prominent, relatively isolated emission features. While the S/N exceeds 100, the spectral  resolution of \spire\ and spatial resolution of \herschel\ limit line identifications. As noted in the introduction (Section \ref{Intro}) and demonstrated in the \pacs\ Data  Analysis (Subsection \ref{PACS}), known structures of the ejecta extend from the sub-arcsecond Weigelt clumps near \ec\ to the 4\arcsec\ cavity of the Butterfly Nebula, the 10\arcsec$ \times$\ 20\arcsec\ Homunculus to external fast moving knots at distances of 30\arcsec from \ec\ \citep{smith08b, Kiminki16}. The foreground lobe of the Homunculus is expanding at $-$600\kms\ and the background lobe at $+$ 600 \kms. Radio observations of the H I recombination lines \citep{Duncan03} constrain the ionized hydrogen region to 0.5\arcsec\ while the \pacs\ studies  of [C II] (subsection \ref{CIIS}) and NH (subsection \ref{NHS}) demonstrate these emissions originate from the Homunculus. 

Improved line identifications hence must wait for future observations of higher angular and spectroscopic resolutions that will reveal resolved structures and thus make possible greatly improved models of the complex ejecta.

The spectral nomenclature in Table \ref{Tbl-SPIRE} and Figure \ref{Sfig} for atomic transitions refers to the principal quantum numbers, n, of the transitions (Example: H I (21$-$20)).For molecular transitions a variety of notations is needed. When there is unresolved fine- and hyperfine structure, only the leading rotational or angular-momentum quantum numbers are listed (Example: NH$_2$ 3(12) $-$ 3(03).

Where blends are obvious, the measurement is placed with the line identification closest to the measured line and the apparently weaker line is noted to be blended. (Example  at 30.350 cm$^{-1}$ in Figure \ref{Sfig6} three lines appear to be blended. The wavelength closest to that measured is 30.244 cm$^{-1}$ identified as CN (8$-$7). The line identifications HCN (10$-$9) and NH$_2$ 2(02)$-$1(11) are listed as blended in the comments column. Fluxes are listed along with measurement error for most lines. Lines considered to be present, but too weak for accurate measure are noted as weak in the comment column.

\label{plotsid}

\onecolumn
\begin{longtable}{llllll}
\caption{Line Identifications from the \spire\ spectrum\label{Tbl-SPIRE}} \\
\hline \multicolumn{1}{l}{Wavenumber} & \multicolumn{1}{c}{Frequency} & \multicolumn{1}{l}{Identification}  & \multicolumn{1}{c}{Measured} & \multicolumn{1}{c}{Line Flux} & \multicolumn{1}{c}{Comments}  \\ 
\multicolumn{1}{c}{(cm$^{-1}$)} & \multicolumn{1}{c}{(GHz)}  & \multicolumn{1}{l}{(cm$^{-1}$)} & \multicolumn{1}{c}{(10$^{-17}$Wm$^{-2}$)} &  \\ \hline 
\endfirsthead

\multicolumn{6}{c}%
{{\tablename\ \thetable{} -- continued from previous page}} \\ \\
\hline \multicolumn{1}{c}{Wavenumber} & \multicolumn{1}{c}{Frequency} & \multicolumn{1}{c}{Identification} & \multicolumn{1}{c}{Measured} & \multicolumn{1}{c}{Line Flux} & \multicolumn{1}{c}{Comments}  \\ 
\multicolumn{1}{c}{(cm$^{-1}$)} & \multicolumn{1}{c}{(GHz)}  & \multicolumn{1}{c}{(cm$^{-1}$)} & \multicolumn{1}{c}{(10$^{-17}$Wm$^{-2}$)} & \\ \hline 
\endhead
\hline \multicolumn{6}{c}{{Continued on next page}} \\ \hline
\endfoot
\hline \hline
\endlastfoot
\input{SPIRE_TableA1_revised_v2.out}
\end{longtable}


\bsp	
\label{lastpage}
\end{document}